%% file: thesis.tex
\documentclass[11pt,english,a4paper] {book}
\usepackage{latexsym,epsfig,cite,amsbsy}

\input{defs.tex}			
\input{layout.tex}			

\begin{document}
\pagenumbering{roman}

\input{titlepage.tex}


\input{foreword.tex}			

\tableofcontents

\chapter{Introduction}
\label{chap:introduction}
\pagenumbering{arabic}
\input{intro.tex}
\input{motivation.tex}			
\input{lagrangeformalism.tex}		
\input{symmetries.tex}			
\input{energy-momentum.tex}		
\input{methods.tex}			

\chapter{Bosonic strings}
\label{chap:bosonic-string}
\input{strings.tex}			
\input{pointparticle.tex}		
\input{NGstring.tex}			
\input{pbrane.tex}			
\input{weylstring.tex}			

\chapter{D-branes}
\input{Dbrane.tex}			

\chapter{Rigid strings}
\label{chap:rigidstring}
\input{rigidstring.tex}			

\chapter{General relativity}
\input{GRgravitation.tex}		

\chapter{Other models}
\label{chap:othermodels}
\input{ym.tex}				
\input{cs.tex}				

\chapter{Conclusion}
\label{chap:conclusion}
\input{conclusion.tex}			

\appendix
\chapter*{Appendix}
\addcontentsline{toc}{chapter}{Appendix}
\addtocounter{chapter}{+1}		
\setcounter{equation}{0}
\chaptermark{Appendix}

\input{densities.tex}			
\input{determinants.tex}		
\input{vielbeins.tex}			
\input{lie.tex}				
\input{curvature.tex}			
 
\raggedright

\bibliographystyle{unsrt}
\bibliography{biblio.bib}

\end{document}

%% file: defs.tex
\newcommand{\nc}{\newcommand}
\nc{\beq}{\begin{equation}}
\nc{\beql}[1]{\begin{equation} \label{eq:#1}}
\nc{\eeq}{\end{equation}}
\nc{\rarr} {\rightarrow}
\nc{\larr} {\leftarrow}
\nc{\Rarr} {\Rightarrow}
\nc{\Larr} {\Leftarrow}
\nc{\lrarr} {\leftrightarrow}
\nc{\lbeq}[1]  {\label{eq:#1}}
\nc{\refeq}[1] {(\ref{eq:#1})}
\def\del{\partial}
\def\12{\frac{1}{2}} 
\nc{\air} {\medskip \noindent}
\nc{\form}[1] {\underline{#1}}

%% file: layout.tex
\usepackage{fancyheadings}

\renewcommand{\chaptermark}[1]{\markboth{\thechapter. #1}{}}

\makeatother

%% file: titlepage.tex

\begin{titlepage}

$ $

\vspace{2cm}

\begin{center}
{\Huge \bf High energy limits}\\
\vspace{0.25cm}
{\Huge \bf of various actions}
\end{center}

\vspace{1.6cm}

\begin{center}
{\large \bf Harald G Svendsen}
\end{center}

\vspace{6cm}

\begin{center}
\mbox{\psfig{figure=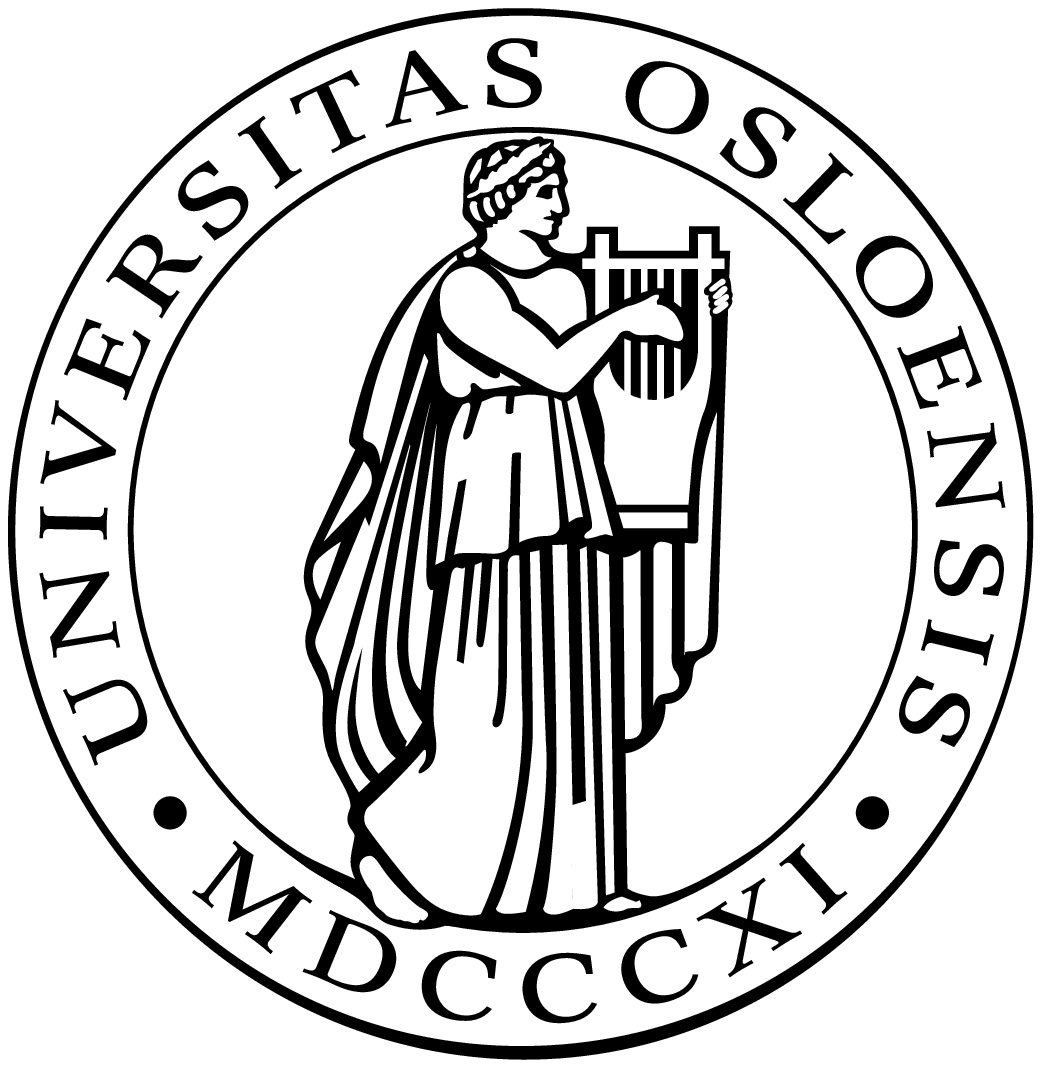,width=4cm,clip=f}}

{\bf Thesis submitted for the degree of Candidatus Scientiarum}\\
{\bf Department of Physics\\
University of Oslo\\
June 1999}

\end{center}

\end{titlepage}

%% file: foreword.tex
\chapter*{Preface}

When people ask me what I \emph{do}, or what kind of subjects I study,
I have sometimes ironically answered that I do abstract art. This
thesis does not have \emph{practical} applications, in the sense
that it can tell you how to wash your clothes, make coffee, or help you
to calculate numbers that can in turn be measured with some apparatus. 
The applications are theoretical -- the thesis
casts some light on mathematical relations within a theoretical
framework. And the results that are obtained are purely analytical.

\emph{Some} people demand from physics that it shall be concerned with
observable things. With such a narrow view of the field this
thesis can hardly be called a work of physics. But at the same time,
I would guess that mathematicians would be horrified by the lack of
stringency, so calling it a piece of mathematics is no less
dangerous.

It is on this basis that I like to call it abstract art. Its value
lies in the aesthetics. One kind of art has an immediate value because
of its beauty to the eye or ear, or because of its direct
associations. Another kind of art is more indirect. Its value may
require some background information, and lies more in the meaning than
in the sensation. 
For me the beauty in the abstract art of physics is of this
kind, and lies in the understanding of certain mathematical relations
-- that in turn are somehow related to the world we live in.

This aesthetics is my prior motivation for the interest in theoretical
physics. I cannot claim to do what I do because I think it will make
the world a better place. It is beauty more than importance that
attracts me. 

\air
For people not so fascinated by art and aesthetical values, I want to
emphasize that this concerns only my subjectice motivation. What is
actually done, the derivations and interpretations, shall of course
satisfy the standards of reliable science.

\subsection*{Acknowledgments}
I would like to express my sincere thanks to Ulf Lindstr\"om,
who gave me this project, and has been a pleasant collaborator. I
also thank my other helpful teachers and fellow students both in Oslo
and during my stay in Stockholm.  
\\

\noindent
Harald G Svendsen
\\June 1999

%% file: intro.tex
The goal of this thesis is twofold. First, it is meant to give a
thorough presentation of two methods for deriving tensionless limits
of strings, and the analogue in other models. Second, the applicability
of these methods are investigated by explicitly going through the
calculations for a variety of models.

We start by going through some background theory in this introductory
chapter about constrained
Hamiltonian systems, and different symmetry considerations. A general
result 
concerning the Hamiltonian of \emph{diff} invariant theories is derived in
section \ref{sec:zerohamilton}. Then the
methods are presented within the general picture, before the simplest 
examples are given in chapter \ref{chap:bosonic-string}. In the
subsequent chapters we study $D$-branes, rigid strings, general
relativity and take a brief look on Yang-Mills and Chern-Simons
theory. A summary and concluding remarks are finally given in chapter
\ref{chap:conclusion}.

\air
Throughout this thesis we use the standard summation notation
$a^ib_i\equiv \sum_i a^ib_i$, and natural units where $c=1$.

%% file: motivation.tex

\section{Motivations}

The question naturally arises of \emph{why} we should bother with the
kind of limits that we are going to study in this thesis.
Do not particles 
have mass, and strings tension? The question is of course
crucial, and worthy of an answer.
If we were not able to give one,
the present undertaking would seem like a meaningless activity beyond
any interest apart from the purely academical one.

But those who are anxious can safely relax: There \emph{is} a
motivation. Actually, there are several different aspects of these
limits that are of interest. 

\paragraph{The limits represent physical situations}

If we start from the action of a massive particle and derive the massless
limit, we will end up with a theory for massless particles describing e.g.
photons (if we disregard spin and charge). And their existence
at least is without doubt.  Whether tensionless strings  also
have direct physical applications is not as easy to tell since strings 
themselves are not yet sufficiently well understood.

But if we want to test the existence of tensionless strings
(or analogous limits of other theories) we need to have a theory
for them. And the methods we will use are ways to arrive at such 
theories.

\paragraph{High energy limit}

For high energies the mass of a particle becomes unimportant compared to
its kinetic energy. 
Therefore the massless limit gives an approximation of the behaviour
at high energies.
The same is supposed to be true for strings: 
The tensionless limit may be viewed as a high energy limit.
And high energy physics is important for 
several reasons. When we study systems at a very small scale we
unavoidably (by the Heisenberg principle) have systems at high energy.
Also, at the earliest stages of the evolution of the universe, the
energy density was very high. An understanding of the ``childhood'' of
the universe thus requires an understanding of high energy physics.

\paragraph{Conformal invariance}

Another characteristic of these limits is that they lead us to
conformally invariant theories. Those are theories with a higher degree
of symmetry and interesting in their own right.

\air
And then of course we still have the purely academical reason to
investigate just for the investigation in itself. For, who knows, we
may find something interesting and important. Or we may at least get
new insight into known theories.

%% file: lagrangeformalism.tex
\section{Lagrange and Hamilton formalism}

When it comes to solving simple problems in classical mechanics, the
formulation by Newton is usually the most natural machinery to
use. However, for analytical discussions the
formalism developed by 18th century
physicists like Lagrange and Hamilton has shown much more
fruitful. 
Since this formalism will be used extensively throughout this
thesis, we begin by giving a short review. A more thorough
introduction is found in textbooks on field theory, e.g.
\cite{froyland:1996,low:1997,soper:1976,weinberg:1995}.

\subsection{Lagrange formalism}

The starting point in this formalism is the \emph{Lagrangian density}
(in the following just called the Lagrangian or Lagrange function),
which is a function
of some fields $\phi^i(x^a)$ and their derivatives 
$\partial_a \phi^i=\frac{\del\phi^i}{\del x^a}$,
\beq 
  L = L(\phi^i, \partial_a \phi^i). 
\eeq
The index $i$ is used, when necessary, to distinguish the different
fields in the Lagrangian. $a$ is an index running over all coordinates. 
$a=0$ denotes the time coordinate, and
we write time derivatives as $\partial_0 \phi = \dot{\phi}$. For 
spatial ($a>0$) derivatives we write $\nabla \phi$. This convention will be
convenient when we work in the Hamilton formalism, since time has
a special role there.

The above Lagrangian is written as a function of first order
derivatives only, and hence we call it a \emph{first order} Lagrangian. 
It is also possible to allow for higher order
derivatives in the Lagrangian, but such Lagrangians can usually be
reduced to first order ones by introduction of extra fields, as discussed
in \cite{lindstrom:1988}. 

{}From the Lagrange function we construct the \emph{action} as the 
integral over the configuration space,
\beql{action}
  S[\phi^i] = \int dx L(\phi^i(x), \partial_a \phi^i(x)),
\eeq
where $x=\{x^a\}$ is usually a set of coordinates, and the integral is
then many-dimensional.
The action integral is a \emph{functional} of the fields $\phi^i(x)$,
which means that it takes functions to numbers (in contrast to
functions, which take numbers to numbers.)
Suppose we make small variations in the arguments of the Lagrangian. In 
other words, consider the transformation 
$\phi^i \to \phi^i + \delta \phi^i$, with $\delta\phi^i=0$ on the
boundaries.  
The values of $\phi^i$ that represent the dynamical behaviour of the 
classical system
are the ones that leave the action unchanged under such infinitesimal
variations. This is the \emph{principle of extremal action}, also
called \emph{Hamilton's principle}.

If we consider a variation $\delta\phi^i$ in the fields, and demand
the action to be extremal (i.e. zero under the transformation), we
arrive at some equations which we
call the \emph{field equations}, \emph{equations of motion} or
\emph{Euler-Lagrange equations}. For Lagrangians that have only first order
derivatives, they can be deduced quite simply as follows.
\beq
  \delta\phi^i \Rarr \delta S = \int dx \left( 
	\frac{\del L}{\del\phi^i}\delta\phi^i
	+\frac{\del L}{\del(\del_a\phi^i)}\delta(\del_a\phi^i)
	\right).
\eeq
We can change the order of variation and derivation and write
$\delta(\del_a\phi^i) = \del_a(\delta\phi^i)$. Together with a partial
integration this gives
\begin{eqnarray}
  \nonumber
  \delta S &=& \int dx \left(
	\frac{\del L}{\del\phi^i}\delta\phi^i
	+\del_a(\frac{\del L}{\del(\del_a\phi^i)}\delta\phi^i)
	-\del_a(\frac{\del L}{\del(\del_a\phi^i)})\delta\phi^i
	\right)
  \\ \label{eq:euler-lagrange1}
  &=& \int dx \left(
	\frac{\del L}{\del\phi^i}
	-\del_a(\frac{\del L}{\del(\del_a\phi^i)}) \right)
	\delta\phi^i,
\end{eqnarray}
where we have disregarded the total derivative. This is allowed since
it will only give rise to a boundary (surface) term
that vanishes since $\delta\phi^i$ are then zero.
Demanding $\delta S=0$ for arbitrary infinitesimal variations
$\delta\phi^i$ we find the field equations,
\beql{euler-lagrange2}
  \psi_i\equiv 
	\del_a (\frac{\del L}{\del(\del_a \phi^i)}) 
	- \frac{\del L}{\del \phi^i} = 0.
\eeq
The fields that satisfy the equations of motion are said to span the
\emph{classical path}.

As long as we are in the classical domain of physics, the action
integral is just a convenient and compact notation that contains the
field equations and the symmetries of the theory. The introduction of
this formalism is, however, crucial when we want to do quantum
mechanics. Classically, two actions that gives rise to the same
equations of motion are equivalent, but this is not true quantum
mechanically. Classically equivalent actions normally lead to
different quantum physics.
This fact may serve as a justification of our eagerness to find classically
equivalent actions. 

One of the major advantages of this formalism is that many symmetries 
are manifest in the Lagrangian. And this is also of great help when
we want to write down the Lagrange function in the first place: It must
have a form that satisfies the symmetries of the theory under consideration.

There is generally no way to \emph{deduce} the Lagrangian. But that is
no weakness of the formalism. Any theory needs a starting point, and
in this formalism we start with the Lagrangian,
and take the extremum of the action as a first principle. The right
attitude is not to try to derive a Lagrangian, but to argue from
general principles (e.g. symmetry) and analogies with other theories
that it should take some specific form.

\subsubsection{Variational derivatives}

Consider a function $F$ constructed from some fields $\phi^i(x)$ and
their first order derivatives, $F=F(\phi,\del\phi)$. 
(The typical example is the Lagrangian.)
Variations in $\phi^i(x)$ will then give a variation in $F(x)$ that we
can write
\begin{eqnarray}
  \delta\phi(x) \Rarr \quad \delta F(x) 
	&=&\int d\tilde{x}\left[ \frac{\del F(x)}{\del\phi^i(\tilde{x})}
	-\del_a\frac{\del F(x)}{\del(\del_a\phi^i(\tilde{x}))} 
	\right]\delta\phi^i(\tilde{x})
  \\
	&\equiv&\int d\tilde{x} \frac{\delta F(x)}{\delta\phi^i(\tilde{x})}
	\delta\phi^i(\tilde{x}).
\end{eqnarray}
The \emph{variational derivative} 
$\frac{\delta F(x)}{\delta\phi^i(\tilde{x})}$ of $F$
gives the contribution to the variation of $F(x)$ from a variation of
$\phi^i(\tilde{x})$. To get the total variation of $F(x)$ we have to sum
over the discrete index $i$, and integrate over the continuous
parameter $\tilde{x}$.

By comparison with \refeq{euler-lagrange1} we see that the
Euler-Lagrange equations can be written by means of the variational
derivative as
\beq
   \frac{\delta L}{\delta\phi^i} = 0.
\eeq

\subsection{Hamilton formalism}
\label{sec:hamilton}

If we already have the Lagrangian, we define the canonical conjugate
\emph{momentum (density)} associated with $\phi^i$ as
\beql{defgeneralmomentum}
   \pi_i = \frac{\partial L}{\partial \dot{\phi}^i},
\eeq
where, again, dot denotes differentiation with respect to the time
parameter.  

With the definition \refeq{defgeneralmomentum} we can perform a
\emph{Legendre transformation} from configuration space to phase
space. First, define the quantity $h$ by
\begin{equation}
  h(\phi, \partial \phi, \pi) = \pi_i \dot{\phi}^i - L(\phi, \partial \phi).
\end{equation}
A variation in $h$ can then be written
\[
 \delta h = \pi_i \delta \dot{\phi}^i + \dot{\phi}^i \delta \pi_i 
            - \frac{\partial L}{\partial \phi^i} \delta \phi^i
            - \frac{\partial L}{\partial \dot{\phi}^i} \delta \dot{\phi}^i
            - \frac{\partial L}{\partial \nabla \phi^i} \delta \nabla \phi^i.
\]
Substitute for the definition of $\pi$ and arrive at
\[
 \delta h = \dot{\phi}^i \delta \pi_i
            - \frac{\partial L}{\partial \phi^i} \delta \phi^i
            - \frac{\partial L}{\partial \nabla \phi^i} \delta \nabla \phi^i.
\]
The variation $\delta h$ can be written by means of variations in $\phi$,
$\pi$ and $\nabla \phi$. This means that we can express $h=\pi\dot{\phi}-L$
as a function of these variables, omitting $\dot{\phi}$. This is true
even though \refeq{defgeneralmomentum} may not always be possible to solve
explicitly for $\dot{\phi}^i$.
Written this way we call $h$ the Hamiltonian, but now denoted with a 
capital $H$,
\beq
  H = H(\phi,\pi,\nabla\phi) = \pi_i\dot{\phi}^i-L(\phi,\del\phi)
\eeq
The transformation from $L$ to $H$ is a Legendre transformation.
We will call the Hamiltonian obtained in this way  the \emph{naive
Hamiltonian},  $H_{naive}$, to distinguish it from
the total Hamiltonian which we introduce in the next section.

Hamilton's modified principle states that the phase space action,
\begin{equation}
  S^{PS} = \int dx(\pi_i \dot{\phi}^i - H(\phi, \pi, \nabla \phi)),
\end{equation}
is unchanged under variations of $\pi$ and $\phi$, 
which are now considered independent of each other. We find\footnote{
We use a trick and write 
$\dot{\phi}^i(x)\delta\pi_i(x) = \int d\tilde{x}\delta(x-\tilde{x})
\dot{\phi}^i(x)\delta\pi_i(\tilde{x})$ (and the same for
$\dot{\pi}_i(x)\delta\phi^i(x)$).}
\begin{eqnarray}
  \nonumber
  \delta S^{PS} &=&\int dx \bigg[ 
	\int d\tilde{x}\delta(x-\tilde{x})
	\Big(\dot{\phi}^i(x)\delta\pi_i(\tilde{x})
	-\dot{\pi}_i(x)\delta\phi(\tilde{x}) \Big)
	\\[-0.1cm] \nonumber && \qquad
	-\int d\tilde{x} \Big( 
	\frac{\delta H(x)}{\delta\phi^i(\tilde{x})}\delta\phi^i(\tilde{x})
	+\frac{\delta H(x)}{\delta\pi_i(\tilde{x})}\delta\pi_i(\tilde{x})
	\Big) \bigg]
  \\[0.1cm] \nonumber&=&
	\int d\tilde{x}\bigg[
	\Big(\dot{\phi}^i(x) - \int dx
	\frac{\delta H(x)}{\delta\pi_i(\tilde{x})} 
	\Big) \delta\pi_i(\tilde{x})
	\\[-0.1cm]&& \qquad
	+\Big(-\dot{\pi}_i(x) - \int dx
	\frac{\delta H(x)}{\delta\phi^i(\tilde{x})}
	\Big) \delta\phi^i(\tilde{x}) \bigg].
\end{eqnarray}
Using Hamilton's modified principle $\delta S^{PS}=0$, this 
leads to
\emph{Hamilton's equations}:
\begin{eqnarray}
  \dot{\phi}^i(x) &=& \int dx' 
	\frac{\delta H(x')}{\delta\pi_i(x)},
  \\
  -\dot{\pi}_i(x) &=& \int dx'
	\frac{\delta H(x')}{\delta\phi^i(x)}.
\end{eqnarray}
These equations are equivalent to the Euler-Lagrange equations.
This is not trivial to prove,
and it is a remarkable result that it is true.

\subsection{Constrained systems}
\label{sec:constrainedsystems}

References for the theory discussed in this section are
\cite{dirac:1964,henneaux:1992,marnelius:1982}.

Suppose that we for a given Lagrangian have found the momenta
$\pi_i = \frac{\partial L}{\partial \dot{\phi}^i}$. The way to the
Hamiltonian picture is easy if we can invert this procedure to 
find $\dot{\phi}^i$
as functions of the fields and momenta, i.e. $\dot{\phi}^i = 
\dot{\phi}^i(\phi, \pi, \nabla \phi)$.  
Then we can just substitute for $\dot{\phi}^i$ in $h$ and immediately arrive
at the naive Hamiltonian $H_{naive}(\phi, \pi, \nabla \phi)$
This is often possible, but not if there
exists some gauge freedom in the theory. 
There are also non-gauge theories which fail to be invertible in this
way \cite{henneaux:1992}.

It is therefore of interest to study this class of systems, which are
called \emph{constrained systems}. The non-invertibility means that
the derived 
momenta will not be independent, and there exist some relations 
between them. These relations can be expressed as functions 
$\theta^I_m(\phi, \pi, \nabla \phi)=0$, $m=1,\dots ,M$, where $M$ is 
the number of such functions. We call these the \emph{primary
constraints} since they follow directly from the definition of the
momenta. 

Suppose that $i$ can take $n$ different values (i.e. there
are $n$ field variables). Then define the $n\times n$ matrix 
$C_{ij}\equiv \frac{\del^2 L}{\del\dot{\phi}^i\del\dot{\phi}^j}$. If
$r$ is the \emph{rank} of this matrix, then the number of independent
primary constraints is $n-r$.

For constrained systems the naive Hamiltonian is not
unique, since we may add to it any linear combination of the constraint
functions $\theta^I$. This fact leads to a modification of Hamilton's
equations. The modified versions are
\begin{eqnarray}
  \label{eq:modhameq1}
  \dot{\phi}^i(x) & = & 
	\int dx' \frac{\delta H_1(x')}{\delta \pi_i(x)}
  \\
  \label{eq:modhameq2}
  -\dot{\pi}_i(x) & = & 
	\int dx' \frac{\delta H_1(x')}{\delta \phi^i(x)},
\end{eqnarray}
where $H_1 \equiv H_{naive} + \lambda_I^m\theta^I_m$ and $\lambda_I^m$ are
coefficients that do not depend on $\phi$ and $\pi$. They are called
\emph{Lagrange multipliers}.

\paragraph{Poisson brackets}

The notation of Poisson brackets is very convenient in this
formalism. Consider two functions $F$ and $G$ that are constructed
from $\phi$ and $\pi$. If we write $F(x)$ as short for 
$F(\phi(x), \pi(x))$, their Poisson bracket is defined as
\beq
  \left\{F(x), G(x')\right\}
   \equiv  \int d\tilde{x}\left(
	\frac{\delta F(x)}{ \delta \phi^i(\tilde{x})}
	\frac{\delta G(x')}{\delta \pi_i(\tilde{x})} -
	\frac{\delta F(x)}{ \delta \pi_i(\tilde{x})}
	\frac{\delta G(x')}{\delta \phi^i(\tilde{x})} \right).
\eeq
The brackets can easily be shown to  satisfy the following relations:
\begin{enumerate}
 \item{Antisymmetry }	 $\{F,G\} = - \{G,F\}$
 \item{Linearity    }	 $\{F+G, H\} = \{F,H\} + \{G,H\}$
 \item{Product law  }	 $\{FG,H\} = F\{G,H\}+\{F,H\}G$
 \item{The Jacobi identity} $\{F,\{G,H\}\} + \{G,\{H,F\}\} + \{H,\{F,G\}\}=0$
\end{enumerate}
It is also easy to show that the fundamental Poisson brackets are
\begin{eqnarray}
  \{\phi^i(x),\pi_j(x')\} &=& \delta^i_j\delta(x-x'),
  \\
  \{\phi^i(x),\phi^j(x')\} &=& \{\pi_i(x),\pi_j(x')\} = 0.
\end{eqnarray}
%
When we work with Poisson brackets in the present circumstance the
following is important: 
\emph{Poisson brackets must be evaluated before we make use of the
constraint equations}. In other words, we should perform the
calculations in phase space, and restrict to the constraint surface
($\theta=0$) at the end.
To emphasize this point we use Dirac's notation
\cite{dirac:1964} and say that the constraint equations are 
\emph{weakly zero}, and
write them with a new \emph{weak equality} sign ``$\approx$'' as:
\beq
  \theta^I_m(\phi,\pi) \approx 0.
\eeq
This makes a difference, for even though $\theta(\phi,\pi)$ is
dynamically zero (i.e. zero when $\phi$ and $\pi$ satisfy 
Hamilton's equations) it is not zero throughout phase space.

\air
Now, let us consider the time evolution of the function 
$F(\phi,\pi)$. By the chain rule we have
\beq
  \dot{F}(x) = \int d\tilde{x} \left(
	\frac{\delta F(x)}{\delta\phi^i(\tilde{x})} \dot{\phi}^i(\tilde{x}) +
	\frac{\delta F(x)}{\delta\pi_i(\tilde{x})} \dot{\pi}_i(\tilde{x})
  \right).
\eeq
Using the (modified) Hamilton's equations \refeq{modhameq1} and
\refeq{modhameq2} we can insert for
$\dot{\phi}$ and $\dot{\pi}$ and get
\begin{eqnarray}
  \nonumber
  \dot{F}(x)
  & = & \int dx' \int d\tilde{x} \left(
	\frac{\delta F(x)}{   \delta \phi^i(\tilde{x})}
	\frac{\delta H_1(x')}{\delta \pi_i(\tilde{x})} -
	\frac{\delta F(x)}{   \delta \pi_i(\tilde{x})}
	\frac{\delta H_1(x')}{\delta \phi^i(\tilde{x})} \right)
  \\
  & = & \int dx' \left\{ F(x), H_1(x') \right\}.
\end{eqnarray}
The constraints must hold (i.e. be weakly equal to zero) for all
times. This means that their time derivatives should vanish (weakly):
\beq
  \dot{\theta}^I_m(\phi,\pi) \approx 0.
\eeq
By putting $F=\theta^I_m$ in the equation above, we thus get the
following consistency conditions:
\beq
  \int dx' \{\theta^I_m(x), H_1(x')\} \approx 0,
\eeq
or more explicitly
\beql{consistency-conditions}
  \int dx' \left[ \{\theta^I_m(x), H_{naive}(x')\} 
	+ \lambda_I^n\{\theta^I_m(x), \theta^I_n(x')\} \right]
	\approx 0.
\eeq
These equations may lead to three different situations.
The first is that they do not give anything new at all, we
merely end up with $0=0$. In this case we have already found the full
constraint structure.

Another possibility is that we arrive at equations not involving the
$\lambda_I$'s. Then we get new constraints on $\phi$ and $\pi$ on the
form
\beq
  \theta^{II}_p(\phi,\pi) \approx  0.
\eeq
These are called \emph{secondary constraints}.

The third kind of equations we may end up with also depend on the
Lagrange multipliers $\lambda_I$. The consistency conditions will then
impose a condition on the $\lambda_I$'s.

If we get secondary constraints we must go on and check the
consistency conditions on them, $\dot{\theta}^{II} \approx 0$ in just the
same way as for the primary constraints. This ``loop'' should be
continued until we get no more new conditions.

The distinction between primary and secondary (and tertiary etc.)
constraints is just a matter of how they appear, and is not physically
important. In fact, different Lagrangians that describe the same
physical system will in general give rise to the constraints in a
different order and different combinations.

When we have found all constraints and conditions on the $\lambda$'s
(if any) we are ready to write down the \emph{total Hamiltonian}:
\beq
  H_T = H' + \lambda^a\theta_a.
\eeq
If we had no conditions imposed on the $\lambda$'s, $H'$ will be
identical to the naive Hamiltonian, and $\lambda^a = \{\lambda_I^m,
\lambda^p_{II},\dots\}$. If, on the other hand, such conditions were
present, $H'$ will be shifted by some factor, and the Lagrange
multipliers can be  redefined so that all the $\lambda^a$'s are
independent. 
This will not be important for our considerations, and is 
explained in more detail in \cite{dirac:1964,henneaux:1992}. 

Several examples of calculations of the kind explained in this section
are given later on in the thesis. 

%% file: symmetries.tex
\section{Symmetries}

Let ${\cal S}$ be some \emph{symmetry}, and let $T_{\cal S}$ be the
\emph{transformation} associated with this symmetry. We say that a theory
has the symmetry ${\cal S}$ if the \emph{action} 
$S=\int L dx$
describing the 
theory is left unchanged under transformations $T_{\cal S}$. We then
say that the action is ${\cal S}$-\emph{invariant}. In this section 
we will see a brief description of some symmetries that we will
frequently encounter.
Most of the examples presented are studied in more detail later on.

\subsection{Diffeomorphism symmetry}
\label{sec:diffeomorphism}

Suppose we have an action integral parameterized by $x$ ($x$ may be one
parameter or a whole set), then we write the action as
\beql{sym-action}
  S=\int dx L(\phi),
\eeq
where $\phi$ symbolizes the field variables. ($\phi$ may be one or a
whole set of fields.)
A diffeomorphism is a general coordinate transformation of the form
\beq
  x \to \tilde{x} = \tilde{x}(x).
\eeq
The name comes from the fact that $\tilde{x}$ is a differentiable
function of $x$ (usually $\in C^\infty$). In the context of particles,
strings etc. where the parameters in the action are not the spacetime
coordinates we usually refer to the  transformation as a
\emph{reparameterization}.
Now, for a
diffeomorphism symmetry to be present in the model, the action has to
be invariant under this transformation. Whether that is the case of
course depends on how the fields transform, 
$\phi(x) \to \tilde{\phi}(\tilde{x})$, 
and of the form of the Lagrangian. For short, diffeomorphism is often
written \emph{diff} (e.g. \emph{diff} invariance).

\paragraph{Example:}

The relativistic point particle. Here the action is (c.f. section
\ref{sec:pointparticle}) 
\beql{pointpartaction}
  S = m\int d\tau \sqrt{-\eta_{\mu\nu}\frac{dX^\mu}{d\tau}
        \frac{dX^\nu}{d\tau}}
\eeq
Consider the reparameterization $\tau \to \tilde{\tau}(\tau)$. The fields are
scalars under this transformation, in the sense that 
$\tilde{X}^\mu(\tilde{\tau}) = X^\mu(\tau)$.
(The position of the particle is obviously independent of the
parameterization.) 
The transformed action (i.e. the action described by the transformed
quantities) is then
\begin{eqnarray}
  \nonumber
  \tilde{S}
	& = &\int d\tilde{\tau} \sqrt{-\eta_{\mu\nu}
	\frac{d\tilde{X}^\mu(\tilde{\tau})}{d\tilde{\tau}}
	\frac{d\tilde{X}^\nu(\tilde{\tau})}{d\tilde{\tau}}}
  \\ \nonumber
	& = &\int d\tau \frac{d\tilde{\tau}}{d\tau} \sqrt{-\eta_{\mu\nu}
	\frac{dX^\mu(\tau)}{d\tau}
	\frac{dX^\nu(\tau)}{d\tau} (\frac{d\tau}{d\tilde{\tau}})^2}
  \\
	& = &\int d\tau \sqrt{-\eta_{\mu\nu}
	\frac{dX^\mu(\tau)}{d\tau}
	\frac{dX^\nu(\tau)}{d\tau} } = S.
\end{eqnarray}
The symbolic form of the action is unchanged, which is precicely the
symmetry criterion.

\subsection{Weyl symmetry}

When we talk about Weyl symmetry we study theories which
involve some metric $g_{ab}$.
A Weyl transformation is then a position-dependent rescaling of this
metric. (We could also imagine rescalings of other fields, but they
are not so interesting.) The transformation can be written as
\beq
  g_{ab}(x) \to \tilde{g}_{ab}(x) = e^{\omega(x)}g_{ab}(x),
\eeq
where $\omega(x)$ is any function. An action that is unchanged by this
transformation is consequently called Weyl invariant.

\paragraph{Example:} The Weyl-invariant string action
\cite{brink:1976,deser:1976}: 
\beq
  S = \int d^2\xi \sqrt{-g} g^{ab} \gamma_{ab}.
\eeq
Here $\gamma_{ab} \equiv G_{\mu\nu} \del_aX^\mu\del_bX^\nu$, where
$G_{\mu\nu}$ is the background (fixed) spacetime metric, and $\del_a
\equiv \frac{\del}{\del\xi^a}$. The field variables in the theory
are the intrinsic worldsheet metric components $g_{ab}(\xi)$ and the
position field $X^\mu(\xi)$. Consider 
then the Weyl transformation $g_{ab}(\xi) \to e^{\omega(\xi)}g_{ab}(\xi).$
Under this rescaling of the metric we will have
\begin{eqnarray}
  \nonumber
  g \equiv \det g_{ab} & \to & e^{2\omega} g
  \\ \nonumber
  g^{ab} & \to & e^{-\omega} g^{ab}.
\end{eqnarray}
And the action will transform as
\beq
  S \to \int d^2\xi\sqrt{-e^{2\omega} g} e^{-\omega}g^{ab}\gamma_{ab}
        = S.
\eeq
So the action is worldsheet Weyl invariant (as its name correctly
announces). Note also that the
dimensionality $D=2$ enters crucially in this derivation.

\subsection{Poincar\'e symmetry}

A Poincar\'e transformation is a coordinate transformation that
consists of the familiar Lorentz transformation plus translation. It
is also called an \emph{inhomogeneous} Lorentz transformation, a name
which is obvious from the form:
\beq
  X^\mu \to \tilde{X}{\mu} = \Lambda^{\mu}_\nu X^\nu + a^{\mu}.
\eeq
$\Lambda^{\mu}_\nu$ are the Lorentz transformation 
coefficients and 
$a^{\mu}$ are constants describing the translational part.

Lorentz transformations can further be split into
into \emph{boosts} and \emph{rotations} in addition to the
discrete transformations of time and space inversion.

A general result worth noting is that 
Lorentz invariance is automatically
achieved if the Lagrangian is written in covariant form as a Lorentz scalar.

One very interesting feature of the Poincar\'e transformation is that
gauge theory based on \emph{local} Poincar\'e invariance (i.e. the
coefficients in the above transformation are position dependent), gives
rise to a theory for gravitation.

\paragraph{Example:} The relativistic point particle.

We start again from the action \refeq{pointpartaction}. The
coefficients $\Lambda_\mu^\nu$ and $a^\nu$ are constants, so the
derivatives of $X^\mu$ are transformed as
\[
  \dot{X}^\mu \equiv \frac{dX^\mu}{d\tau} \to \Lambda_\nu^\mu \dot{X}^\nu.
\]
This is the only quantity that is changed under the Poincar\'e
transformation, so the action will transform as
\begin{eqnarray}
  \nonumber
  S &\to& \int d\tau \sqrt{-\eta_{\mu\nu} \Lambda_\rho^\mu
  \Lambda_\sigma^\nu \dot{X}^\rho\dot{X}^\sigma}
  \\ \nonumber
  & = & \int d\tau
  \sqrt{-\eta_{\rho\sigma}\dot{X}^\rho \dot{X}^\sigma} = S,
\end{eqnarray}
where we have used the general result
$\eta_{\rho\sigma}=\eta_{\mu\nu} \Lambda_\rho^\mu \Lambda_\sigma^\nu$.
Thus, we see that the relativistic point particle action has
Poincar\'e symmetry.

\subsection{Conformal symmetry}
\label{sec:sym-conf}

A conformal transformation is a mapping from flat space (Minkowski or
Euclidian) onto itself, such that the flat metric $\eta_{ab}$
is left invariant up to a rescaling. In other words, it has the same
effect as the substitution $\eta_{ab}\to\Omega(x)\eta_{ab}$ for some
positive definite function $\Omega(x)$. Because of this, the line
element will transform as $ds^2\to\Omega ds^2$, showing that the
causal structure is conserved. In particular, light cones will 
transform into light cones.

Alternatively, the conformal transformation may be viewed as a
composite \emph{diff} and Weyl
transformation that leaves the flat metric invariant. (The conformal
symmetry group is a subgroup of the \emph{diff$\times\!\!$Weyl} symmetry
group.) 

In general relativity terminology, what is here called Weyl
transformation is often named conformal transformation.
But they are not to be considered as the same.

It is also important not to confuse conformal symmetry with the
\emph{diff} invariance of general relativity. Conformal symmetry is a
symmetry in 
flat space theory, with no independent metric fields to vary. Hence,
conformal transformations in general actually change the distances
between points. From this it follows that conformally symmetric theories
have no length scale. 

A full conformal transformation involves a Poincar{\'e} transformation
(Lorentz + translation), a dilatation and a special conformal
transformation. It acts on the spacetime coordinates, and
the infinitesimal transformation can be written 
$x^\mu\to \tilde{x}^\mu = x^\mu+\delta x^\mu$, with
\begin{eqnarray}
  \mbox{Lorentz:}&	\delta_\omega x^\mu &=\omega^\mu_{~\nu} x^\nu,
  \\
  \mbox{translation:}&	\delta_a x^\mu &=a^\mu,
  \\
  \mbox{dilatation:}&	\delta_c x^\mu &=cx^\mu,
  \\
  \mbox{special conformal:}&	
			\delta_b x^\mu &=b^\nu x_\nu x^\mu 
			- \12 x^\nu x_\nu b^\mu.
\end{eqnarray}
In 4 dimensions there are 15 independent parameters of this symmetry
group:  
6 Lorentz ($\omega_{\mu\nu}=-\omega_{\nu\mu}$); 
4 translations ($a^\mu$); 1 dilatation ($c$); 4 special ($b^\mu$). 

\paragraph{Example:} The massless scalar field. 

Consider the action
\beq
  S = \int dx \del_\mu\phi\del_\nu\phi \eta^{\mu\nu}.
\eeq
Since $\phi$ is a scalar we have 
\beq
  \tilde{\phi}(\tilde{x})=\phi(x);\quad
  \tilde{\del}_\mu\tilde{\phi}(\tilde{x})
	=\frac{\del x^\nu}{\del \tilde{x}^\mu}
	\del_\nu\phi(x)=(\delta^\mu_\nu -\del_\nu\delta
	x^\mu)\del_\mu\phi(x),
\eeq
where $\tilde{\del_\mu}\equiv \frac{\del}{\del\tilde{x}^\mu}$.
The transformed action can then be written as
\begin{eqnarray}
  \tilde{S} &=& \int d\tilde{x} \tilde{\del}_\mu\tilde{\phi}
	\tilde{\del}_\nu\tilde{\phi}\eta^{\mu\nu}
  \\
  &=&\int dx \det(\frac{\del \tilde{x}^\mu}{\del x^\nu})
	(\delta^\rho_\mu- \del_\mu\delta x^\rho)
	(\delta^\sigma_\nu-\del_\nu\delta x^\sigma)
	\del_\rho\phi\del_\sigma\phi \eta^{\mu\nu}.
\end{eqnarray}
The Jacobi determinant is to first order 
$\det(\frac{\del \tilde{x}^\mu}{\del x^\nu})=1+\del_\alpha\delta
x^\alpha$. This gives
\begin{eqnarray}
  \nonumber
  \tilde{S}= S+\delta S; \quad& 
	\delta S &= \int dx \del_\mu\phi\del_\nu\phi X^{\mu\nu},
  \\ \nonumber
   &X^{\mu\nu}&\equiv -\eta^{\mu\nu} \del_\alpha \delta x^\alpha
	+\eta^{\alpha\nu}\del_\alpha \delta x^\mu
	+\eta^{\mu\alpha}\del_\alpha \delta x^\nu.
\end{eqnarray}
Let us now consider the conformal transformations one by one.

\subparagraph{Lorentz}

Consider first the Lorentz transformation $\delta x^\mu =
\omega^\mu_{~\alpha}x^\alpha$.  We find then
$\del_\mu \delta x^\nu = \omega^\nu_{~\mu}$, which together with the
antisymmetry of $\omega_{\mu\nu}$  gives
$X^{\mu\nu}= 0$. So the action is Lorentz invariant. As mentioned
earlier, this could be concluded solely from the fact that the
Lagrangian is written covariantly as a Lorentz scalar.

\subparagraph{Translation}

For translations we have $\delta x^\mu = a^\mu$,
which gives $\del_\nu \delta x^\nu = 0$ and immediately
$X^{\mu\nu}=0$. Thus the action is always Poincar\'e invariant.

\subparagraph{Dilatation}

Turn then to the dilatation, $\delta x^\mu = cx^\mu$. We find 
$\del_\mu \delta x^\nu = c \delta^\nu_\mu$, and
$X^{\mu\nu}=\eta^{\mu\nu}c(\delta^\alpha_\alpha-2)$ which is zero in
\emph{two} dimensions ($D=2$). For general spacetime dimensions $D$,
we get
$\tilde{S}= \int dx \del_\mu\phi\del_\nu\phi \eta^{\mu\nu}\Omega$,
with $\Omega = 1+c(D-2)\approx 1 >0$. Thus we
find that the effect of a dilatation is the same as a rescaling of the
metric.

\subparagraph{Special conformal}

Special conformal transformations have 
$\delta x^\mu =b^\nu x_\nu x^\mu - \12 x^\nu x_\nu b^\mu$, which
gives 
$\del_\mu \delta x^\nu = b_\mu x^\nu+b\cdot x \delta^\nu_\mu 
- x_\mu b^\nu$. This gives 
$X^{\mu\nu}=\eta^{\mu\nu}b\cdot x(\delta^\alpha_\alpha - 2)$, which is
again zero for $D=2$. In general we find
$\tilde{S} = \int dx \del_\mu\phi\del_\nu\phi \eta^{\mu\nu}\Omega'$,
with $\Omega'=1+b \cdot x(D-2)\approx 1>0$.

\air
We have now seen explicitly that a conformal transformation on the
massless scalar field has the effect of a rescaling of the flat
metric ($\eta^{\mu\nu}\to \eta^{\mu\nu}(\Omega+\Omega')$, with
$\Omega$ and $\Omega'$ as defined above).
Furthermore, we found that it is conformal invariant in two
dimensions.

%% file: energy-momentum.tex
\section{More on coordinate transformations}

In this section we will use symmetry principles to derive some
important general results in field theory.

Consider a general action integral, 
\beq
  S=\int dx L(\phi^i, \del \phi^i),
\eeq
where $\phi^i$ are some general fields; scalars, vectors or
whatever. We have assumed that there is no explicit coordinate 
dependence.
In the following we will see what happens if we make the
coordinate transformation,
\beql{inf-diff-transl}
  x^a \to \tilde{x}^a = x^a + a^a,
\eeq
where $a^a=a^a(x)$ is infinitesimal. 
We now define the \emph{Jacobi matrix} $J^a_b$ and the \emph{Jacobi
determinant} $J$ as
\begin{eqnarray}
  \label{eq:def-jacobimatrix}
  J^a_b &\equiv& \frac{\del x^a}{\del\tilde{x}^b}
  	=\delta^a_b-\del_b a^a,
  \\
  \label{eq:def-jacobideterminant}
  J &\equiv& \det(J^a_b) 
	=\det(\delta^a_b-\del_b a^a) \cong 1-\del_ca^c.
\end{eqnarray}
Whit this definition the integral measure transforms as 
$dx\to d\tilde{x}=dxJ^{-1}$.

Under the transformation \refeq{inf-diff-transl} the
action will transform to
\beq
  S\to \tilde{S} = \int d\tilde{x} L(\tilde{\phi}(\tilde{x}), 
	\tilde{\del} \tilde{\phi}(\tilde{x})).
\eeq
Let the transformation of the fields be written
\beq
  \phi^i(x) \to \tilde{\phi}^i(\tilde{x}) = \phi^i(x) +
	\epsilon^i(x)
\eeq
This defines $\epsilon^i$. Note that the fields are taken at different
points on the left and right hand side. This is not the usual way to
compare fields, but convenient for the moment. The notation, and the
subsequent calculation, is inspired by Fr\o yland \cite{froyland:1996}.
For scalars we have $\epsilon=0$.

The derivatives of the fields will transform as
\begin{eqnarray}
  \nonumber
  \tilde{\del}_b\tilde{\phi^i}(\tilde{x}) &=& 
	\frac{\del x^c}{\del\tilde{x}^b} \del_c (\phi^i(x) + \epsilon^i)
  \\
  &\cong& \del_b\phi^i(x) + \del_b\epsilon^i(x) - \del_b a^c\del_c\phi(x).
\end{eqnarray}
The transformed action integral may now be Taylor expanded and
rewritten in the following way:
\begin{eqnarray}
  \nonumber
  \tilde{S} &=& \int dx J^{-1} L(\phi^i + \epsilon^i,
	\del_a\phi^i + \del_a\epsilon^i-\del_a a^b\del_b\phi^i)
  \\ \nonumber
  &=& \int dx (1+\del_c a^c)\left(
	L(\phi,\del\phi^i) + \frac{\del L}{\del\phi^i}\epsilon^i
	+\frac{\del L}{\del(\del_a\phi^i)}
	(\del_a\epsilon^i-\del_a a^b\del_b\phi^i) \right)
  \\ \nonumber
  &=& \int dx \bigg( L +\del_c a^c L + 
	\frac{\del L}{\del\phi^i}\epsilon^i
	+\underbrace{\frac{\del L}{\del (\del_a\phi^i)}\del_a\epsilon^i}
	_{-\del_a(\frac{\del L}{\del(\del_a\phi^i)})\epsilon^i}
	-\frac{\del L}{\del(\del_a\phi^i)}\del_b\phi^i\del_a a^b
	\bigg)
  \\ \nonumber
  &=& \int dx \bigg( L - \underbrace{ \left[
	\del_a(\frac{\del L}{\del(\del_a\phi^i)}) 
	- \frac{\del L}{\del\phi^i}\right]}_{=\psi_i} \epsilon^i
	+ \del_a a^b \underbrace{ \left[
	\delta^a_b L - \frac{\del L}{\del (\del_a\phi^i)}\del_b\phi^i
	\right]}_{\equiv T^a_{~b}} \bigg)
  \\ \label{eq:diff-varaction}
  &=& S + \int dx \left(\del_a a^b T^a_{~b}- \psi_i\epsilon^i\right),
\end{eqnarray}
where we have performed a partial integration and assumed the fields
to vanish at infinity.
For fields that satisfy the Euler-Lagrange equations, $\psi^i=0$, we
find
\beql{ep-tensor1}
  \delta S = \int dx \del_a a^b T^a_{~b}.
\eeq
Another partial integration gives
\beql{ep-tensor2}
  \delta S = -\int dx (\del_a T^a_{~b})a^b.
\eeq
In the calculations above we have assumed that $a^a$ is position
dependent. But consider now the \emph{global} transformation
(i.e. $a^a=const$), which is an infinitesimal translation.
This is usually a symmetry of the action, in which case we have
$\delta S=0$. Equation \refeq{ep-tensor2} then gives the condition 
\beq
  \del_a T^a_{~b}=0.
\eeq
In other words, the (global) symmetry leads to a conserved 
\emph{translation current} $T^a_{~b}$. This is a special case of
\emph{Noether's theorem} which states that any symmetry implies a
conserved current.

\subsection{Energy-momentum tensor}

The translation current is often called the \emph{canonical
energy-momentum tensor}, and we defined it as
\beql{ep-can}
  {\cal T}^a_{~~b} = \delta^a_b L 
	- \frac{\del L}{\del (\del_a\phi^i)}\del_b\phi^i.
\eeq
However, the tensor ${\cal T}^{ab}=\eta^{bd}{\cal T}^a_{~~d}$ 
is not symmetric
and therefore cannot be used on the right hand side of Einstein's
field equations for general relativity.
Neither is it always possible to generalize to curved spacetime.

We will now present another way of defining the energy-momentum
tensor, which avoids these problems.
Consider the action for general relativity coupled to matter, which
can be written (c.f. chapter \ref{chap:GR} and \cite{carroll:1997})
\beq
  S=\int d^4x \sqrt{-g}(\frac{1}{\kappa} R + {\cal L}_M),
\eeq
where $\kappa$ is a constant, $g=\det(g_{ab})$ is the determinant of
the metric, $R$ is the Ricci curvature scalar and ${\cal L}_M$ is the
term describing the matter field. It is the same as the Lagrangian
expressed in flat spacetime with the flat metric exchanged by the
general metric $g_{ab}$. (This is not valid for spinors.)
If we define $L_M\equiv \sqrt{-g}{\cal L}_M$,
a variation $\delta g_{ab}$ leads to the field equations
\beq
  -\frac{1}{\kappa}\sqrt{-g}(R^{ab}-\12  g^{ab}R) 
	+ \frac{\delta L_M}{\delta g_{ab}} = 0.
\eeq
We recover the Einstein field equations, 
$R^{ab}-\12 g^{ab}R = \frac{\kappa}{2} T^{ab}$,
if we use
\beql{energy-momentum}
  T^{ab} = \frac{2}{\sqrt{-g}}\frac{\delta L_M}{\delta g_{ab}}.
\eeq
This tensor is manifestly symmetric, and gives a convenient 
definition of the \emph{energy-momentum tensor}. 
In the following we will go through the necessary calculations to
prove that the two definitions of the energy-momentum tensor are
equivalent in flat spacetime, provided that $\phi^i$ couples to
gravity via $g_{ab}$.

\subsubsection{Equivalence of \refeq{ep-can} and
	\refeq{energy-momentum}}

Consider again the infinitesimal transformation
$x^a\to\tilde{x}^a + a^a(x)$. The metric $g_{ab}$ transforms as a
second rank tensor, i.e.
\begin{eqnarray}
  g_{ab}\to\tilde{g}_{ab}(\tilde{x}) &=& \Lambda^c_a\Lambda^d_b g_{ab}(x),
  \\
  \label{eq:def-lambda}
  \Lambda^c_a &\equiv& \frac{\del x^c}{\del\tilde{x}^a} 
	= J^c_a = \delta^c_a-\del_a a^b.
\end{eqnarray}
This gives
\beq
  \tilde{g}_{ab}(x) = g_{ab}(x) + \delta g_{ab}; \quad
	\delta g_{ab}= -(a^c\del_c g_{ab} + g_{ac}\del_b a^c +
	g_{cb}\del_a a^c).
\eeq
We recognize $\delta g_{ab}$ as a Lie derivative. Using the results
from appendix \ref{app:lie-derivative} we find
\beql{var-g}
  \delta g_{ab} =  -\pounds_{\vec{a}} g_{ab} 
	= -\nabla_a a_b - \nabla_b a_a
	= -\nabla_{(a} a_{b)}.
\eeq 
Now, consider the flat space action $S=\int dx L(\phi)$. Coupling to
gravity gives 
\beql{phigravity}
  S_G = \int dx L_G(\phi,g); \quad L_G = \sqrt{-g}L(\phi,g).
\eeq
The infinitesimal diffeomorphism $x^a\to x^a+a^a$ gives a variation in
the action, which we write
\beq
  S_G\to S_G + \delta S_G; \quad 
	\delta S_G = \delta_\phi S_G + \delta_g S_G. 
\eeq
The first term in $\delta S_G$ is proportional to $\delta\phi$ and the
second is proportional to $\delta g_{ab}$
Since any gravity-coupled action is generally coordinate invariant,
i.e. \emph{diff} invariant, we must have
\beq
  \delta S = \delta_\phi S_G + \delta_g S_G = 0,
\eeq
and as a special result
\beql{equat}
  (\delta_\phi S_G + \delta_g S_G)|_{g=\eta} = 0.
\eeq
If we use equation \refeq{ep-tensor1} we get 
\beq
  \delta_\phi S_G|_{g=\eta}= \delta_\phi S 
	= \int dx \del_a a^b {\cal T}^a_{~~b}
	= \int dx \del_a a_b {\cal T}^{ab}
\eeq
where ${\cal T}^{ab}$ is the canonical energy-momentum tensor.
Furthermore, by use of \refeq{var-g} we find
\begin{eqnarray}
  \nonumber
  \delta_g S_G|_{g=\eta} &=& \int dx \frac{\delta L_G}{\delta g_{ab}}
	\delta g_{ab} |_{g=\eta} 
	= -\int dx \frac{\delta L_G}{\delta g_{ab}}\nabla_{(a}a_{b)}
	|_{g=\eta}
  \\
	&=& -2\int dx \frac{\delta L_G}{\delta g_{ab}}|_{g=\eta}
	\del_a a_b
\end{eqnarray}
Equation \refeq{equat} gives then the following relation in flat space:
\beql{Ep-equiv}
  {\cal T}^{ab} = 2 \frac{\delta L_G}{\delta g_{ab}}|_{g=\eta}
	= T^{ab}|_{g=\eta},
\eeq
which is exactly what we wanted to show. It says that the two
definitions \refeq{ep-can} and \refeq{energy-momentum} are the same in
flat spacetime for models that couple to gravity according to
\refeq{phigravity}.  
Since the two definitions are the same, we immediately find that even
the canonical energy-momentum tensor ${\cal T}^{ab}$ is symmetric. As
noted earlier, this is not a general result, but comes here as a
consequence of the assumpsion \refeq{phigravity} that $\phi^i$ is a
kind of field that couples to gravity via $g_{ab}$.

\subsubsection{Energy-momentum tensor for spinors}

A description of gravity models with spinors is most easily done in
the vielbein formalism (see appendix \ref{sec:vielbeins}). Denote the
vielbeins $e_a^{~A}(x)$ and their determintants $\det(e_a^{~A})\equiv
e$. We then have $\sqrt{-g}=e$, and a gravity-coupled model can be
written $L_G = e L(\phi^i, e_a^{~A})$, where $L(\phi^i)$ is the
non-coupled theory. $\phi^i$ may now be spinors, but also tensors.

We define again the energy-momentum tensor 
$T^a_{~A}=\frac{\delta L_G}{\delta e_a^{~A}}$ as the
translation current:
\beq
  \delta x^a \Rarr \qquad
	\delta S_G = \int dx\left[ \frac{\delta L_G}{\delta
	e_a^{~A}}\delta e_a^{~A} + \frac{\delta L_G}{\delta
	\phi^i}\delta \phi^i \right].
\eeq
For \emph{Lorentz transformations} $\Lambda_B^{~~A}$ we have 
\beq
  \delta e_a^{~A}=e_a^{~B}\Lambda_B^{~~A}; 
	\qquad \Lambda^{AB}=-\Lambda^{BA},
\eeq
which gives
\begin{eqnarray}
  \nonumber
  \delta_\Lambda S_G &=& 
	\int dx \left[ e_a^{~B}\Lambda_B^{~A} T^C_{~A}e_C^{~~a}
	+\frac{\delta L_G}{\delta\phi^i}\delta_\Lambda\phi^i\right]
  \\ &=& \nonumber
	\int dx \left[ \Lambda^{BA}T_{BA} 
	+\frac{\delta L_G}{\delta\phi^i}\delta_\Lambda\phi^i\right].
\end{eqnarray}
The Lorentz transformation is a symmetry of the theory, so
$\delta_\Lambda S = 0$. Furthermore, if the fields $\phi^i$ satisfy
the equations of motion $\frac{\delta L_G}{\delta\phi^i}=0$, we 
get
$\int dx \Lambda^{AB}T_{AB}=0$ which means that the antisymmetic part
of the energy-momentum tensor is zero, i.e. $T_{[AB]}=0$. In other
words, the energy-momentum tensor is symmetric for fields that satisfy
the equations of motion, but not generally.

\subsubsection{Energy-momentum tensor for conformally invariant theories}

As noted in section \ref{sec:sym-conf}, 
a conformal transformation has the same effect as a rescaling of the
metric. Thus we may consider the variation of the action $S$ as a result
of a variation in $g_{ab}$ with $\delta g_{ab} = \Omega g_{ab}$. This
means that we can write
\beq
  \delta S = \int dx \frac{\delta L}{\delta g_{ab}}\delta g_{ab}
	= \int dx 2T^{ab} \Omega g_{ab}
	= 2 \int dx T^{ab}g_{ab}\Omega.
\eeq
Invariance means $\delta S=0$ so \emph{we have for conformally invariant
theories that the energy-momentum tensor is traceless}, i.e.
\beq
  T^a_{~a} = T^{ab}g_{ab} = 0.
\eeq
This is indeed a simple way to determine conformal invariance.

\paragraph{Example:} We now return to the massless scalar field we
considered in section \ref{sec:sym-conf}, with the Lagrangian
\beq
  L = \del_\mu\phi\del_\nu\phi\eta^{\mu\nu}.
\eeq
The canonical energy-momentum tensor is found to be
\begin{eqnarray}
  \nonumber
  {\cal T}^\mu_{~~\nu} &\equiv& \delta^\mu_\nu L 
	- \frac{\del L}{\del \del_\mu\phi}\del_\nu\phi
  \\ \nonumber
  &=&\del_\alpha\phi \del_\beta\phi (
	\delta^\mu_\nu \eta^{\alpha\beta}
	-2\delta^\beta_\nu \eta^{\mu\alpha}),
  \\
  {\cal T}^{\mu\nu} &=& \del_\alpha\phi \del_\beta\phi (
	\eta^{\mu\nu}\eta^{\alpha\beta}
	-2\eta^{\mu\alpha}\eta^{\nu\beta}).
\end{eqnarray}
To use the other definition of the energy-momentum tensor, we couple
the model to gravity, and get
\beq
  L_G = \sqrt{-g}\del_\mu\phi\del_\nu\phi g^{\mu\nu}.
\eeq
Then we find
\begin{eqnarray}
  \nonumber
  T^{\mu\nu} &\equiv& \frac{2}{\sqrt{-g}} 
	\frac{\delta L_G}{\delta g_{\mu\nu}}
  = \frac{2}{\sqrt{-g}}\frac{\del L_G}{\del g_{\mu\nu}}
 \\
  &=& \del_\alpha\phi \del_\beta\phi (
	g^{\mu\nu}g^{\alpha\beta}
	-2g^{\mu\alpha}g^{\nu\beta}).
\end{eqnarray}
We see immediately that ${\cal T}^{\mu\nu} =
T^{\mu\nu}|_{g=\eta}$. Furthermore, the trace is
\beq
  T^\mu_{~\mu} = g_{\mu\nu} T^{\mu\nu} 
	= \del_\alpha\phi \del_\beta\phi (D-2)g^{\alpha\beta},
\eeq
which says that the energy-momentum tensor is traceless in two
dimension, $D=2$. This is in complete agreement with the fact that
the massless scalar field is conformal invariant in two dimensions.

\subsection{Naive Hamiltonian for \emph{diff} invariant theories}
\label{sec:zerohamilton}

If we let $a$ be position dependent, and arbitrary, the transformation
\refeq{inf-diff-transl} is
identical to an infinitesimal diffeomorphism. If we demand the action
to be \emph{diff} invariant,
but do \emph{not} restrict to fields that satisfy the equations of motion,
we get from \refeq{diff-varaction} the condition
\beq
  \int dx \left[ T^a_{~b}\del_a a^b - \psi_i\epsilon^i\right] = 0.
\eeq
We want to show how this can give us an expression for the
Hamiltonian. To do so we need to know
the form of
$\epsilon$. Let us consider scalar, vector and second rank tensor
fields, $\phi^i = \{\phi, A_a, A_{ab}\}$, and define
$\Lambda^b_a\equiv \frac{\del x^b}{\del \tilde{x}^a} 
=\delta^b_a-\del_a a^b$.

\paragraph{Scalars}
For scalar fields we have simply
\beq
  \epsilon^\phi = 0.
\eeq

\paragraph{Vectors}
For vector fields we have
\beq
  \tilde{A}_a(\tilde{x})= \Lambda^b_a A_b(x) = (\delta^b_a-\del_a a^b) A_b(x),
\eeq
which gives
\beq
  \epsilon^A_b = -\del_b a^a A_a.
\eeq

\paragraph{Second rank tensors}
In this case we have
\beq
  \tilde{F}_{ab}(\tilde{x}) = \Lambda^c_a\Lambda^d_b F_{cd}(x),
\eeq
which gives
\beq
  \epsilon^F_{ab} = -\del_c a^d (\delta^c_a F_{db} + \delta^c_b F_{ad})
\eeq

\air
If the action depends on both scalars, vectors and second rank
tensors, their contributions will just add up. The \emph{diff} symmetry
criterion is then
\beq
  \int dx \del_a a^b \left[ T^a_{~b}
	+ \psi^{a} A_b
	+ \psi^{ac}F_{bc}+\psi^{da}F_{dc}
	\right]=0,
\eeq
where $\psi^a$ are the Euler-Lagrange equations associated with $A_a$
and $\psi^{ab}$ are the Euler-Lagrange equations associated with
$F_{ab}$. For the equation to be true for arbitrary $a^a$ we must have
\beq
  T^a_{~b} = 
	- \psi^{a} A_b
	- \psi^{ac}F_{bc}-\psi^{da}F_{dc}.
\eeq
Furthermore, we recognize $T^0_{~0}$ as the naive Hamiltonian with
opposite sign. In other words we have
\beql{diffhamilton}
  h = -T^0_{~0} = 
	 \psi^{0} A_0
	+ \psi^{0c}F_{0c}+\psi^{d0}F_{d0}
\eeq
We get the Hamiltonian $H_{naive}$ by elimination of time
derivatives in favour of momenta in the expression for
$h$. Immediately, we see that if the action depends only on scalar
fields, the Hamiltonian will be zero. This is an important result that
we can state as a theorem:
\newtheorem{theorem}{Theorem}
\begin{theorem}
  \label{the:scalar}
  Diffeomorphism invariant theories
  with Lagrangians that depend on scalar-transforming fields and their
  first derivatives have vanishing (naive) Hamiltonian.
\end{theorem}
The same result has been proved by von Unge in \cite{unge:1994}.

\air
The significance of the more general result \refeq{diffhamilton} is
perhaps not obvious.
For the case
of scalar fields it is of course simple and easily applicable. If we
also have vectors or higher rank tensors, \refeq{diffhamilton}
gives certainly not any simpler route for a calculation of the
Hamiltonian than its definition itself. On the other hand, it
allows us to make an interesting interpretation. Observe that the
Hamiltonian is proportional to the $\psi$'s, which by the equations of
motions (Euler-Lagrange equations) are zero. This means that at 
any point on the classical path the Hamiltonian will be zero, since
the $\psi$'s are then zero. In this sense the we say that the
Hamiltonian is \emph{dynamically zero}.
\begin{theorem}
  \label{the:general}
  Diffeomorphism invariant theories with a Lagrangian that depends
  only on tensor fields (of any rank) and their derivatives have a
  Hamiltonian that is dynamically zero.
\end{theorem}
This result is not completely general, since we have still considered
only tensor fields. It is not necessarily true if the Lagrangian
depends on e.g. spinors.
In this thesis, however, we will consider only fields of the first
kind.

An example that validates the result \refeq{diffhamilton} is given 
in section \ref{sec:ex-theorem2}. There we will derive the Hamiltonian
for the Polyakov string both directly from its definition, and using
the result in this section.

%% file: methods.tex
\section{Methods}

The main purpose of this thesis is to describe and apply two methods
for deriving high energy limits of various actions. This section is
devoted to a general description of these methods. 
The first is the simplest.
It can be applied to any model, although it does not always lead o any
interesting field equations.
The second requires more calculations, but makes it at the same time
possible to derive several limits. The limit found by the first method
is usually one of these.

The starting point is an action of the form
\beql{genaction}
  S = T\int dx {\cal L}(\phi, \partial \phi),
\eeq
where $T$ is some dimensionful constant, like mass or string tension.
It is a basic assumption that we can write the  Lagrangian as
$L=T{\cal L}$.  The quantity ${\cal L}$ can be called the
\emph{reduced} Lagrangian, since we have taken out the constant $T$. 
This action is clearly not very suitable for studying the $T \to 0$
limit. 
The philosophy now is to search for an action that is classically
equivalent to \refeq{genaction} as long as $T\neq 0$, but also well 
defined
for $T=0$. We will then treat this new action (with $T=0$ inserted) as
a $T\to 0$ limit of the original model.
The methods described below are systematic ways for finding such
actions.

\subsection{Method I: Auxiliary field}
\label{sec:method-I}

This is the simplest approach, and involves the introduction of an 
auxiliary field $\chi$. 
A reference for the this method is Karlhede and Lindstr\"om 
\cite{karlhede:1986}.
We use the ${\cal L}$ from the original
action and write
\beql{chiaction}
  S_\chi = \frac{1}{2} \int dx (\chi {\cal L}^2 + \frac{T^2}{\chi}).
\eeq
This action 
is equivalent to \refeq{chiaction}. To show this explicitly, we solve
the equations of motion for $\chi$:
\begin{eqnarray}
  \nonumber
  \delta \chi \Rarr \quad \delta S_\chi & = 
  & \frac{1}{2}\int dx (\delta\chi {\cal L}^2 - \frac{T^2}{\chi^2}\delta\chi)
  \\ \nonumber
  & = & \frac{1}{2} \int dx ({\cal L}^2 - \frac{T^2}{\chi^2})\delta\chi.
\end{eqnarray}
Using Hamilton's principle and demanding
$\delta S_\chi = 0$ for arbitrary variations $\delta \chi$ gives
\begin{eqnarray}
  \nonumber
  {\cal L}^2 - \frac{T^2}{\chi^2} & = & 0
  \\
  \chi & = & \frac{T}{{\cal L}}; \qquad \mbox{when}\quad T \neq 0.
\end{eqnarray}
If we put this back into \refeq{chiaction} we get
\[
  S_\chi = \frac{1}{2}\int dx(\frac{T}{{\cal L}}{\cal L}^2 
  + \frac{{\cal L}}{T}T^2) = T\int dx {\cal L} = S.
\]
Thus the two actions $S$ and $S_\chi$ are equivalent for $T \neq 0$. In
addition $S_\chi$ allows us to take the $T \to 0$ limit simply
by setting $T=0$ in the action. This gives
\beql{chi0action}
  S_\chi^{T=0} = \frac{1}{2} \int dx \chi {\cal L}^2.
\eeq
You may ask what this new field $\chi$ really is. From the current
point of view we cannot say anything more than we already have -- that
it helps us in our calculations. Hence the name \emph{auxiliary} field.

Note however, that in the simplest case of a massless particle in
section \ref{subsec:pointparticleI} we are lead to interpret $\chi$ as
the einbein.

A general remark on symmetry properties can already be made. 
Consider diffeomorphism invariance. We know that the integral
measure transforms as $dx\to dxJ^{-1}$, where $J$ is the Jacobi determinant
as defined by equation \refeq{def-jacobideterminant}. 
If the original action is to be \emph{diff} invariant, the
Lagrangian must transform as a density, i.e. 
${\cal L} \to J {\cal L}$. 
We see then that $S_\chi$ is also \emph{diff} invariant if we demand
$\chi$ to be an inverse density (i.e. scalar density of weight $-1$,
c.f. appendix \ref{sec:densities}). 
And since $\chi$ was introduced as an
auxiliary field with no \emph{a priori} physical interpretation, this
transformation property is something we can impose on $\chi$.

\subsubsection{Dynamics}

A variation of $\chi$ gives one equation of motion,
\beql{chi-motion}
  \delta\chi \Rarr\qquad {\cal L}^2=0 \quad\Rarr\quad {\cal L}=0.
\eeq
The effect of a variation in $\phi^i$, on the other hand, depends on
the form of ${\cal L}$:
\begin{eqnarray}
  \nonumber
  \delta\phi^i \Rarr\qquad \delta S &=&
	\12\int dx \chi 2{\cal L}\delta{\cal L}
  \\ &=& \nonumber
  \int dx \chi {\cal L}\left[ \frac{\del {\cal
	L}}{\del\phi^i}\delta\phi^i
	+\frac{\del{\cal
	L}}{\del(\del_a\phi^i)}\del_a\delta\phi^i\right]
  \\ &=& \nonumber
  \int dx \left[\chi{\cal L}\frac{\del{\cal L}}{\del\phi^i} -
  \del_a\left(\chi{\cal L}\frac{\del{\cal
  L}}{\del(\del_a\phi^i)}\right)\right]\delta\phi^i.
\end{eqnarray}
The field equation is found from demanding $\delta S=0$ for arbitrary
$\delta\phi^i$. The result is
\beql{phi-motion}
  \delta\phi \Rarr\qquad \chi{\cal L}\frac{\del{\cal L}}{\del\phi^i} -
  \del_a\left(\chi{\cal L}\frac{\del{\cal
  L}}{\del(\del_a\phi^i)}\right) = 0.
\eeq
Often, this equation will reduce to an identity by use of equation
\refeq{chi-motion}, ${\cal L}=0$. But it \emph{does} give non-trivial 
equations in cases where  
$\frac{\del{\cal L}}{\del\phi^i}\sim \frac{1}{{\cal L}}$
or $\frac{\del{\cal L}}{\del(\del_a\phi^i)}\sim \frac{1}{{\cal
L}}$.  (Then the factors ${\cal L}$ are eliminated from equation
\refeq{phi-motion}.) 
However, this is not the general situation, so method I has
limited applicability.

\subsection{Method II: Phase space}

This method of arriving at an action that admits taking the $T \to 0$ 
limit is designed for constrained systems. Demonstrations of the method 
can be found in 
\cite{marnelius:1982,karlhede:1986,isberg:1994,lindstrom:1991,lindstrom:1991b,lindstrom:1991c,hassani:1994}, but
also later in this thesis.
Again we start from the action
\refeq{genaction}. We derive the canonical conjugate momenta
\beq
  \pi_i = \frac{\del L}{\del \dot{\phi^i}},
\eeq
and find the total Hamiltonian as in section \ref{sec:hamilton},
\beq
  H = H' + \lambda^m\theta_m.
\eeq
The derivations of the total Hamiltonian $H$ involves working out the
constraint structure, which can be a cumbersome task. But since we are
interested only in the limit $T=0$, these can be simplified by
putting $T=0$ as early as possible.

\begin{figure}[t]
  \begin{center}
    \fbox{\epsfig{file=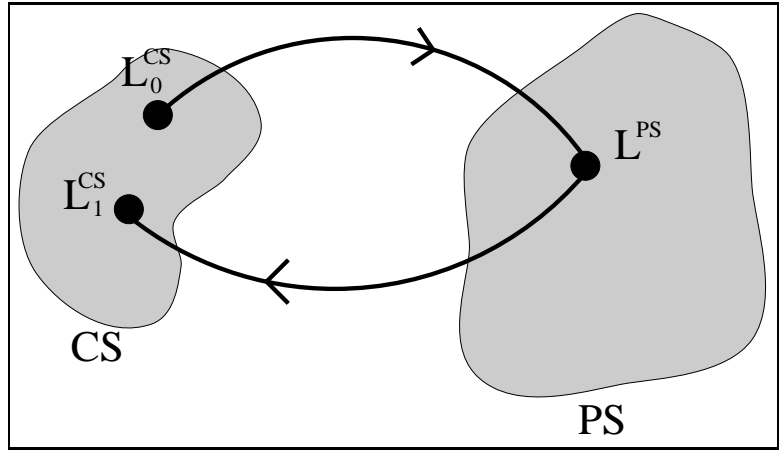, width=10cm}}
  \end{center}
  \caption{{\small We start off with a Lagrangian $L^{CS}_0$ in configuration
    space (CS), and perform a Legendre transform to the phase space
    (PS) Lagrangian $L^{PS}$. When we go back to configuration space
    we may, for constrained systems, end up
    with a new (but equivalent) Lagrangian
    $L^{CS}_1\neq L^{CS}_0$.
    And although $L^{CS}_0$ is not defined in the limit $T=0$,
    $L^{CS}_1$ may be.}} 
  \label{fig:method2}
\end{figure}

Now, having found the total Hamiltonian, we write down the 
phase space action
\beq
  S^{PS} = \int dx \left( \pi_i\dot{\phi^i} 
	- H(\phi, \pi, \nabla\phi) \right).
\eeq
The momenta 
$\pi_i$ can then be eliminated by solving their equations of
motion. (This is often called ``integrating out the momenta'', a
notation which is natural in the context of path integrals. In that
case we go from phase space to configuration space by literary
integrating out the momenta from the functional integral.)
Substituting for the solutions of $\pi_i$, we arrive at the
configuration space action
\beql{genCSction}
  S^{CS} = \int dx \left[ \pi(\phi, \del \phi)\dot{\phi} - H(\phi, 
   \pi(\phi, \del \phi), \nabla \phi) \right] .
\eeq
Unless the system under study is non-constrained (giving $H=H_{naive}$)
this action will contain something new compared to the one we started
with. In other words, it is different from the original configuration
space action, but still equivalent (see figure \ref{fig:method2}) to it.
Hopefully the new aspects make it possible to take the $T \to 0$ 
limit. What is new are the Lagrange multipliers, which are now
independent fields.  As discussed later, these may often be 
reinterpreted as components of some metric, or as (degenerated)
vielbeins.

In cases where there are no constraints imposed from the definition of
the momentum,
the calculations above will give only a circle where we end
up at the point we started. However, it will show possible
(c.f. \ref{sec:weylstring}) 
via some redefinitions in phase space to allow for the $T\to0$ limit
in a sensible way. This situation is better discussed when it appears.

%% file: strings.tex
\section{Introduction to strings and branes}

\subsubsection{Short history of strings}

Strings originated in the late sixties as a model describing strong
interactions \cite{veneziano:1968}. Quarks are known always to exist
in bound states, and the string approach was a proposal for explaining
this quark confinement. Very simplified, the picture was that of quarks
attached to strings.

This theory was pushed aside by the successful QCD (quantum
chromo-dynamics) theory.
But in 1974, Scherk and Schwarz \cite{scherk:1974} made the
remarkable suggestion that string theory was a correct mathematical
theory of a different problem, the unification of elementary particle
interactions with gravity.

After this the theory attained much attention, but had no real
breakthrough. Many properties made it attractive, but the problems were
too serious. However, with the introduction of \emph{supersymmetry} 
(a symmetry between bosons and fermions) into the \emph{superstring}
theory \cite{green:1984,green:1985,green:1987} 
(in contrast to the old \emph{bosonic}
string theory), a lot of the problems disappeared. 
And this is the theory that has attracted enormous attention over the
last fifteen years. 
In these years there have appeared different types of superstring
theories, but they are now thought to be limits of one fundamental
theory, which is called M-theory (or matrix theory), and is
for the moment under constant investigation.

\subsubsection{String theory}

Strings are one-dimensional objects with a length of the order of the
Planck length, $10^{-34}$m. They are free to vibrate, much like 
ordinary guitar strings. The possible modes of vibration are determined
by the string tension, which is the only fundamental parameter in
string theory.

We know that at larger scales the strings must behave as particles
with certain masses. The model is that the different vibration modes
of the one fundamental string, give rise to the whole zoo of particles
we know from elementary particle physics. Different vibration modes
mean different frequencies or energies, and hence, different masses of
the particles. Thus,
in principle, string theory could be used to \emph{derive} the mass
spectrum of all particles. This is one of the ultimate goals, but is in
practice very difficult.

As was mentioned above, this theory is a promising candidate for a
unification of quantum field theory (elementary particle physics) and
general relativity (gravitation). 
Actually, string theory is not
consistent without gravity.
This aspect of string theory is probably the most important, but there
are also other sides that make such a large number of theoretical physicists
talk warmly about it.

Another of its advantages is that it avoids altogether the
divergences that unavoidably appear in quantum field theories. This
is due to the fact that strings are not pointlike, but have extension
in space. Also in contrast to quantum field theories, string theory
involves no arbitrary choice of gauge symmetry group and choice of
representation: string theory is essentially unique.

The last comment is an example of a general attractive feature of
string theory, namely that there is very little freedom. Once the
basic theory is formulated, important results follow directly or by
consistency. The spacetime dimension is also fixed in this
way. Superstring theory is consistent only with 10 ($1+9$) spacetime
dimensions. (For bosonic strings there must be 26 dimensions.)
At first this may sound like a catastrophe, since we know by
everyday experience that the world is 4-dimensional ($1+3$). The way
to handle this difficulty, is to say that the extra 6 dimensions are
compactified, or curled up so that they play no role at large
scales. This is very much an \emph{ad hoc} assumption, but at least
string theory gives a way to understand the spacetime dimensionality.

\air
In this thesis we will not consider the supersymmetric string theory
(superstrings), but only different models for \emph{bosonic}
strings. Bosonic string theory does not have fermions, and is not a
realistic theory. However, it is a good starting point, and gives
insight in crucial aspects of the more realistic models as well.

\subsubsection{What is real?}

\emph{To be or not to be, that is the question.} Hamlet certainly had
other things than elementary particle physics in mind, but his
question is indeed fundamental for our discussion as well:
Do strings \emph{really}
exist, or do they not? And are strings the \emph{fundamental}
building blocks of the universe?

The traditional view is that the world
is built up of pointlike particles. These are ``things'' with certain
properties like mass and spin, but without extension in space. In other
words, they are thought to be \emph{zero}-dimensional. String theory
changes this view only insofar as the strings are not points, but have
extension in one dimension, i.e. are \emph{one}-dimensional.

With such views we are immediately faced with the question of model
versus reality. Is our model only a mathematical construction that by
coincidence happens to resemble the physical reality, or does the
success of the model give a deeper understanding of \emph{reality}
itself? 

To meet this question, we must know what we mean by reality. 
What does it actually mean that something is real?
One answer to this is to say that the real world is the
\emph{observable} world. This description works quite well in everyday
life. However, when it comes to elementary particles, the task of
observing 
becomes very difficult. For instance, our understanding of
\emph{shape} is useless at such small scales. 
The quantum theory of physics, with its Schr\"odinger equation and
Heisenberg relation, leads us to view the elementary particles as
rather diffuse objects that are somehow smeared out in spacetime.

Returning then to the question of the being of the fundamental building
blocks, one possible answer is to say that they \emph{are} neither
particles nor strings.
But they \emph{behave} at very small scales as
strings, and at larger scales as points. And someone has also said
that \emph{you are what you do}.

\subsubsection{Branes}

As we have accepted the leap from points to strings, it is
natural to go further and consider even higher-dimensional
objects. Perhaps the fundamental objects are membranes, two-dimensional
surfaces. Or why not $p$-dimensional \emph{$p$-branes}?

$p$-Brane\footnote
{
  The notation is such that a point particle is called a
  $0$-brane, and a string is called a $1$-brane.
}
theory is the obvious generalization from string theory. However,
strings seem to be special among the branes with their success as a
fundamental model. 
For instance, increasing the world surface dimensionality increases
the probability of finding divergences (from integrations on the world
surface) similar to those found in quantum field theories.

\air
A non-technical introduction to string theory is found in
\cite{green:1986,waldrop:1985,peat:1988}. 
Thorough textbooks on string theory are
\cite{green:1987,polchinski:1998}. 

%% file: pointparticle.tex
\section{The point particle}
\label{sec:pointparticle}

It is now time to do some real calculations to demonstrate how
everything we have said so far applies.
And we naturally start with the simplest possible case, the
relativistic point particle.

\subsection{The action}

The action of a relativistic point particle can be written as its mass
times the length of its world line:
\beql{point1}
  S = m\int ds.
\eeq
This length is a distance in spacetime.
We let $\tau$ parameterize the world line, and let $X^\mu(\tau)$ be
the particle's position at some moment.
Then we can write
\beql{ds2}
   ds^2 = -G_{\mu\nu}dX^\mu dX^\nu
        = -dX^\mu dX_\mu
        = -\frac{dX^\mu}{d\tau} \frac{dX_\mu}{d\tau} d\tau^2. 
\eeq
$G_{\mu\nu}$ denotes the spacetime metric. 
With the metric signature $(-,+,\dots,+)$, a timelike vector $v^\mu$
has negative norm, $v^2<0$. And since $\frac{dX^\mu}{d\tau}$ is
timelike, the overall minus sign above is a
conventional choice to make $ds^2$ a positive quantity.
%
We take the square root of \refeq{ds2}
and plug back in \refeq{point1} to get the 
more familiar expression for the relativistic point particle action:
\beql{pointparticleaction}
  S =  m\int d\tau\sqrt{ -\dot{X}^\mu \dot{X}_\mu} 
	= m\int d\tau\sqrt{-\dot{X}^2},
\eeq
where $\dot{X}^\mu \equiv \frac{\partial X^\mu}{\partial \tau}= 
\frac{dX^\mu}{d\tau}$.
This action is reparameterization (\emph{diff}) invariant by
construction. Since it is covariantly written as a scalar, 
and only contains derivatives of $X^\mu$, it easy to see that it is
also Poincar\'e invariant.

To show that \refeq{pointparticleaction} is indeed an appropriate
action for the point particle, and
to give an example of how Hamilton's principle can be applied, we will
now deduce the equations of motion from this action (in Minkowski
space).
Consider a small
variation $\delta X^\alpha$ in $X^\alpha$. This will give a small
variation $\delta S$ in the action, which we find as follows.
\begin{eqnarray}
  \delta X^\alpha \Rarr \delta S 
  & =& m\delta \int d\tau  \sqrt{-\dot{X}^2}
  = m\int d\tau \delta \sqrt{-\dot{X}^2}
  \\ \nonumber 
  & =& -m\int d\tau \12(-\dot{X}^\mu \dot{X}^\nu)^{-\12}
        2 \dot{X}_\alpha \delta \dot{X}^\alpha
\end{eqnarray}
The order of variation and differentiation can be interchanged, so we have
$\delta\dot{X}^\alpha=\frac{d}{d\tau}(\delta X^\alpha)$. This,
together with a partial integration gives
\begin{eqnarray} 
  \\ \nonumber
  \delta S & = & -m\int d\tau 
	\frac{-\dot{X}_\alpha}{\sqrt{\dot{X}^2}}
        \frac{d}{d\tau}(\delta X^\alpha)
  \\ \nonumber
  & = & -m\int d\tau \bigg( \underbrace{ \frac{d}{d\tau} \left[ 
         \frac{-\dot{X}_\alpha}{\sqrt{\dot{X}^2}}\delta X^\alpha
        \right] }_{\to 0}
        - \frac{d}{d\tau} \left[ 
	\frac{\dot{X}_\alpha}{\sqrt{-\dot{X}^2}} \right]
        \delta X^\alpha \bigg)
\end{eqnarray}
The first part leads to a boundary term, which gives zero contribution
since the fields are held fixed at the boundaries. Hamilton's
principle states that the action must be extremal for the dynamically
allowed fields. In other words, we must have $\delta S=0$ for
arbitrary $\delta X^\alpha$. Thus we end up with the equation
\beq
  \frac{d}{d\tau} \left[ 
	\frac{m\dot{X}_\alpha}{\sqrt{-\dot{X}^2}} \right]=0.
\eeq
We recognize the quantity within the
brackets as the relativistic \emph{momentum} for a point particle. 
(We will soon recover it by direct calculation.) The equation then
says that the momentum is conserved, which is a well known consequence
of translational invariance.

Furthermore, if we let $\tau$ be 
proper time, we have $\eta_{\mu\nu}\dot{X}^\mu\dot{X}^\nu = -1$. 
In this case the equations of motion read
\beq
  \ddot{X}^\mu = 0,
\eeq
i.e. the acceleration is zero. This is the familiar result for a free
point particle.

We would deduce the same equation of motion if we started from the
alternative 
action $S=m\int d\tau \dot{X}^2$. But this action has the disadvantage
that the parameter $\tau$ \emph{has} to be the proper time, i.e. it is
not reparameterization invariant. For this reason we will
in the following consider the action \refeq{pointparticleaction}.

\paragraph{Massless limit}

{}From now on we will focus on the massless limit ($m\to 0$) of the
action  
\refeq{pointparticleaction} above. We read off the Lagrangian and
find
\beq
  L = m{\cal L}(X,\dot{X}) = m\sqrt{-\dot{X}^\mu\dot{X}_\mu}.
\eeq
This will now be our starting point for a discussion on the massless
limit. We know that a particle's energy can be split into the rest
energy, which is a constant ($E_0=m$) and a kinetic energy. If the
total energy becomes very high, it is a good approximation to neglect
the rest energy. Thus the massless limit represents a high energy
limit of the point particle.

The derivations presented here for the point particle are also found
in  
\cite{lindstrom:1991}.

\subsection{Method I}
\label{subsec:pointparticleI}

As explained before, the easiest way to find a massless limit is by
introduction of an auxiliary field $\chi$. We write, according to the
general theory,
\begin{eqnarray}
  \nonumber
  S_\chi &=& \frac{1}{2}\int d\tau(\chi {\cal L}^2+ \frac{m^2}{\chi})
  \\ \label{eq:pointIaction}
  &       =& \frac{1}{2}\int d\tau(-\chi \dot{X}^\mu\dot{X}_\mu + 
           \frac{m^2}{\chi}).
\end{eqnarray}
The massless limit is obtained directly by putting $m=0$.
\beql{pointI-lim}
  S_\chi^{m=0} = -\12 \int d\tau \chi \dot{X}^\mu\dot{X}_\mu.
\eeq

It was mentioned as part of the motivation for studying the massless (or
tensionless) limits that they lead to
conformal invariant theories. We will now see an example of
this. First, we note that the action \refeq{pointI-lim} is 
obviously \emph{Poincar\'e} invariant. 

Under the \emph{dilatation} $X^\mu \to (1+c)X^\mu$, we get
$\dot{X}^2 \to (1+c)^2\dot{X}^2= (1+2c)\dot{X}^2$ to first order in
$c$.
The action will then be invariant under dilatations if $\chi$
transforms as 
\beql{point-dil}
  \delta_c X^\mu: \quad \chi \to (1-2c)\chi
\eeq
to first order.

The \emph{special conformal} transformations
$X^\mu \to X^\mu + b\cdot X X^\mu - \12 X^2 b^\mu$ give to first
order in $b$: 
$\dot{X}^2 \to (1+2b\cdot X)\dot{X}^2$. Thus the action will be
invariant if $\chi$ transforms as
\beql{point-spec}
  \delta_b X^\mu: \quad \chi \to (1-2b\cdot X)\chi.
\eeq
The field $\chi$ is an auxiliary field, so the transformation
properties \refeq{point-dil} and \refeq{point-spec} are something we
can \emph{impose} on $\chi$. And given these properties, we see that
the action is conformally invariant.
Thus, in the massless limit, we have that the original Poincar\'e 
symmetry is enlarged to full conformal symmetry.

\air
The equations of motion derived from \refeq{pointI-lim} are
\begin{eqnarray}
  \delta\chi \Rarr &\quad& \dot{X}^2 = 0,
  \\
  \delta X_\mu \Rarr && \frac{d}{d\tau}(\chi\dot{X}^\mu)=0.
\end{eqnarray}
$\dot{X}^\mu$ is a tangent to the world line, so the first equation
says that the particle follows a lightlike (or null-) curve. With the
conformal symmetry in mind, this is a natural result, since conformal
transformations are transformations that preserve the light cone
(c.f. section \ref{sec:sym-conf}).

An interesting observation is that the action \refeq{pointIaction}
leads us to identify $\chi$ as an inverse \emph{einbein} field
(c.f. appendix \ref{sec:vielbeins}). In 
\cite{brink:1976b} it is shown that the action for a point-particle
coupled to one-dimensional gravity through the einbein field 
$e=e^1_1$ can be written
\beq
S=\12\int d\tau\left(\frac{1}{e}\dot{X}^2-em^2\right),
\eeq
if we disregard the spin. 
And by comparison with \refeq{pointIaction} we have the identification
$\chi = -\frac{1}{e}$. Let us now see how this compares to the
results of method II.

\subsection{Method II}

{}From the Lagrangian $L(X, \dot{X}) = m\sqrt{-\dot{X}^2}$ we find
the canonical momenta
\beq
  P_\mu = \frac{\del L}{\del \dot{X}^\mu} = 
           -\frac{m\dot{X}_\mu}{\sqrt{-\dot{X}^2}}.
\eeq
This is the familiar relativistic expression for the momentum of a
free particle.
Notice also that this form sesults independently of which metric is 
used. This is true
because a metric is a function of positions only, not of their
derivatives;
$G_{\mu\nu} = G_{\mu\nu}(X).$

The Hamiltonian is defined as
\beq
  H_{naive} = P_\mu \dot{X}^\mu - L(X, \dot{X}).
\eeq
The fields $X^\mu$ are scalars under diffeomorphisms, so the naive
Hamiltonian vanishes, as discussed in section
\ref{sec:zerohamilton}. This can also easily be seen explicitly by
noting that
\beq
  P_\mu\dot{X}^\mu = \frac{-m\dot{X}_m}{\sqrt{-\dot{X}^2}}\dot{X}^\mu
    = m\sqrt{-\dot{X}^2} = L.
\eeq

The expression for $P_\mu$ is not invertible so in accordance with what we
said in section \ref{sec:constrainedsystems} there must exist some
constraints. And indeed we find
\[
  P^2 \equiv P^\mu P_\mu = \frac{m \dot{X}^\mu}{\sqrt{-\dot{X}^2}}
                           \frac{m \dot{X}_\mu}{\sqrt{-\dot{X}^2}}
      = m^2\frac{\dot{X}^2}{-\dot{X}^2} = -m^2
\]
\beq
  \Rarr P^2 + m^2 = 0.
\eeq

Since the naive Hamiltonian vanishes, the total Hamiltonian is made
from the constraint as follows
\beq
  H = \lambda(P^2 + m^2),
\eeq
where the coefficient $\lambda$ is a Lagrange multiplier. We 
write the phase space action as described earlier,
\beql{pointPSaction}
  S^{PS}=\int d\tau\left(P_\mu\dot{X}^\mu - \lambda(P^\mu P_\mu + m^2)\right).
\eeq
Now it is time to start simplifying and return to configuration space.
To do so we need to eliminate the momenta. 
A variations in $P_\mu$ gives
\beq
  \delta P_\mu \Rarr \delta S^{PS} = \int d\tau( \dot{X}^\mu 
           - 2\lambda P^\mu) \delta P_\mu.
\eeq
For $\delta S^{PS}$ to be zero for arbitrary (though infinitesimal)
variations $\delta P_\mu$, we need to have
\begin{eqnarray}
  \nonumber
  \dot{X}^\mu - 2\lambda P^\mu &=& 0
  \\
  P^\mu &=& \frac{\dot{X}^\mu}{2\lambda}.
\end{eqnarray}
Plugging this back into \refeq{pointPSaction} we end up with  a configuration 
space action,
\begin{eqnarray}
  \nonumber
  S^{CS} &=& \int d\tau\left( \frac{\dot{X}_\mu}{2\lambda}\dot{X}^\mu
         -\lambda(\frac{\dot{X}^\mu\dot{X}_\mu}{4\lambda^2}+m^2)\right)
  \\ \label{eq:pointCSaction}
  & = & \frac{1}{2}\int d\tau\left( \frac{1}{2\lambda} \dot{X}^\mu\dot{X}_\mu
         - 2\lambda m^2\right).
\end{eqnarray}
Comparison with the action \refeq{pointIaction} we found by using
method I,  
reveals that the two methods give exactly the same result. We just
have 
to identify the auxiliary field $\chi$ with the Lagrange
multiplier as $\chi = (-2\lambda)^{-1}$.

A discussion of the massless limit $m=0$ was given in the previous
section. 

%% file: NGstring.tex
\section{The Nambu-Goto string}
\label{sec:NGstring}

\subsection{The action}

\label{sec:stringNGaction}

The action for the point particle is proportional to the length of its world
line. This suggests that we can generalize to a string which sweeps out a
world \emph{surface} and say that its action is proportional to the area of
this surface. In mathematical terms, we have
\begin{equation}
  S = T \int dA.
\end{equation}
$T$ will have have the dimension of $energy/length$, or
$(length)^{-2}$ in natural units. We thus call it the string 
tension. It plays a role analogous to the particle mass.

Let us denote the spacetime metric by $G_{\mu\nu}$. If $\xi^a; a=0,1$
are world sheet coordinates that parameterize the world surface, we
can write the spacetime points of the surface as
$X^\mu=X^\mu(\xi)$.
The \emph{induced metric} $\gamma_{ab}$ is then given by 
(see appendix \ref{sec:curvature})
\beql{def-gamma}
  \gamma_{ab} = G_{\mu\nu}\frac{\del X^\mu}{\del \xi^a}
        \frac{\del X^\nu}{\del \xi^b}
	= G_{\mu\nu}\del_a X^\mu\del_b X^\nu.
\eeq
We write the inverse of this matrix as $\gamma^{ab}$, i.e. 
$\gamma^{ab}\gamma_{bc}=\delta^a_c$, and its determinant simply as 
$\gamma \equiv \det(\gamma_{ab})$.

With the introduction of the induced metric, an infinitesimal area
element of a 2-dimensional surface embedded in spacetime can be
written as (see e.g. \cite{hatfield:1992})
\beq
  dA = \sqrt{-\det(\gamma_{ab})}d\xi^0 d\xi^1.
\eeq
Then we can write the string action as
\beql{stringNGaction}
  S = T\int d^2\xi \sqrt{-\gamma}.
\eeq
This is the famous Nambu-Goto form 
\cite{nambu:1970,goto:1971}
of the action for a (bosonic) string.

\paragraph{Tensionless limit}
The high-energy limit of the strings has been studied with
different approaches. For a short review, and a list of references,
the reader may consult \cite{isberg:1994}.

In analogy with the point particle, we expect the tensionless limit
$T\to 0$ of
strings to give insight into the high-energy behaviour, just as the
massless limit of particles does.  Schild \cite{schild:1977} was the
first to study this limit. Later, the tensionless limit of strings
(not just the Nambu-Goto string) has
been studied by several authors
\cite{karlhede:1986,isberg:1994,lindstrom:1991,lindstrom:1991b,lindstrom:1991c,zheltukhin:1988,barcelos-neto:1989,lizzi:1986,gamboa:1990}.

We will in the following go through the derivation of different
tensionless limits, starting from the Nambu-Goto string
\refeq{stringNGaction}.

\subsection{Method I} 
\label{sec:nullstring}

The (reduced) Lagrangian is ${\cal L}(X,\del X)= \sqrt{-\gamma}$, and
following the general recipe, we write
\begin{eqnarray}
  \nonumber
  S_\chi &=& \12\int d^2\xi(\chi {\cal L}^2 + \frac{T^2}{\chi})
  \\ \label{eq: stringIaction}
  &=& -\12 \int d^2\xi(\chi\gamma - \frac{T^2}{\chi}),
\end{eqnarray}
where $\chi$ is the auxiliary field.
This action is equivalent to \refeq{stringNGaction}, and allows us to
take the $T=0$ limit. We find simply
\beql{stringIT0action}
  S_\chi^{T=0}=-\12\int d^2\xi~\chi\gamma
\eeq
Variations in $\chi$ and $X^\mu$ give the equations of motion:
\begin{eqnarray}
  \delta\chi \Rarr &\quad& \gamma \equiv \det(\gamma_{ab})=0,
  \\
  \delta X_\mu \Rarr &&
  \del_b\left[\chi\epsilon^{ac}\epsilon^{bd}\gamma_{cd}\del_a
	X^\mu\right] = 0.
\end{eqnarray}
The induced metric is degenerate, which means that the surface is
a \emph{null surface}. It means that the world surface has tangent 
vectors $v^a$ that
are null (lightlike), i.e. $v^2=0$. Tensionless strings are for
this reason often referred to as null-strings.

A difference from the point particle is that we cannot give
$\chi$ a geometric interpretation.

\paragraph{Conformal invariance}
As was the case for the point particle, this action is conformally
invariant given that $\chi$ transforms in a special way. If we now for
a moment generalize to $D$-dimensional surfaces, the induced metric
(which is then $D$-dimensional)
transforms under
Poincar\'e, dilatation and special conformal transformations as 
\begin{eqnarray}
  \label{eq:gamma-ct}
  \nonumber
  \delta_{\omega,a}X^\mu: &\quad& \gamma_{ab} \to \gamma_{ab},
  \\
  \delta_c X^\mu: &\quad& \gamma_{ab} \to (1+2c)\gamma_{ab},
  \\ \nonumber
  \delta_b X^\mu: &\quad& \gamma_{ab} \to (1+2b^\nu X_\nu)\gamma_{ab}.
\end{eqnarray}
Thus, the combination $\chi\gamma$ in \refeq{stringIT0action} 
is conformally invariant provided that $\chi$
transforms as
\begin{eqnarray}
  \label{eq:chi-ct}
  \nonumber
  \delta_{\omega,a} X^\mu: &\quad& \chi \to \chi,
  \\
  \delta_c X^\mu: &\quad& \chi \to (1-2Dc)\chi,
  \\ \nonumber
  \delta_b X^\mu: &\quad& \chi \to (1-2Db^\nu X_\nu)\chi.
\end{eqnarray}
Putting $D=2$ gives the string result, which we are interested in
here.

\subsection{Method II} 

As for a point particle we start by deriving the momenta. Defining 
$\dot{X}^\mu\equiv \del_0X^\mu=\frac{\del X^\mu}{\del\xi^0}$ and
$\acute{X}^\mu\equiv\del_1X^\mu=\frac{\del X^\mu}{\del\xi^1}$, we get
\begin{eqnarray}
  \nonumber
  P_\mu &\equiv& \frac{\del L}{\del \dot{X}^\mu} = 
   T\frac{\del}{\del \dot{X}^\mu}\sqrt{-\gamma} = 
   \frac{T}{2\sqrt{-\gamma}} \frac{\del (-\gamma)}{\del \dot{X}^\mu}
  \\ \nonumber
  &=& \frac{T}{2\sqrt{-\gamma}}\frac{\del}{\del\dot{X}^\mu}
   \left( (G_{\alpha\beta}\dot{X}^\alpha\acute{X}^\beta)^2 - 
     (G_{\alpha\beta}\dot{X}^\alpha\dot{X}^\beta)
      (G_{\alpha\beta}\acute{X}^\alpha\acute{X}^\beta)\right)
  \\
  &=&\frac{T}{\sqrt{-\gamma}}\left( 
     (\dot{X}^\alpha\acute{X}_\alpha)\acute{X}_\mu
     -(\acute{X}^\alpha\acute{X}_\alpha)\dot{X}_\mu\right).
\end{eqnarray}
As was the case for the point particle, we cannot use this expression
for $P_\mu$ to solve for $\dot{X}^\mu$, i.e. it is not invertible. 
Therefore we look for constraints. The constraints should not be functions
of time derivatives ($\dot{X}^\mu$), and this limits the number of
possible candidates. Anyway, we find
\begin{eqnarray}
  \nonumber
  P^2 &=& \frac{T^2}{-\gamma} ( (\dot{X}\cdot \acute{X})\acute{X}^\mu
           -(\acute{X}^\mu)^2 \dot{X}^\mu)
          ((\dot{X}\cdot \acute{X})\acute{X}_\mu - {\acute{X}_\mu}^2 
          \dot{X}_\mu)
  \\ \nonumber
  &=& \frac{T^2}{-\gamma} \underbrace{\left( \dot{X}^2\acute{X}^2 - 
         (\dot{X}\cdot \acute{X})^2 \right)}_\gamma
  = -T^2 \acute{X}^2
  \\
  P_\mu\acute{X}^\mu &=& \frac{T}{\sqrt{-\gamma}}
    ( (\dot{X}\cdot \acute{X})^2 \acute{X}^2-\acute{X}^2 \dot{X}\cdot 
    \acute{X}) = 0.
\end{eqnarray}
Thus, we have these primary constraints
\beq
  \theta_0\equiv P^2 + T^2\acute{X}^2 \approx 0; \qquad
  \theta_1\equiv P^\mu \acute{X}^\mu \approx 0.
\eeq

\noindent
Also, we notice that
\beq
  P_\mu\dot{X}^\mu = \frac{T}{\sqrt{-\gamma}}\left( (\dot{X}\cdot \acute{X})^2
    -\acute{X}^2\dot{X}^2 \right) = T\sqrt{-\gamma} = L.
\eeq
So the naive Hamiltonian vanishes, in accordance with the general discussion 
in section \ref{sec:zerohamilton}.  The primary constraints do not
give rise to secondary constraints.
Thus, the total Hamiltonian can be written as a sum of $\theta_0$ and
$\theta_1$,
\beql{NGstringHamilton}
  H=\lambda (P^2+T^2\acute{X}^2) + \rho P_\mu\acute{X}^\mu,
\eeq
where $\lambda$ and $\rho$ are Lagrange multipliers. This gives the
phase space action
\beql{stringPSaction}
  S^{PS} = \int d^2\xi \left[P_\mu\dot{X}^\mu - \lambda (P_\mu P^\mu
    + T^2\acute{X}^\mu\acute{X}_\mu) - \rho P_\mu \acute{X}^\mu \right].
\eeq
The equations of motion for $P_\mu$ gives
\begin{eqnarray}
  \nonumber
  \dot{X}^\mu - 2\lambda P^\mu - \rho \acute{X}^\mu &=&0
  \\
  P^\mu &=& \frac{\dot{X}^\mu - \rho \acute{X}^\mu}{2\lambda}.
\end{eqnarray}
Substituted back into \refeq{stringPSaction} we find after a few
rearrangements 
\beql{stringCSaction}
  S^{CS}  = \int d^2\xi \frac{1}{4\lambda}\left[ 
	\dot{X}^2 - 2\rho\dot{X}\cdot \acute{X}
    +(\rho^2-4T^2\lambda^2)\acute{X}^2\right].
\eeq

\paragraph{A Weyl-invariant action} 
\label{sec:NG-Weyl}
We can now identify $\lambda$ and
$\rho$ as components of a metric field in this way:
\beql{lagr-metric}
  g^{ab} = \Omega \left( 
	\begin{array}{cc}
		1 & -\rho \\
		-\rho & \rho^2-4T^2\lambda^2
	\end{array}
	\right),
\eeq
where $\Omega$ is any scaling function.
We have then $g=\det(g_{ab}) = \frac{1}{4T^2\lambda^2\Omega^2}$, 
and we can write \refeq{stringCSaction} as
\beql{weylstring}
  S = \frac{T}{2}\int d^2\xi \sqrt{-g}g^{ab}\gamma_{ab}.
\eeq
This is the Weyl-invariant string action, which is further discussed
in section \ref{sec:weylstring}.

However, by this rewriting we are in no better position to study the
limit $T\to 0$. To do so, we go back to \refeq{stringCSaction}, and
make another interpretation of the Lagrange multipliers.

\subsubsection{Limit one}

Following \cite{lindstrom:1991} we introduce an auxiliary vector density
(of weight $-\frac{1}{2}$ -- see appendix \ref{sec:densities} for an
explanation of what is meant by \emph{density}), 
\beq
  V^a = \frac{1}{2\sqrt{\lambda}}(1,-\rho).
\eeq
Noting that we can write
\[
  \lambda^2\acute{X}^2 = \frac{\gamma_{11}}{4V^0V^0} 
    = \frac{\gamma \gamma^{00}}{4V^0V^0}
\]
we see that we can write the action \refeq{stringCSaction} in 
the simpler form:
\beql{NG-1}
  S^{CS} = \int d^2\xi\left( V^aV^b\gamma_{ab} - 
    \frac{T^2\gamma \gamma^{00}}{V^0V^0} \right).
\eeq
The tensionless limit is then easily found if we set $T=0$. We end up with
\beql{NG-1-lim}
  S^{T=0} = \int d^2\xi~V^aV^b\gamma_{ab}.
\eeq
The equations of motion follow from variations in $V^a$ and $X^\mu$,
\begin{eqnarray}
  \delta V^a &\Rarr& \quad V^b\gamma_{ab}=0,
  \\
  \delta X^\mu &\Rarr& \quad \del_b(V^aV^b\del_aX^\mu) = 0.
\end{eqnarray}
The first equation means that $\gamma_{ab}$ has eigenvectors with 
zero eigenvalue,
which again means that the determinant of the induced metric is zero,
i.e. $\det(\gamma_{ab})=0$. We have thus the same situation as we
found using method I. We may say that the two limits represent the
same physical situation, but are differently formulated.
And accepting that there must be a \emph{formal} difference between the two 
results is not difficult, 
as we have in the present case one extra variable.
($V^a$ are \emph{two} variables, while $\chi$ is \emph{one}.) 

\air
Owing to the \emph{diff} invariance, we can choose a gauge (the transverse
gauge) where $V^a = (v,0)$, where $v$ is a constant. 
The equations of motion then reduce to 
\beq
  \ddot{X}^\mu=0; \qquad 
  \dot{X}^2=\dot{X}^\mu\acute{X}_\mu=0.
\eeq
We conclude that the tensionless string behaves as a collection of
massless particles moving  transversally to the direction of the string. 

\air
We said above that $V^a$ are \emph{densities}, which is
necessary to preserve worldsheet \emph{diff} invariance.
We can interpret the $V^a$ fields further by comparing with the
Weyl-invariant form \refeq{weylstring}, which by introduction of
inverse \emph{zweibeins} $e_A^a$ (c.f. appendix \ref{sec:vielbeins}) can 
be written
\beq
  S= \frac{T}{2}\int d^2\xi ~e \eta^{AB}e_A^{~~a}e_B^{~~b}\gamma_{ab}
	=\12 \int d^2\xi \eta^{AB} \tilde{e}_A^{~~a}\tilde{e}_B^{~~b}
		\gamma_{ab},
\eeq
where $\tilde{e}_A^{~~a}\equiv\sqrt{Te}e_A^{~~a}$ are densities, and 
$a,b,A,B=0,1$.
Clearly the limit $T\to 0$ \refeq{NG-1-lim} corresponds to the case
when
the zweibeins has become parallel, and we can make the substitution
$\tilde{e}_A^{~~a}\to V^a$.
%

{}From the transformation properties \refeq{gamma-ct} of
$\gamma_{ab}$, it is easy to check that \refeq{NG-1-lim} is
conformally invariant provided $V^a$ transform under dilatations and
special conformal transformations as
\begin{eqnarray}
  \delta_c X^\mu: &\quad& V^a \to (1-c)V^a,
  \\
  \delta_b X^\mu: &\quad& V^a \to (1-b^\nu X_\nu) V^a.
\end{eqnarray}

Another important symmetry observation is 
that, with this interpretation of the Lagrange multipliers, the
manifest covariance is broken in \refeq{NG-1}, but
recovered in the $T=0$ limit, \refeq{NG-1-lim}.

\subsubsection{Limit two}

To find a limit that resembles the result of method I, it is clear
that we must at least eliminate one degree of freedom.  This will now
be done, and we start from \refeq{stringCSaction} and eliminate $\rho$
by solving its equation of motion. The result is immediate:
\beq
  \rho = \frac{\dot{X}\cdot\acute{X}}{\acute{X}^2}.
\eeq
Substituted back into the action, this gives
\begin{eqnarray}
  \nonumber
  S & = & \int d^\xi\frac{1}{4\lambda} \left(
        \frac{\dot{X}^2\acute{X}^2-(\dot{X}\cdot\acute{X})^2}{\acute{X}^2}
        - 4T^2\lambda^2\acute{X}^2 \right)
  \\ 
  & = & \int d^2\xi \frac{1}{4\lambda} \left(
        \frac{1}{\gamma^{00}} - 4T^2\lambda^2\gamma\gamma^{00} \right),
\end{eqnarray}
where we use that $\acute{X}^2 = \gamma_{11} = \gamma \gamma^{00}$.
If we define $V \equiv 2\lambda \gamma\gamma^{00}$, the action can be
written in the form
\beq
  S^{CS}= \12\int d^2\xi(\frac{\gamma}{V}-T^2V).
\eeq
This is identical to what we obtained using method I, by the
identification $\chi = -V^{-1}$, and was discussed in the previous
section. It is interesting to note that we can arrive at the same
formulation of the limit with both methods. We also note that method
II gives an additional possible limit, and is thus more general.

%% file: pbrane.tex
\section{$p$-Branes}
\label{sec:p-brane}
The discussion in the previous section, can quite straight forwardly
be generalized to $p$-branes. This is what we will do in this
section. Most of what is here written is also found in
\cite{hassani:1994} and 
\cite{isberg:1994}.

\subsection{The action}

The $p$-brane action is a direct generalization of 
the Nambu-Goto string action \refeq{stringNGaction}. The action is now
taken to be proportional to the ``area'' of the $(p+1)$-dimensional
world ``surface''. (We use the words area, surface or volume even
though they may refer to higher dimensional objects.) 
We get
\beql{pbraneNGaction}
  S= T\int d^{p+1}\xi\sqrt{-\gamma},
\eeq
where $\gamma_{ab}=\del_aX^\mu\del_bX_\mu$ 
is the induced metric as in the previous
section. The indices $a,b$ now take values $0,\dots,p$.
The action is invariant under the reparametrization 
$\xi^a \to \sigma^a(\xi)$, which must be true from the geometrical
interpretation of the action. It is also easy to demonstrate, by
noting the transformation properties of the integral measure and the
induced metric. The Jacobi determinant \refeq{def-jacobideterminant} is 
$J=\det(\frac{\del \xi^a}{\del\sigma^b})$, and we get
\begin{eqnarray}
  d^{p+1}\xi & \to & J^{-1} d^{p+1}\xi,
  \\
  \gamma_{ab} \equiv 
	\frac{\del X^\mu}{\del\xi^a}\frac{\del X_\mu}{\del\xi^b}
  & \to & (\frac{\del\xi^a}{\del\sigma^c})
 	(\frac{\del\xi^b}{\del\sigma^d})
	\gamma_{cd}
  \\
  \sqrt{-\gamma} & \to & J\sqrt{-\gamma}.
\end{eqnarray} 
Thus the action transforms as
\beq
  S \to T\int d^{p+1}\xi J^{-1} \sqrt{-\gamma} J = S,
\eeq
which proves the invariance.

\air
In the $p$-brane action, the constant $T$ has natural dimensions of
$(length)^{-(p+1)}$, which means \emph{mass} for $p=0$ (points) and
\emph{tension} for $p=1$ (strings). Although the dimensionality
varies with $p$, we generally refer to $T$ as the $p$-brane
\emph{tension}. And we will now turn to the problem of finding models
where this constant is allowed to be zero.

\subsection{Method I}

Formally this first method will give exactly the same as it did for
strings, since the 
action \refeq{pbraneNGaction} looks the same. The only difference is that
$\gamma_{ab} = \del_a X^\mu \del_b X_\mu$
is no longer a $2\times 2$ matrix, but a $(p+1)\times(p+1)$ matrix.
The result is
\beql{p-braneIaction}
  S_\chi = -\12\int d^{p+1\xi} (\chi\gamma - \frac{T^2}{\chi}),
\eeq
and the tensionless limit:
\beq
  S_\chi^{T=0} = -\12\int d^{p+1}\xi~\chi\gamma.
\eeq
It is found to be conformally invariant in the same way as the string,
but now with $D=p+1$ in the $\chi$ transformations \refeq{chi-ct}.
As before, the $\chi$ equation of motion, $\det(\gamma_{ab})=0$, 
says that the world surface is degenerate.

\subsection{Method II}

As the attentive reader may already have guessed, we start by deriving
the momenta. The Lagrangian is read off from \refeq{pbraneNGaction} to
give $L=T\sqrt{-\gamma}$. Then we find
\beq
  P_\mu \equiv \frac{\del L}{\del(\del_0 X^\mu)} 
  = -\12 T\sqrt{-\gamma}\gamma^{ab}
     \frac{\del \gamma_{ab}}{\del (\del_0 X^\mu)},
\eeq
where we have used the result in appendix \ref{sec:det-derivative}.
Furthermore we have
\begin{eqnarray}
  \nonumber
  \gamma_{ab} &=& G_{\alpha\beta}\del_a X^\alpha \del_b X^\beta
  \\
  \frac{\del}{\del(\del_0 X^\mu)}\gamma_{ab} 
  &=& \delta^0_a \del_b X_\mu + \delta^0_b \del_a X_\mu
  \\ \nonumber
  \gamma^{ab} \frac{\del \gamma_{ab}}{\del(\del_0 X^\mu)}
  &=& 2\gamma^{a0}\del_a X_\mu.
\end{eqnarray}
So the momenta can finally be written
\beq
  P_\mu = T\sqrt{-\gamma} \gamma^{a0}\del_a X_\mu,
\eeq
and are not invertible.
Following the usual route, we go on by searching for constraints, and
find
\begin{eqnarray}
  P^2 &=& -T^2 \gamma \gamma^{a0} \del_a X_\mu \gamma^{b0} \del_b X^\mu
	=-T^2 \gamma \gamma^{00},
  \\[0.2cm] \nonumber
  P_\mu \del_k X^\mu &=& T\sqrt{-\gamma} \gamma^{a0} \del_a X_\mu \del_k X^\mu
  \\ 
  &=& T\sqrt{-\gamma} \delta^0_k
  =0 \quad \mbox{when}\quad k > 0.
\end{eqnarray}
We know that
$\gamma$ and $\gamma^{00}$ contain time derivatives, so the first of
these equations does not look
like a proper constraint at first sight. But the time derivatives cancel
in this special combination, as we easily see:
\beq
  \gamma \gamma^{00}= \gamma \frac{\det \gamma_{ik}}{\det \gamma_{ab}}
  = \det \gamma_{ik},
\eeq
where $\det \gamma_{ik}$ is the determinant of the spatial part of the 
$\gamma$-matrix. It contains no time derivatives. So we have found
these constraints:
\beql{pbrane-constraints}
  \theta_0 \equiv P^2 + T^2\gamma \gamma^{00} \approx 0; \qquad
  \theta_k \equiv P_\mu \del_k X^\mu \approx 0 \quad \mbox{for}\quad k>0.
\eeq
Since the action is \emph{diff} invariant, and the theory contains only the
scalar (under world volume diffeomorphisms) field $X^\mu$, the naive
Hamiltonian vanishes 
(c.f. theorem \ref{the:scalar} in section \ref{sec:zerohamilton}).
The total Hamiltonian is thus a sum of the constraints:
\beql{pbrane-hamilton}
  H = \lambda(P^2 + T^2\gamma\gamma^{00})
  +\rho^k P_\mu\del_k X^\mu.
\eeq
Now we are ready to write the phase space action:
\beq
  S^{PS} = \int d^{p+1}\xi\left[ P_\mu \del_0 X^\mu
    -\lambda(P^2+T^2\gamma\gamma^{00})
    -\rho^k P_\mu\del_k X^\mu \right].
\eeq
The equations of motion for $P_\mu$ reads
\beq
  P^\mu = \frac{1}{2\lambda}(\del_0 X^\mu - \rho^k\del_k X^\mu).
\eeq
Inserted into the phase space action this gives a new configuration
space action
\beql{pbraneCSaction}
  S^{CS} = \int d^{p+1}\xi \frac{1}{4\lambda}\left[
    \gamma_{00} - 2\rho^k\gamma_{k0} +\rho^i\rho^k\gamma_{ik}
    -4\lambda^2 T^2\gamma\gamma^{00} \right].
\eeq

\subsubsection{Limit one}
In analogy to the string case we introduce the vector density
\beq
  V^a \equiv \frac{1}{2\sqrt{\lambda}} (1, -\rho^k),
\eeq
which is now $(p+1)$-dimensional. Using this variable we can write the
action as
\beq
  S_1 = \int d^{p+1}\xi \left[ V^a V^b \gamma_{ab} - 
    T^2\frac{\gamma\gamma^{00}}{V^0 V^0} \right].
\eeq
The $T \to 0$ limit may now be taken and we end up with
\beq
  S_1^{T=0} = \int d^{p+1}\xi V^a V^b \gamma_{ab}.
\eeq
We see that, in this sense, the tensionless $p$-brane is a simple
generalization from the string.

\subsubsection{Limit two}
Another way to a tensionless limit is to eliminate $\rho^k$ in the 
action \refeq{pbraneCSaction}. The equations of motion for $\rho^k$
gives
\beq
  \rho^i\gamma_{ik} = \gamma_{k0}.
\eeq
If we define $G_{ij}$ to be the spatial part of $\gamma_{ab}$, i.e.
$G_{ij} = \gamma_{ij}$, $i,j = 1,2\dots, p$, and $G^{ij}$ as its
inverse, we can write
\beq
  \rho^i = G^{ik}\gamma_{0k}.
\eeq
Inserted back into the action, this gives
\beq
  S_2 = \int d^{p+1}\xi \frac{1}{4\lambda}\left[
    \gamma_{00} - G^{kj}\gamma_{0j}\gamma_{0k}
    -4\lambda^2 T^2\gamma\gamma^{00} \right].
\eeq
Furthermore, we find
$\gamma^{00}(\gamma_{00} - G^{kj}\gamma_{0j}\gamma_{0k}) = 1$, which
means that the action can be written
\beq
  S_2 = \int d^{p+1}\xi \frac{1}{4\lambda}
  \left(\frac{1}{\gamma^{00}} - 4\lambda^2 T^2\gamma\gamma^{00} \right).
\eeq
Defining the scalar $V \equiv 2\lambda\gamma\gamma^{00}$ we end up with
\beq
  S_2 = \12 \int d^{p+1}\xi \left( \frac{\gamma}{V} - T^2 V\right).
\eeq
Again, this is exactly the same action as we found by using method I 
\refeq{p-braneIaction}, if we let $\chi = -V^{-1}$.

\air
It should be noted that as long as we are interested only in the
tensionless limit, we could set $T=0$ already in the Hamiltonian
\refeq{pbrane-hamilton}. The subsequent calculations would then be a
little simpler while giving the same result.

%% file: weylstring.tex

\section{The string in the Polyakov form}
\label{sec:weylstring}

\subsection{The action}
\label{sec:weylstring-action}

A Weyl-invariant form of the $p$-brane action can be written
\cite{lindstrom:1988}
\beql{ULp-brane}
  S = d^{-\frac{d}{2}} T\int d^d\xi \sqrt{-g} (g\cdot\gamma)^\frac{d}{2},
\eeq
where $d=p+1$ is the dimension of the world volume of the membrane. The
fields to be taken as the independent variables are $X^\mu$ and $g_{ab}$.

The action is also world volume \emph{diff} invariant, which is easy to
check. Consider the reparametrization $\xi\to\sigma(\xi)$. Then the
fields will transform as
\begin{eqnarray}
  g_{ab}(\xi) &\to& \tilde{g}_{ab}(\sigma) = 
	\frac{\del\xi^a}{\del\sigma^c}\frac{\del\xi^b}{\del\sigma^d}
	g_{cd}(\xi),
  \\
  X^\mu(\xi) &\to& \tilde{X}^\mu(\sigma) = X^\mu(\xi).
\end{eqnarray}
The Jacobi matrix is $J^a_b = \frac{\del\xi^a}{\del\sigma^b}$, and if
we let its determinant be written $J$, we have
\begin{eqnarray}
  d^d\xi &\to&  d^d\sigma = \frac{1}{J} d^d\xi,
  \\
  \sqrt{-g} &\to& \sqrt{-\tilde{g}} =  J\sqrt{-g},
  \\
  g\cdot\gamma &\to& \tilde{g}\cdot\tilde{\gamma} = g\cdot\gamma.
\end{eqnarray}
The action will transform as
\begin{eqnarray}
  S\to \tilde{S}&=& \int d^d\sigma\sqrt{-\tilde{g}}
	\tilde{g}\cdot\tilde{\gamma}
  \\
  &=& \int d^d\xi\sqrt{-g}g\cdot\gamma = S,
\end{eqnarray}
which proves the statement of diffeomorphism invariance.

The classical equivalence between the Weyl-invariant and Nambu-Goto
type actions
is shown by elimination of the metric fields $g_{ab}$. Start from the
Weyl-invariant action \refeq{ULp-brane},
and demand it to extremal with respect to $g^{ab}$, 
\[
  \delta g^{ab}\Rarr \quad -\12\sqrt{-g} g_{ab} 
	(g\cdot\gamma)^{\frac{d}{2}}
	+\frac{d}{2}\sqrt{-g}(g\cdot\gamma)^{\frac{d}{2}-1}
	\gamma_{ab}
	=0
\]
which gives
\begin{eqnarray}
  \nonumber
  d^{-1}g_{ab}(g\cdot\gamma)&=&\gamma_{ab}
  \\
  \nonumber
  \mbox{det} \Rarr\qquad
  d^{-d} g (g\cdot\gamma)^{d} &=&\gamma
  \\
  d^{-\frac{d}{2}}\sqrt{-g}(g\cdot\gamma)^{\frac{d}{2}} 
	&=&\sqrt{-\gamma}.
\end{eqnarray}
Substituted into the action, this gives
\beq
  S=T \int d^d\xi \sqrt{-\gamma},
\eeq
which is exactly the Nambu-Goto action for a $p$-brane.

In the string case when $d=2$, the Weyl-invariant action is
particularly neat. Since we have
in mind to derive momenta, we are interested in the
$\dot{X}^\mu$-dependence of 
this action.  The only place we find this dependence is in $\gamma_{ab}$;
$\gamma_{00}$ is quadratic and $\gamma_{0i}$ are linear in $\dot{X}$. This
means that the derivative of $\gamma \cdot g$ will depend linearly on
$\dot{X}$. Thus, for $d=2$ we will find a linear relationship between the
momenta and $\dot{X}$, whereas $d \not = 2$ gives something more 
complicated because of the exponent $\frac{d}{2}$. These complications will
make the calculations considerably more difficult.
For this reason we will in the following consider only the string case
($d=2$):
\beql{Pol-string}
  S = \frac{T}{2} \int d^2\xi \sqrt{-g} g\cdot\gamma.
\eeq
This action was first described by Brink--Di Vecchia--Howe and by 
Deser--Zumino \cite{brink:1976,deser:1976}, 
but is often named after Polyakov, who
has given important contributions to its investigation. 
A generalization of this action to $p$-branes 
(which is, in contrast to \refeq{ULp-brane}, \emph{not}
Weyl-invariant) is the Howe-Tucker action \cite{howe:1977} 
\beql{howe-tucker}
  \frac{T}{2}\int d^{p+1}\xi \sqrt{-g}\left(g\cdot\gamma - (p-1)\right).
\eeq
In the following we will consider the two-dimensional case, the
\emph{Polyakov action} 
\refeq{Pol-string}, and try to derive tensionless limits of it. Since
it is classically equivalent to the Nambu-Goto action, we expect the
results here also to be equivalent to those we found in section
\ref{sec:NGstring}. 

\subsection{Method I}
Following the usual logic we get an equivalent action,
\beq
  S_\chi = \12\int d^2\xi\left(-g (g\cdot\gamma)^2\chi + 
    \frac{T^2}{4\chi}\right),
\eeq
where again $\chi$ is an auxiliary field. In the tensionless limit we
have: 
\beql{weylstringIaction}
  S_\chi^{T=0} = -\12\int d^2\xi~ \chi g(g\cdot\gamma)^2.
\eeq
Variations of the fields give the equations of motion
\begin{eqnarray}
  \delta \chi &\Rarr& \quad g(g^{ab}\gamma_{ab})^2 = 0,
  \\
  \delta g^{ab} &\Rarr& \quad \chi\left[
	-gg_{ab}(g\cdot\gamma)^2 +
	2g(g\cdot\gamma)\gamma_{ab}\right]=0,
  \\
  \delta X_\mu &\Rarr& \quad \del_b\left[
	\chi g (g\cdot\gamma) g^{ab}\del_a X^\mu\right]=0.
\end{eqnarray}
To see exactly what this means, we use the totally antisymmetric
symbol\footnote{
In two dimensions we have $\epsilon^{00}=\epsilon^{11}=0$,
$\epsilon^{01}=-1$, $\epsilon^{10}=1$.
}
$\epsilon^{ab}$ to
write the inverse of $g_{ab}$ as
$g^{ab}=g^{-1}\epsilon^{ac}\epsilon^{bd}g_{cd}$, which gives
\begin{eqnarray}
  \label{eq:pol-1}
  g^{-1}(\epsilon^{ac}\epsilon^{bd}g_{cd}\gamma_{ab})^2&=&0,
  \\
  \label{eq:pol-2}
  \chi\left[ -g^{-1} g_{ab}
  (\epsilon^{ec}\epsilon^{fd}g_{cd}\gamma_{ef})^2
  +2(\epsilon^{ec}\epsilon^{fd}g_{cd}\gamma_{ef})\gamma_{ab}\right]&=&0,
  \\
  \label{eq:pol-3}
  \del_b\left[\chi g^{-1}(\epsilon^{eg}\epsilon^{fh}g_{ef}\gamma_{gh})
	\epsilon^{ac}\epsilon^{bd}g_{cd}\del_aX^\mu \right] &=&0.
\end{eqnarray}
If $g^{-1}=0$ we get from \refeq{pol-2} the solution
$\chi ((\epsilon^{ec}\epsilon^{fd}g_{cd}\gamma_{ef})\gamma_{ab}=0$,
i.e. $\epsilon^{ec}\epsilon^{fd}g_{cd}\gamma_{ef}=0$ or
$\chi=0$. However, $g^{-1}=0 \Leftrightarrow \det(g_{ab})=g\to\infty$,
which is a situation we are not interested in. So we assume that
$g^{-1}\neq 0$, in which case we find
\beql{pol-lim1}
  \epsilon^{ac}\epsilon^{bd}g_{cd}\gamma_{ab}= 0,
\eeq
which satisfies all the equations above.
The fact that the other equations of motion reduce to identities by
use of \refeq{pol-lim1} stems from the form of the action
\refeq{Pol-string}, which is not as it should be for method I to work
well (c.f. section \ref{sec:method-I}).

By two-dimensional \emph{diff} invariance we can choose a
parameterization that fixes $g_{ab}$ to $g_{ab}=\Omega\gamma_{ab}$,
with $\Omega$ as a positive definite function.
Then, by Weyl-invariance, we may rescale the metric to get
$g_{ab}=\gamma_{ab}$.  (This represents a gauge choice.)
Then \refeq{pol-lim1} gives
$\det(\gamma_{ab})=0$. And since the induced metric is degenerate in one
gauge, it is degenerate in all systems. So the tensionless Polyakov
action has generally the solution
\beq
  \det(\gamma_{ab})=0.
\eeq
In the above gauge the $X^\mu$ equation \refeq{pol-3} becomes
\beq
  \del_b\left[\chi\epsilon^{ac}\epsilon^{bd}
	\gamma_{cd}\del_aX^\mu\right]=0.
\eeq
which is what we found for the $T\to 0$ limit of the 
Nambu-Goto string. Hence we conclude that method I gives exactly the
same tensionless limits for the Nambu-Goto and Polyakov strings. In
the next section we will see that this is true also for the second
method.

\subsection{Method II}
We start again by deriving the momenta:
\begin{eqnarray}
  \nonumber
  P_\mu & \equiv & \frac{\del L}{\del \dot{X}^\mu} 
  \\ \nonumber
  & = & \12 T\sqrt{-g} g^{ab}\frac{\del \gamma_{ab}}{\del \dot{X}^\mu}
  \\ 
  & = & T\sqrt{-g}(g^{00}\dot{X}_\mu + g^{10}\acute{X}_\mu),
  \\
  \Pi^{\mu\nu} & \equiv & \frac{\del L}{\del \dot{g}_{\mu\nu}} = 0.
\end{eqnarray}
Since $\Pi^{\mu\nu}$ vanishes everywhere, we know that  $g_{\mu\nu}$ are 
not really
dynamical variables. We will later see explicitly that they play 
(with a suitable identification) the role of Lagrange multipliers. 

{}From the equation for $P_\mu$ we see easily that the transformation 
from configuration space to phase space is invertible. This means that we
can obtain $\dot{X}^\mu$ from $P^\mu$. Simple rearrangement indeed gives
\beql{PolinverseP}
  \dot{X}_\mu = \frac{1}{g^{00}}\left( \frac{P_\mu}{T\sqrt{-g}}-
    g^{10}\acute{X}_\mu \right).
\eeq
The fact that this is possible further means that the momenta are
independent functions of $\dot X^\mu$. Thus we will have no functions 
connecting them by $\theta_m(P,X,\acute{X})=0$, and hence there will be 
no constraints in the theory. For higher dimensional Weyl-invariant
$p$-branes we would not find the momenta to be invertible. In this
sense the string case is special.

Now, as we do not have to think about constraints, we go on by explicitly 
deriving the Hamiltonian. The total Hamiltonian will in this case equal 
the naive Hamiltonian. The way to do this is first to define the quantity
\beq
  h = h(X,\del X, P) \equiv P_\mu \dot{X}^\mu - L(X,\del X).
\eeq
If we manage to eliminate time derivatives of $X$ (i.e. $\dot{X}^\mu$), 
we arrive at the Hamiltonian $H$ which is a function of $X$, $P$ and 
\emph{spatial} derivatives of $X$ (i.e. $\acute{X}^\mu$). Thus, if we 
start from $h$ and substitute for $\dot{X}^\mu$ we find:
\begin{eqnarray}
  \nonumber
  h & = & P_\mu\dot{X}^\mu - L
  \\ \nonumber
  & = & P_\mu \dot{X}^\mu - \12 T\sqrt{-g}(g^{00}\dot{X}^\mu\dot{X}_\mu 
  + 2g^{10}\dot{X}^\mu\acute{X}_\mu
  +g^{11}\acute{X}^\mu\acute{X}_\mu)
  \\ \nonumber
  & = & \frac{1}{2Tg^{00}\sqrt{-g}} P^2 - \frac{g^{01}}{g^{00}}P\cdot \acute{X}
  + \12 T\sqrt{-g} (\frac{(g^{10})^2}{g^{00}}-g^{11}) \acute{X}^2
  \\ \nonumber
  & \equiv & H(P, X, \acute{X}).
\end{eqnarray}
Rearranging the terms we can write this as
\beq
  H = \frac{1}{2Tg^{00}\sqrt{-g}} (P^2 + T^2 \acute{X}^2)-
  \frac{g^{01}}{g^{00}} P\cdot \acute{X}.
\eeq
Here we have $T$ in the denominator of the first term. If we let $T\to0$ that
term will blow up, and the limit will not be well defined. To overcome this
problem we can make the redefinitions
\begin{eqnarray}
  \lambda & \equiv & \frac{1}{2Tg^{00}\sqrt{-g}},
  \\
  \rho & \equiv & -\frac{g^{10}}{g^{00}}.
\end{eqnarray}
This is the same identification as we did in \refeq{lagr-metric}.
Interpreting $\lambda$ and $\rho$ as Lagrange multipliers, this gives
exactly the Hamiltonian for the Nambu-Goto string \refeq{NGstringHamilton}.

\paragraph{Alternative calculation of the Hamiltonian}
\label{sec:ex-theorem2}
In section \ref{sec:zerohamilton} (theorem \ref{the:general})
we saw that diffeomorphism invariant
models made from tensor fields $g_{ab}$ will have a naive Hamiltonian
\beq
  H = \psi^{0a}g_{0a} + \psi^{a0}g_{a0},
\eeq
where $\psi^{ab}=0$ are the Euler-Lagrange equations associated with
$g_{ab}$. To show that
this gives indeed the same result, we will now present an explicit
calculation. We first have to find the $\psi$'s.
\begin{eqnarray}
  \nonumber
  \psi^{ab} &\equiv& \del_c\frac{\del L}{\del(\del_cg_{ab})}
	-\frac{\del L}{\del g_{ab}}
  \\
  &=&-\frac{\del L}{\del g_{ab}}
	= -\12\sqrt{-g}( g^{ab} g^{cd}\gamma_{cd} 
	- g^{ca}g^{db}\gamma_{cd}).
\end{eqnarray}
This gives
\begin{eqnarray}
  \nonumber
  h&=& 2\psi^{0b}g_{0b}
  \\ \nonumber
  &=& -T\sqrt{-g}(\12 g^{cd}\gamma_{cd} - g^{c0}\gamma_{c0})
  \\
  &=& -T\sqrt{-g}(-\12 g^{00}\gamma_{00} + \12 g^{11}\gamma_{11}).
\end{eqnarray}
Remember that $\gamma_{00}=\dot{X}^2$ and $\gamma_{11}=\acute{X}^2$,
and use equation \refeq{PolinverseP} to eliminate 
$\dot{X}$ in favour of the momenta. Then we end up with
\beq
  H = \frac{1}{2\sqrt{-g}g^{00}T}(P^2 +T^2\acute{X}^2) 
	-\frac{g^{10}}{g^{00}},
\eeq
which is just what was found above.

\air
We find the same Hamiltonian as in the Nambu-Goto case, and hence the
two models lead to the same phase space action, and of course the same
limits. This result should not come as a chock, as the two actions are
classically equivalent. We also remember that in the phase space
formulation we were able to deduce the Weyl-invariant action from the
Nambu-Goto action (c.f. section \ref{sec:NG-Weyl}).

One interesting observation concerns the degrees of freedoms. Naively,
$g_{ab}$ has three degrees of freedom, while $\lambda$ and $\rho$ are
only two degrees of freedom. But owing to the symmetries, there is no
real freedom in $g_{ab}$ (remember that it is an auxiliary field). 
The 2D \emph{diff} invariance ``kills'' two degrees of freedom, and
Weyl-invariance ``kills'' the last. Similarly, in the Nambu-Goto case
$\lambda$ and $\rho$ represent no real degrees of freedom, due to the 
two-parametric \emph{diff} invariance.
The \emph{superficial} difference in degrees of freedom is then understood
from the fact that the Nambu-Goto action has no Weyl-symmetry. 

%% file: Dbrane.tex

D-branes\footnote{
  The \emph{D} is an abbreviation for
  \emph{Dirichlet}.
}
\cite{polchinski:1995}
are soliton-like, extended objects (so-called topological
defects) that appear naturally in string theory. They are defined
by the property that strings can end on them, but have their own
dynamics. 

\section{Actions}

In \cite{polchinski:1998} the action for a Dp-brane is  given as
\beql{Dp-brane}
  S= T_p\int d^{p+1}\xi~ e^{-\phi}
	\sqrt{-\det(\gamma_{ab}+B_{ab}+F_{ab})},
\eeq
where $T_p$ is a constant, $\phi$ is the dilation field, 
and\footnote{ 
  Remember that $\del_a = \frac{\del}{\del\xi^a}$ and 
  $\del_{[a}A_{b]}=\del_a A_b- \del_b A_a$.
}
$F_{ab}= 2\pi\alpha'\del_{[a}A_{b]}$, $A_a$ being a gauge field and
$2\pi\alpha'$ the inverse of the fundamental string tension.
Furthermore,
\beq
  \gamma_{ab}(\xi)\equiv \del_aX^\mu\del_bX^\nu G_{\mu\nu}(X); \qquad
  B_{ab}(\xi) \equiv \del_aX^\mu\del_bX^\nu {\cal B}_{\mu\nu}(X)
\eeq
are the induced metric and antisymmetric tensor on the brane.
$G_{\mu\nu}$ is the background (symmetric) metric, and 
${\cal B}_{\mu\nu}$ is the background (antisymmetric) Kalb-Ramond field.
The indices take values $\mu =0,\dots,D-1$; $a=0,\dots,p$.

\paragraph{The Born-Infeld action} 
The original 
Born-Infeld action\cite{born:1934} was (unsuccessfully)
invented to describe nonlinear electrodynamics, and has the form
\beq
  S = \int d^4x\sqrt{-\det(\eta_{\mu\nu} + f_{\mu\nu})},
\eeq
where $\eta_{\mu\nu}$ is the Minkowski metric, and 
$f_{\mu\nu}=\del_{[\mu} A_{\nu]}$ is the electromagnetic field
strength. 

If we exchange $\eta_{\mu\nu}$ for a general metric $g_{\mu\nu}$ we
will have a gravity-coupled model. Furthermore, if we consider the
\emph{two}-dimensional case and use the \emph{induced} metric
$\gamma_{ab}$ we will get a kind of string. For higher dimensions we
may interpret the action as describing some kind of $p$-brane:
\beq
  S = T \int d^{p+1}\xi \sqrt{-\det(\gamma_{ab} + F_{ab})},
\eeq
where T is some constant. This resembles very much the D-brane action
\refeq{Dp-brane}. Thus, we will call \refeq{Dp-brane} the Born-Infeld
action for D-branes. 

The dilation $e^{-\phi}$ makes no difference to what
concerns the dynamics, and can for our purposes be disregarded (or
taken together with the constant $T_p$). This will be done in the
following. 
The independent field variables are the embedding $X^\mu(\xi)$ and the
gauge fields $A_a(\xi)$.

\paragraph{A Weyl-invariant action}
It is shown in \cite{lindstrom:1997} that the Born-Infeld action can
be written in the two classically equivalent forms
\begin{eqnarray}
  \label{eq:Dp-weylbrane}
  S&=&T_p\int d^{p+1}\xi \sqrt{-s}
  (s^{ab}(\gamma_{ab}+B_{ab}+F_{ab}))^\frac{p+1}{2},
  \\
  \label{eq:Dp-tuckerbrane}
  S&=&\frac{T}{2}\int d^{p+1}\xi \sqrt{-s}\left[ 
	s^{ab}(\gamma_{ab}+B_{ab}+F_{ab})-(p-1)\right],
\end{eqnarray}
where $s_{ab}$ is an auxiliary tensor field with no symmetry assumed.
In the usual way we have defined $s$ as the determinant, 
$s\equiv\det(s_{ab})$ and $s^{ab}$ as the inverse, 
$s^{ab}s_{bc} = \delta^a_c$.
Elimination of $s_{ab}$ gives back the original form \refeq{Dp-brane}.
The first of these \refeq{Dp-weylbrane} is Weyl-invariant (under
rescalings of $s_{ab}$), while the
second \refeq{Dp-tuckerbrane} is simpler when it comes to calculations.

In two dimensions ($p=1$) the two actions are the same.
For the same reason as we investigated only the Weyl-invariant
\emph{string} in section \ref{sec:weylstring}, we will for the moment
consider only the two-dimensional case of the Weyl-invariant D-brane
action. 

A reference where the second of the alternative formulations of the
D-brane action has been used is \cite{abouzeid:1997}.

\paragraph{$T_p\to0$ limit}
The Dp-brane tension is given by
$T_p=\frac{1}{g(2\pi)^p\sqrt{\alpha'}^{p+1}}$, where $g$ is the string
coupling. The $T_p\to 0$ limit can thus be viewed as a
\emph{strong coupling} limit where $g\to\infty$ and $\alpha'$ held
fixed. 
We will focus on this limit in what follows.

\section{The Born-Infeld action}

Defining $M_{ab}\equiv \gamma_{ab}+B_{ab}+F_{ab}$ and $M\equiv \det(M_{ab})$
we write the Born-Infeld Dp-brane action as
\beql{DBI-action}
  S = T\int d^{p+1}\xi \sqrt{ -M},
\eeq
What is new compared to the usual $p$-brane action is the addition of the 
antisymmetric terms $B_{ab}$ and $F_{ab}$. We will get the old
$p$-brane in the limit $A_a=0$ and ${\cal B}_{\mu\nu}=0$. 

\subsection{Method I}

Introduction of an auxiliary field $\chi$, and taking the $T=0$
limit gives, in the usual way,
\beq
  S^{T=0}_\chi = -\12\int d^{p+1}\xi~ \chi \det(M_{ab}).
\eeq
The equation of motion for $\chi$ is found from a variation
$\delta\chi$:
\beq
  \delta \chi \Rarr \quad
	\det(M_{ab}) = 0.
\eeq 
This is similar to what we found for the strings and p-branes. But in
the present case the degeneracy does not imply that the world volume is
a null surface (c.f. section \ref{sec:nullstring}). It only gives a
relation between the
$X^\mu$ and $A_a$ fields.

\subsection{Method II}

The calculations presented here are given in more detail by 
Lindstr\"om and von Unge in \cite{lindstrom:1997}. 
In the following we have for simplicity set ${\cal B}_{\mu\nu}=0$,
which means $M_{ab}= \gamma_{ab}+F_{ab}$.
Calculations with the antisymmetric ${\cal B}_{\mu\nu}$ field included
are given in \cite{gustafsson:1998}. The Lagrangian is given from
\refeq{DBI-action} as $L=T\sqrt{-M}$.

The field variables are $X^\mu(\xi)$ and $A_a(\xi)$, and the canonical
conjugate momenta associated with them are 
\begin{eqnarray}
  \label{eq:D-pi}
  \Pi_\mu &\equiv& \frac{\del L}{\del \dot{X}^\mu} 
	= \frac{T}{2}\sqrt{-M} M^{(a0)}\del_a X_\mu,
  \\
  \label{eq:D-p}
  P^a &\equiv& \frac{\del L}{\del \dot{A}_a}
	= \frac{T}{2}\sqrt{-M} M^{[a0]}2\pi\alpha',
\end{eqnarray}
where $M^{ab}$ is the inverse of $M_{ab}$. Round parenthesis and
brackets around the indices denote symmetrization and
antisymmetrization respectively, i.e. $M^{(ab)} = M^{ab} + M^{ba}$;
$M^{[ab]} = M^{ab} - M^{ba}$.

The equations (\ref{eq:D-pi}, \ref{eq:D-p}) are not invertible, and give rise
to the following primary constraints:
\begin{eqnarray}
  \Theta_i &\equiv& \Pi_\mu\del_i X^\mu 
	+ \frac{P^j}{2\pi\alpha'} F_{ij} \approx 0,
  \\
  \Theta_A &\equiv& \Pi_\mu\Pi^\mu 
	+ \frac{P^i\gamma_{ij}P^j}{(2\pi\alpha')^2} + T^2\det(M_{ij})
	\approx 0,
  \\
  \Theta_B &\equiv& P^0 \approx 0,
\end{eqnarray}
where $i$ takes spatial values $i=1,\dots,p$.

The naive Hamiltonian can be calculated straight forwardly, and reads
\beq
  H_{naive} = P^a\del_a A_0.
\eeq
(Equivalently, we could use theorem \ref{the:general} of section
\ref{sec:zerohamilton} and write $h=\psi_A^0 A_0$, where $\psi_A^0$ is
the Euler-Lagrange equation associated with $A_0$. 
This would, up to a total derivative, $\del_aX^a$, yield the same result.)
Consistency conditions on the primary constraints give us one
secondary constraint
\beq
  \Theta_C \equiv \del_c P^c \approx 0,
\eeq
but there are no tertiary constraints. The total Hamiltonian will
then be
\beql{DBI-hamilton}
  H = P^a\del_a A_0 
	+ \lambda\Theta_A + \rho^i \Theta_i + \sigma\Theta_B 
	+ \tau\Theta_C.
\eeq
The phase space action is
\begin{eqnarray}
  \nonumber
  L^{PS} &=& \Pi_\mu \del_0 X^\mu + P^a\del_0 A_a
	-P^a\del_a A_0
	-\lambda\Theta_A
	-\rho^i\Theta_i
	-\sigma P^0
	-\tau \del_a P^a
  \\
  &=& \Pi_\mu \del_0 X^\mu
	+P^i\del_0A_i
	-P^i\del_i(A_0-\tau)
	-\lambda\Theta_A
	-\rho^i\Theta_i.
	-(\sigma-\del_0\tau)P^0
\end{eqnarray}
We can redefine\footnote{
Note that in the path integral picture, a shift in the fields does not
make any difference, as
$\int{\cal D}A_0 F[A_0-\tau] = \int{\cal D}A_0 F[A_0]$,
in analogy with the simple result
$\int dx f(x-a) = \int dx f(x)$.
}
$A_0-\tau\to A_0$ and $\sigma-\del_0\tau\to \sigma$.
Thus, we get
\beq
  L^{PS}=  \Pi_\mu \del_0 X^\mu
	+\frac{P^i F_{ij}}{2\pi\alpha'}
	-\lambda\Theta_A
	-\rho^i\Theta_i
	-\sigma P^0.
\eeq
Elimination of $\sigma$ is trivial, and elimination of the momenta
gives a configuration space action
\begin{eqnarray}
  \nonumber
  S^{CS}&=& \int d^{p+1}\xi \frac{1}{4\lambda} \bigg[
	\gamma_{00} - 2\rho^i\gamma_{0i}
	+\rho^i\rho^j\gamma_{ij}
  \\&&
	+ \hat{\gamma}^{ij}(F_{0i}-\rho^kF_{ki})
	(F_{0j}-\rho^lF_{lj})
	-4\lambda^2T^2\det(M_{ij}) \bigg],
\end{eqnarray}
where $\hat{\gamma}^{ij}$ is the inverse of the spatial part of
$\gamma_{ab}$, i.e. $\hat{\gamma}^{ij}\gamma_{jk}=\delta^i_k$.
The $T\to 0$ limit is now easily taken by dropping the last term.
And as shown in \cite{lindstrom:1997} this gives rise to two different
tensionless limits of the D-brane:
\begin{eqnarray}
  \label{eq:DBI-limit1}
  S_1^{T=0} &=& \frac{1}{4} \int d^{p+1}\xi V\det{M},
  \\
  \label{eq:DBI-limit2}
  S_2^{T=0} &=& \frac{1}{4} \int d^{p+1}\xi V^aW^bM_{ab},
\end{eqnarray}
$V$, $V^a$ and $W^a$ are scalar and vector fields that are defined by
means of the Lagrange multipliers, but may be treated as independent
fields.  The result would be the same if we included the background
field ${\cal B}_{\mu\nu}$ (with the proper modification of $M_{ab}$).
The first action \refeq{DBI-limit1} is identical to what we found
using method I.

If we define $V^a = e_0^{~a}+e_1^{~a}$ and $W^a=-e_0^{~a}+e_1^{~a}$,
equation \refeq{DBI-limit2} can be rewritten as\footnote{ 
In conventions where $\eta^{AB}=diag(-1,+1)$ and $\epsilon^{10}=+1$.
}
\beql{DBI-limit3}
  S_3^{T=0} = \frac{1}{4}\int d^{p+1}\xi\left(
	\eta^{AB}e_A^{~a}e_B^{~b}-\epsilon^{AB}e_A^{~a}e_B^{~b}\right)M_{ab},
\eeq
where $A,B=0,1$.
We may identify $e_A^{~a}$ as zweibeins, and the form above then shows
that the dynamics of the tensionless D-brane is governed by an action
that involves a degenerate metric, $g^{ab}=\eta^{AB}e_A^{~a}e_B^{~b}$, of
rank 2.
%
%

The equations of motion derived from this action can be shown
\cite{lindstrom:1997} to imply that the world volume of the brane
generally splits into a collection of tensile strings or, in special
cases, massless particles. Thus, it leads to a parton picture of
D-branes in this limit.

\section{Weyl-invariant form}
\label{sec:weylDbrane}

In the two-dimensional case we found the Weyl-invariant D-brane to be
\beql{D2-weylaction}
  S = \frac{T}{2}\int d^2\xi \sqrt{-s} s^{ab}M_{ab}.
\eeq
Throughout this section we will set ${\cal B}_{\mu\nu}=0$, giving
$M_{ab}=\gamma_{ab} + 2\pi\alpha'\del_{[a}A_{b]}$. 
The subsequent calculations of tensionless limits resembles much those
in the previous section and those for the Polyakov string.
 
%

\paragraph{Method I}
As discussed in the introduction (section \ref{sec:method-I}) the
action \refeq{D2-weylaction} is not on the form that we need for
method I to give dynamical equations for the tensionless limit.
This was also true for the Polyakov string action in section
\ref{sec:weylstring}, but in that case we could nonetheless use method
I to make interesting interpretations concerning the tensionless
limit. An important property that made that possible was the
symmetries that allowed us to choose a gauge where the auxiliary
metric was equal to the induced metric, $g_{ab}=\gamma_{ab}$
(proportionality would be enough).
In the present case, however, the auxiliary tensor $s_{ab}$ is not
symmetric and hence is 4-parametric. And since the Weyl+\emph{diff}
symmetry is still only 3-parametric we cannot by a gauge choice set
$s_{ab}=M_{ab}$ similar to what we did for the Polyakov string.

\subsection{Method II}

The fields to be considered as independent variables in the action
\refeq{D2-weylaction} are $X^\mu$, $A_a$ and $s_{ab}$. Let us first derive
the canonical conjugate momenta associated with these fields.
\begin{eqnarray}
  \Pi_\mu & \equiv & \frac{\del L}{\del \dot{X}^\mu}
        = \frac{T}{2}\sqrt{-s}s^{cd}\frac{\del\gamma_{dc}}{\del\dot{X}^\mu}
   =  T \sqrt{-s} \left( s^{00} \dot{X} + \12(s^{01}+s^{10})\del_1X_\mu 
        \right)
  \label{eq:BImomentumPi}
  \\
  P^a & \equiv & \frac{\del L}{\del \dot{A}_a} 
        = \frac{T}{2}\sqrt{-s}s^{cd}\frac{\del F_{dc}}{\del (\del_0A_a)}
   =  -T \sqrt{-s} \12(s^{0a} - s^{a0})2\pi\alpha'
  \label{eq:BImomentumP}
  \\
  \Sigma^{ab} & \equiv & \frac{\del L}{\del \dot{s}_{ab}} = 0
  \label{eq:BImomentumSigma}
\end{eqnarray}
The first equation \refeq{BImomentumPi} is invertible which means that 
we can find an explicit 
expression for $\dot{X}^\mu$:
\beq
  \dot{X}_\mu = \frac{1}{s^{00}} \left(\frac{\Pi_\mu}{T\sqrt{-s}}
        -\12(s^{01}+s^{10})\del_1X_\mu \right) 
\eeq
The second equation \refeq{BImomentumP} is obviously not
invertible. We have actually found a momentum $P$ that is completely
independent of the fields $A$.
Its definition gives then immediately rise to the constraints
\begin{eqnarray}
  \Theta_0 &\equiv& P^0 \approx 0,
  \\
  \Theta_1 &\equiv& P^1 +
	  \frac{T}{2}\sqrt{-s}(s^{01}-s^{10})2\pi\alpha' \approx 0.
\end{eqnarray}
The last equation \refeq{BImomentumSigma} says that the conjugate
momenta to $s_{ab}$ are
identically zero. This follows the pattern of previous results,
and $s_{ab}$ are non-dynamical variables to be treated on the same
footing as Lagrange multipliers.

We are now ready to derive the naive Hamiltonian. Disregarding 
$\Sigma^{ab}$, we have:
\begin{eqnarray}
  \nonumber
  H_{naive} & = &
        \Pi_\mu\dot{X}^\mu + P^a\dot{A}_a - 
        \frac{T}{2}\sqrt{-s}s^{ab}(\gamma_{ab} 
	+ 2\pi\alpha'(\del_aA_b - \del_bA_a))
  \\ \nonumber
  & = & f(\Pi, \dot X, \del_1 X) + g(P, \dot A, \del_1 A)
\end{eqnarray}
To arrive at a proper Hamiltonian we have to eliminate all time derivatives.
Consider first $g$:
\begin{eqnarray}
  \nonumber
  g & = & P^a\del_0A_a - T\sqrt{-s} \12 s^{ab}
	(\del_aA_b-\del_bA_a)2\pi\alpha'
  \\ \nonumber
  & = &P^a\del_0A_a - T\sqrt{-s} \12[s^{01}(\del_0A_1-\del_1A_0)
   + s^{10}(\del_1A_0 - \del_0A_1)]2\pi\alpha'
  \\ \nonumber
  & = & P^a\del_aA_0. 
\end{eqnarray}
Now remains only to rewrite $f$. If we insert the expression for 
$\dot{X}^\mu$ and simplify we find:
\begin{eqnarray}
  \nonumber
  f & = & \Pi_\mu \dot{X}^\mu - T\sqrt{-s}\12 s^{ab}\gamma_{ab}
  \\ \nonumber
  & = & \frac{1}{2Ts^{00}\sqrt{-s}} \Pi^2 
        - \frac{s^{01}-s^{10}}{2s^{00}}\del_iX^\mu\Pi_\mu
  \\ \nonumber          
  & &   + \frac{T\sqrt{-s}}{2s^{00}} \left( \frac{1}{4}(s^{01}+ s^{10})
        (s^{01}+ s^{10}) - s^{00}s^{11} \right)\gamma_{11}
  \\ \nonumber
\end{eqnarray}
If we calculate $P^1P^1$ we see that we can still simplify the expression
within the large brackets of the last term:
\beq
  \frac{1}{4}(s^{01}+ s^{10}) (s^{01}+ s^{10}) - s^{00}s^{11}
        = \frac{P^1P^1}{-T^2s(2\pi\alpha')^2} - \frac{1}{s}.
\eeq
If we put together terms with the same coefficients we can now write the
naive Hamiltonian in its simplest form:
\begin{eqnarray}
  \nonumber     
  H_{naive} &=& \frac{1}{2Ts^{00}\sqrt{-s}}\left( \Pi_\mu\Pi^\mu 
        + \frac{P^1\gamma_{11}P^1}{(2\pi\alpha')^2} 
	+ T^2\gamma_{11}\right) 
  \\    
  & &   - \frac{s^{01}+s^{10}}{2s^{00}}\Pi_\mu\del_1X^\mu
        + P^a\del_aA_0.
\end{eqnarray}
The consistency condition on the primary constraint $\Theta_0$ gives a
secondary ``Gauss law'' constraint 
\beq
  \Theta_2\equiv \del_aP^a\approx 0,
\eeq
while $\Theta_1$ gives nothing new.
There are no tertiary constraints.
The four component fields of $s_{ab}$ are Lagrange multipliers which
can be redefined as
\begin{eqnarray}
  \lambda & \equiv & \frac{1}{2Ts^{00}\sqrt{-s}},
  \\
  \rho & \equiv & -\frac{s^{01}+ s^{10}}{2s^{00}},
  \\
  \varphi & \equiv & \frac{T}{2}\sqrt{-s}(s^{01}-s^{10}).
\end{eqnarray}
Including the constraints, we can then write the total Hamiltonian as
\begin{eqnarray}
  \nonumber
  H &=& 
	\lambda (\Pi^2 + \frac{P^1\gamma_{11}P^1}{(2\pi\alpha')^2}
	+T^2\gamma_{11})
	+\rho \Pi_\mu\del_1X^\mu
	+ P^a\del_aA_0
	\\&&	
	+\sigma_0P^0
	+\sigma_1(P^1+\varphi)
	+\tau\del_aP^a
\end{eqnarray}
The phase space Lagrangian is 
$L^{PS}=\Pi_\mu\del_0X^\mu+P^a\del_0A_a-H$, and a variation of $\varphi$
gives $\sigma_1=0$, which means that the constraint
$\Theta_1=P^1+\varphi$
in fact makes no difference.
Then we see that we have exactly the same Hamiltonian and phase space
Lagrangian as we derived
from the Born-Infeld action for the two-dimensional D-brane (D-string)
\refeq{DBI-hamilton}.
Hence, we get the same tensionless limits \refeq{DBI-limit1} and
\refeq{DBI-limit2}/\refeq{DBI-limit3}.

We have thus the same situation as we found for the Polyakov string
versus the Nambu-Goto string.
This should not really come as a surprise, since the string
action can be seen as a special case of the D-string action when we
let $A_a= {\cal B}_{\mu\nu}= 0$, and $s_{ab}$ be symmetric.

%% file: rigidstring.tex

Rigid strings are strings with an extra curvature term that depends on
the spacetime embedding. They are also referred to as \emph{smooth}
strings, because this extra term (provided that it has the right sign)
makes it energetically favourable for them to be less creased.

The rigid string was introduced in 1986 as an attempt to find a string
that corresponds to QCD (quantum chromo-dynamics)
\cite{polyakov:1986}, and, independently, to
study the string near a phase transition \cite{kleinert:1986}.
Different aspects of rigid strings have later been investigated in
\cite{curtright:1986,forster:1986,olesen:1987,pisarski:1987,kleinert:1987,ambjorn:1987,curtright:1986b,curtright:1987,lindstrom:1987,lindstrom:1988b,pisarski:1988,itoi:1989}.

\section{The action}
The intrinsic curvature of a surface $\Sigma$ embedded in a flat background
space $S$ can be written by means of the extrinsic curvatures $K^{ib}_a$
generally (up to a total derivative) as
(c.f. appendix \ref{sec:curvature})
\beql{rig-curvature}
  R = K^{ia}_a K^{~~b}_{ib} - K^{ib}_a K^{~~a}_{ib},
\eeq
where $a,b=0,1$ are worldsheet indices and $i=2,3,\dots, D-1$ refer
to directions normal to the surface. 
$D$ is the spacetime dimension.
Indices are raised and lowered by the induced metric $\gamma_{ab}$ and
its inverse $\gamma^{ab}$.
The intrinsic curvature is a total derivative in two dimensions, but
the separate terms in \refeq{rig-curvature} are not. A generalization
of the Nambu-Goto action to include curvature can therefore be written
\cite{polyakov:1986}
\beql{rig-pol-action}
  S = T\int d^2\xi\sqrt{-\gamma} + \frac{1}{2\alpha}\int
	d^2\xi\sqrt{-\gamma}K^{ia}_aK^{~~b}_{ib},
\eeq
where, again, $\gamma=\det(\gamma_{ab})$ and $T$ is the string
tension. The coupling
constant $\alpha$ is referred to as the \emph{rigidity parameter}.

The term describing the extrinsic curvature contains double
derivatives. To get an action with only first derivatives, we
introduce an extra field $B^\mu$, and write the action as in
\cite{itoi:1989}, 
\beql{rigidstringaction}
  S = T\int d^2\xi \sqrt{-\gamma}\left(
	1-\frac{\alpha T}{2} B^\mu B_\mu - \gamma^{ab}\del_aX^\mu\del_bB_\mu
	\right).
\eeq
This action is seen to be equivalent to the original one by elimination
of $B$:
\begin{eqnarray}
  \\ \nonumber
  \delta B_\mu \Rarr \delta S 
  & = & T\int d^2\xi\left( -\sqrt{-\gamma}\alpha T B^\mu\delta B_\mu
	- \sqrt{-\gamma}\gamma^{ab}\del_aX^\mu\del_b(\delta B_\mu)\right)
  \\ \nonumber
  & = & T\int d^2\xi\left(-\sqrt{-\gamma}\alpha T B^\mu 
	+ \del_b(\sqrt{-\gamma}\gamma^{ab}\del_aX^\mu)\right)\delta B_\mu.
\end{eqnarray}
The covariant d'Alembertian operator is
$\Box\equiv \gamma^{ab}\nabla_a\del_b =
\frac{1}{\sqrt{-\gamma}}\del_a\sqrt{-\gamma}\gamma^{ab}\del_b$, 
where $\nabla$ is the covariant derivative with respect to the metric
$\gamma^{ab}$. Demanding the action to be extremal, we find immediately
\beq
  B^\mu = \frac{1}{\alpha T}\Box X^\mu
\eeq
Inserting this solution for $B$ in the action
\refeq{rigidstringaction} we get
\beq
  S = T\int d^2\xi \sqrt{-\gamma} 
	+\frac{1}{2\alpha}\int d^2\sqrt{-\gamma} \xi \Box X^\mu\Box X_\mu.
\eeq 
Furthermore, we have the identity
$\del_a\del_bX^\mu = \{^c_{ab}\}\del_cX^\mu + K^i_{ab}n_i^\mu$,
where $\{^c_{ab}\}$ is the Christoffel symbol associated with 
$\gamma_{ab}$, whereas $K^i_{ab}$ is the
extrinsic curvature and $n_i^\mu$ are normal vectors to the 
worldsheet. Using this together with
the definition of $\Box X^\mu$ we find
$\Box X^\mu = \gamma^{ab}K_{ab}^i n_i^\mu$,
which gives $\Box X^\mu\Box X_\mu = K^{ia}_aK_{ib}^{~~b}$. Hence 
we recover the ``second order'' action \refeq{rig-pol-action}.

The form \refeq{rigidstringaction}
of the action may further be derived from a membrane action
\cite{lindstrom:1989}.

\air
In the following we will focus on the tensionless limit of the rigid
string. 

\section{Method I}

The action \refeq{rigidstringaction} is not appropriate for method I
(see section \ref{sec:method-I}). For
this reason we will here instead consider the
action \cite{lindstrom:1987}
\beq
  S=T\int d^2\xi\sqrt{-\det(\gamma_{ab}+H_{ab})}; \qquad
	H_{ab} \equiv \alpha^{-1}
	(\nabla_a\del_c X^\mu\nabla_b\del_d X_\mu)\gamma^{cd},
\eeq
which is equivalent to \refeq{rig-pol-action} to first order in
$\alpha^{-1}$. The tensionless limit is
\beq
  S_\chi^{T=0} = -\12\int d^2\xi~\chi\det(\gamma_{ab}+H_{ab}),
\eeq
and a variation in $\chi$ gives
\beq
  \delta\chi\Rarr\qquad \det(\gamma_{ab}+H_{ab})=0.
\eeq
This equation together with the equation we find from a variation in
$X^\mu$ are the field equations for this model.

%
%

\section{Method II}


More interesting than method I is the calculations and results we
obtain from the phase space method. It is a more general method, and
usually makes it easier to see what is going on as we take the limit
$T\to 0$. 

We define $N\equiv \frac{\alpha T}{2}$ and allow
for the possibility of $N$ to remain finite as $T\to 0$.
Our starting point is the Lagrangian 
\beq
  L = T\sqrt{-\gamma}(1-NB^2-\gamma^{ab}\del_aX^\mu\del_bB_\mu).
\eeq
As usual, we start by deriving the canonical conjugate momenta
(~$\dot{}=\frac{\del}{\del\xi^0}$, 
$\acute{} =\frac{\del}{\del\xi^1}$),
\begin{eqnarray}
  \Pi_\mu &\equiv& \frac{\del L}{\del \dot{B}^\mu}
	= -T\sqrt{-\gamma}\gamma^{a0}\del_aX_\mu,
  \\
  \nonumber
  P_\mu & \equiv& \frac{\del L}{\del \dot{X}^\mu}
  = T\sqrt{-\gamma}\Big[ 
	\gamma^{d0}\left((1-NB^2)
	\del_dX_\mu-\del_dB_\mu\right)
  \\
  & &\qquad\qquad
	+ (\gamma^{a0}\gamma^{bd}
	+\gamma^{b0}\gamma^{ad}
	-\gamma^{ab}\gamma^{d0})
	(\del_aX^\nu\del_bB_\nu)\del_dX_\mu\Big].
\end{eqnarray}
Neither of these equations are invertible, and we find the following
primary constraints:
\begin{eqnarray}
  \Theta_1 &\equiv& \Pi^2 + T^2\acute{X}^2 \approx 0,
  \\
  \Theta_2 &\equiv& \Pi_\mu\acute{X}^\mu \approx 0,
  \\
  \Theta_3 &\equiv& P_\mu\acute{X}^\mu + \Pi_\mu\acute{B}^\mu 
	\approx 0,
  \\ \label{eq:rig-theta4}
  \Theta_4 &\equiv&  P_\mu\Pi^\mu + T^2\acute{X}_\mu\acute{B}^\mu
	+(1-NB^2)\Pi^2
	\approx 0.
\end{eqnarray}
The diffeomorphism invariance ensures that the naive Hamiltonian is
zero, which is also easy to check by direct calculation. This means that
the consistency conditions \refeq{consistency-conditions}
on the primary constraints take the form
\beql{rig-consistency}
  \int d^2\xi' ~\lambda^n \{\Theta_m(\xi),\Theta_n(\xi')\} \approx 0,
\eeq
where $\lambda^n$ is the
Lagrange multiplier associated with the constraint $\Theta_n$. Working
out these conditions will give us the secondary constraints.

\paragraph{Secondary constraints} Since we have so far never really
done any thorough calculations to find secondary constraints, we will
now do this in great detail. Refer back to section
\ref{sec:constrainedsystems} for the general theory.
We first have to find the variational derivatives of the fields. This
is easily done, with the results:
\begin{eqnarray*}
  \lefteqn{
  \frac{\delta\Theta_1(\xi)}{\delta X^\mu(\tilde{\xi})}
	=2T^2\acute{X}_\mu\del_1\delta(\xi-\tilde{\xi}),
  } \hspace{5cm}
  &&\frac{\delta\Theta_1(\xi)}{\delta B^\mu(\tilde{\xi})}=0,
  \\ \lefteqn{
  \frac{\delta\Theta_1(\xi)}{\delta P_\mu(\tilde{\xi})}=0,
  } \hspace{5cm}
  &&\frac{\delta\Theta_1(\xi)}{\delta \Pi_\mu(\tilde{\xi})} 
	= 2\Pi^\mu(\xi)\delta(\xi-\tilde{\xi});
  \\[0.4cm]
  \lefteqn{
  \frac{\delta\Theta_2(\xi)}{\delta X^\mu(\tilde{\xi})}
	=\Pi_\mu(\xi)\del_1\delta(\xi-\tilde{\xi}),
  } \hspace{5cm}
  &&\frac{\delta\Theta_2(\xi)}{\delta B^\mu(\tilde{\xi})}=0,
  \\ \lefteqn{
  \frac{\delta\Theta_2(\xi)}{\delta P_\mu(\tilde{\xi})}=0,
  } \hspace{5cm}
  &&\frac{\delta\Theta_2(\xi)}{\delta \Pi_\mu(\tilde{\xi})} 
	= \acute{X}^\mu(\xi)\delta(\xi-\tilde{\xi});
  \\[0.4cm]
  \lefteqn{
  \frac{\delta\Theta_3(\xi)}{\delta X^\mu(\tilde{\xi})}
	=P_\mu(\xi)\del_1\delta(\xi-\tilde{\xi}),
  } \hspace{5cm}
  &&\frac{\delta\Theta_3(\xi)}{\delta B^\mu(\tilde{\xi})} 
	= \Pi_\mu(\xi)\del_1\delta(\xi-\tilde{\xi}),
  \\ \lefteqn{
  \frac{\delta\Theta_3(\xi)}{\delta P_\mu(\tilde{\xi})}
	= \acute{X}^\mu(\xi)\delta(\xi-\tilde{\xi}),
  } \hspace{5cm}
  &&\frac{\delta\Theta_3(\xi)}{\delta \Pi_\mu(\tilde{\xi})} 
	=\acute{B}^\mu(\xi)\delta(\xi-\tilde{\xi});
  \\[0.4cm]
  \lefteqn{
  \frac{\delta\Theta_4(\xi)}{\delta X^\mu(\tilde{\xi})}
	=T^2\acute{B}(\xi)\del_1\delta(\xi-\tilde{\xi}),
  } \hspace{5cm}
	&&\frac{\delta\Theta_4(\xi)}{\delta B^\mu(\tilde{\xi})} 
	= T^2\acute{X}_\mu(\xi)\del_1\delta(\xi-\tilde{\xi})
  \\[-0.3cm]
	&&\hspace{2cm} -2N\Pi^2(\xi)B_\mu(\xi)\delta(\xi-\tilde{\xi}),
  \\ \lefteqn{
  \frac{\delta\Theta_4(\xi)}{\delta P_\mu(\tilde{\xi})}
	= \Pi^\mu(\xi)\delta(\xi-\tilde{\xi}),
  } \hspace{5cm}
	&&\frac{\delta\Theta_4(\xi)}{\delta \Pi_\mu(\tilde{\xi})} 
	=[~P^\mu(\xi)+2(1-NB^2(\xi))
  \\[-0.3cm]
 	 &&\hspace{2cm} \times \Pi^\mu(\xi)~]~\delta(\xi-\tilde{\xi}).
\end{eqnarray*}
Next, we must calculate the Poisson brackets between the
constraints, which are in general given by
\begin{eqnarray}
  \nonumber
  \{\Theta_m(\xi),\Theta_n(\xi')\}
	&=& \int d^2\tilde{\xi} \bigg[
	 \frac{\del\Theta_m(\xi)}{\del X^\mu(\tilde{\xi})}
	 \frac{\del\Theta_n(\xi')}{\del P_\mu(\tilde{\xi})}
	+\frac{\del\Theta_m(\xi)}{\del B^\mu(\tilde{\xi})}
	 \frac{\del\Theta_n(\xi')}{\del \Pi_\mu(\tilde{\xi})}
	\\ && \qquad
	-\frac{\del\Theta_m(\xi)}{\del P_\mu(\tilde{\xi})}
	 \frac{\del\Theta_n(\xi')}{\del X^\mu(\tilde{\xi})}
	-\frac{\del\Theta_m(\xi)}{\del \Pi_\mu(\tilde{\xi})}
	 \frac{\del\Theta_n(\xi')}{\del B^\mu(\tilde{\xi})}
	\bigg].
\end{eqnarray}
Performing the required calculations, we find\footnote{
We use the notation $\xi = (\xi^0,\xi^1)$ and 
$\del_1 = \frac{\del}{\del\xi^1}$.}
\begin{eqnarray}
  \label{eq:pois-12}
  \{\Theta_1(\xi),\Theta_2(\xi')\} &=& 0,
  \\ 
  \label{eq:pois-13}
  \{\Theta_1(\xi),\Theta_3(\xi')\} &=&
	2\left[ \Pi^\mu(\xi)\Pi_\mu(\xi') 
	+ T^2\acute{X}^\mu(\xi)\acute{X}_\mu(\xi')
	\right]\del_1\delta(\xi-\xi'),
  \\ \nonumber
  \label{eq:pois-14}
  \{\Theta_1(\xi),\Theta_4(\xi')\} &=& 
	2T^2\left[ \Pi_\mu(\xi)\acute{X}^\mu(\xi') 
	+\Pi_\mu(\xi')\acute{X}^\mu(\xi)\right]\del_1\delta(\xi-\xi')
	\\[-0.1cm] && \qquad
	+4N\Pi^2\Pi_\mu B^\mu \delta(\xi-\xi'),
  \\[0.1cm]
  \label{eq:pois-23}
  \{\Theta_2(\xi),\Theta_3(\xi')\} &=&
	\left[\Pi_\mu(\xi)\acute{X}^\mu(\xi')
	+\Pi_\mu(\xi')\acute{X}^\mu(\xi)\right]
	\del_1\delta(\xi-\xi'),
  \\ \nonumber
  \label{eq:pois-24}
  \{\Theta_2(\xi),\Theta_4(\xi')\} &=& 
	\left[ \Pi_\mu(\xi)\Pi^\mu(\xi') 
	+ T^2\acute{X}^\mu(\xi)\acute{X}_\mu(\xi') \right]
	\del_1\delta(\xi-\xi')
	\\[-0.1cm] && \qquad
	+2N \Pi^2 \acute{X}^\mu B_\mu \delta(\xi-\xi'),
  \\[0.1cm] \nonumber
  \label{eq:pois-34}
  \{\Theta_3(\xi),\Theta_4(\xi')\} &=&
	2(1-NB^2(\xi'))\Pi_\mu(\xi')\Pi^\mu(\xi)
	\del_1\delta(\xi-\xi')
	\\ && \qquad
	+2N \Pi^2\acute{B}^\mu B_\mu \delta(\xi-\xi').
\end{eqnarray}
When put inside an integration, we can integrate by parts
to get rid of the derivatives of the delta function. This will give
rise to a vanishing surface term, and a term that vanishes (weakly) by
use of the primary constraints. Let us show this for
$\{\Theta_2(\xi),\Theta_3(\xi')\}$. The expression has meaning only
within an integral, and we find
\begin{eqnarray}
  \nonumber \lefteqn{
  \int d^2\sigma \varphi(\sigma)\{\Theta_2(\xi),\Theta_3(\sigma)\} }\qquad&&
  \\ \nonumber
  &=& \int d^2\sigma \varphi(\sigma) 
	\Big[ \Pi_\mu(\xi)\acute{X}^\mu(\sigma) 
	+ \Pi_\mu(\sigma)\acute{X}^\mu(\xi)
	\Big]\del_1\delta(\xi-\sigma)
  \\ \nonumber
  &=& \int d^2\sigma \del_1\left(\varphi(\sigma)
	\left[\Pi_\mu(\xi)\acute{X}^\mu(\sigma) 
	+ \Pi_\mu(\sigma)\acute{X}^\mu(\xi)\right]\delta(\xi-\sigma)\right)
  \\[-0.2cm] \nonumber
  &&-\int d^2\sigma \del_1\left(\varphi(\sigma)
	\left[\Pi_\mu(\xi)\acute{X}^\mu(\sigma) 
	+\Pi_\mu(\sigma)\acute{X}^\mu(\xi)\right]\right)\delta(\xi-\sigma)
  \\[0.1cm] \nonumber
  &=& \del_1\int d^2\sigma \varphi(\sigma)
	\left[\Pi_\mu(\xi)\acute{X}^\mu(\sigma) 
	+\Pi_\mu(\sigma)\acute{X}^\mu(\xi)\right]\delta(\xi-\sigma)
  \\[-0.2cm] \nonumber
  && - \int d^2\sigma\varphi(\sigma)
	\left[\acute{\Pi}_\mu(\xi)\acute{X}^\mu(\sigma) 
	+\Pi_\mu(\sigma)\del_1\del_1X^\mu(\xi)\right]\delta(\xi-\sigma)
  \\ \nonumber
  &=&\del_1\left(\varphi(\xi) 2\Pi\acute{X}\right)
	-\varphi(\xi)\del_1(\Pi\acute{X})
  \\
  &\stackrel{\Theta_2}{\approx}& 0.
\end{eqnarray}
The same can be shown to hold for all the other terms including the
factor $\del_1\delta(\xi-\xi')$.  

Thus, the consistency conditions \refeq{rig-consistency} yield in
general three secondary constraints\footnote{
This is different from what is found in \cite{itoi:1989}. In this
article the authors use, instead of \refeq{rig-theta4},
$\Theta_4= P\Pi + T^2\acute{X}\acute{B} + 
(1-NB^2)(\Pi^2-T^2\acute{X}^2)$ and find 
$\{\Theta_3,\Theta_4\}=0$. 
This is obviously not equivalent with our results.
},
\begin{eqnarray}
  \label{eq:rig-2nd-1}
  4N \Pi^2 \Pi^\mu B_\mu &\approx& 0,
  \\ \label{eq:rig-2nd-2}
  2N \Pi^2 \acute{X}^\mu B_\mu &\approx& 0,
  \\ \label{eq:rig-2nd-3}
  2N \Pi^2 \acute{B}^\mu B_\mu &\approx& 0.
\end{eqnarray}
The fields are here to be evaluated at the same world-sheet points $\xi$.

\paragraph{The limit $B=0$}
From the original action integral \refeq{rigidstringaction} we see
that in this limit we recover the Nambu-Goto action for a string. It is
instructive to check that this will be true also in the phase space
picture. If we let $B=0$ in the expressions for the momenta we find
that $P^\mu = -\Pi^\mu = T\sqrt{-\gamma}\gamma^{a0}\del_aX^\mu$, which
is of course the same as for the Nambu-Goto string. Furthermore,
$\Theta_4\approx 0$ will be reduced to an identity, and $\Theta_3$
will be identical to $-\Theta_2$. The remaining two constraints will be
the same as in the Nambu-Goto case,
\begin{eqnarray}
  \Theta_1 & = & P^2 + T\acute{X}^2, \\
  \Theta_2 & = & P\cdot\acute{X}.
\end{eqnarray}
Since the naive Hamiltonian is zero, this immediately tells us that the
phase space action will also be the same.

\paragraph{The limit $T=0$}
This is the tensionless case that we are really interested in. 
The primary constraints are now reduced to
\begin{eqnarray}
  \Theta_1 &=& \Pi^2, \\
  \Theta_2 &=& \Pi^\mu\acute{X}_\mu, \\
  \Theta_3 &=& P^\mu\acute{X}_\mu + \Pi^\mu\acute{B}_\mu, \\
  \Theta_4 &=& P^\mu\Pi_\mu.
\end{eqnarray}
Since $\Pi^2\approx 0$ the consistency conditions
(\ref{eq:rig-2nd-1}--\ref{eq:rig-2nd-3}) reduce to identities, so we
have no secondary constraints in the tensionless limit.

The Hamiltonian is then just the sum of primary constraints,
\beq
  H = a\Pi^2 + b \Pi\cdot\acute{X} 
	+ c[ P\cdot\acute{X} + \Pi\cdot\acute{B}]
	+ dP\cdot\Pi,
\eeq
where $a$, $b$, $c$ and $d$ are Lagrange multipliers. 
Note that the parameter $N$ did also vanish as we put $T=0$.
What we now want
to do is to write the phase space action, and eliminate the momenta to
arrive at a new configuration space action. The phase space action is
\beq
  S = \int d^2\xi \left[ P\dot{X} + \Pi\dot{B} - H \right].
\eeq
This action is only linear in $P$, and a variation $\delta P$ gives
\begin{eqnarray}
  \delta P_\mu \Rarr \quad \dot{X}^\mu - c\acute{X}^\mu - d\Pi^\mu 
	&=& 0, \\ 
  \Rarr \quad \Pi^\mu 
	&=&\frac{1}{d}(\dot{X}^\mu - c\acute{X}^\mu).
\end{eqnarray}
Inserted back into the action, and defining $\rho = \frac{a}{d}$,
this gives the configuration space action
\begin{eqnarray}
  \nonumber
  S & = & \int d^2\xi \frac{1}{d}\Big[ 
	-\rho\gamma_{00}
	+ (2c\rho-b)\gamma_{01}	-(c^2\rho - cb)\gamma_{11} \\
  & & \qquad\qquad
	+\beta_{00} - c\beta_{10} -c\beta_{01}+ c^2\beta_{11} 
	~\Big].
  \label{eq:rig-CS-action}
\end{eqnarray}
(Remember that $\gamma_{ab}=\del_aX^\mu\del_bX_\mu=\gamma_{ba}$ and
$\beta_{ab}=\del_aX^\mu\del_bB_\mu\neq\beta_{ba}$.)
Let us define the vector density
\beq
  V^a \equiv \frac{1}{\sqrt{d}}\left(
	\begin{array}{c} 1\\-c
	\end{array}\right),
\eeq
which means 
$V^0V^0=\frac{1}{d}$, 
$V^0V^1=-\frac{1}{d}c$ and
$V^1V^1=\frac{1}{d}c^2$. Redefining $\frac{1}{\sqrt{d}}b\to b$, we can
write the Lagrangian in \refeq{rig-CS-action} as
\beq
  L=-\rho V^aV^b\gamma_{ab}
	-bV^a\gamma_{a1}
	+V^aV^b\beta_{ab}.
\eeq
If we also define
\beq
  W^a \equiv \left(
	\begin{array}{c}\rho V^0\\ \rho V^1+b
	\end{array}\right),
\eeq
we can write the action for the tensionless rigid string in a manifestly
covariant form,
\beql{rig-lim-action}
  S = \int d^2\xi\left[V^a\del_aX^\mu(V^b\del_bB_\mu -
	W^b\del_bX_\mu)\right].
\eeq
The four degrees of freedom of the Lagrange multipliers
are now replaced by the four degrees of freedom in the two vector
densities. 

Variations in the fields give the equations of motion,
\begin{eqnarray}
  \label{eq:rig-1}
  \delta B_\mu \Rarr &\quad&
	\del_a(V^aV^b\del_bX^\mu)=0,
  \\ \label{eq:rig-2}
  \delta W^b \Rarr &&
	V^a\gamma_{ab}=0,
  \\ \label{eq:rig-3}
  \delta V^b \Rarr &&
	V^a\del_{(a}X_\mu\del_{b)}B^\mu = W^a\gamma_{ab},
  \\ \label{eq:rig-4}
  \delta X_\mu \Rarr&&
	\del_a\left[V^{(a}W^{b)}\del_bX^\mu-V^aV^b\del_bB^\mu\right]=0.
\end{eqnarray}
The first two equations are the same as we found for the tensionless
Nabu-Goto string. What is special for the rigid string must then be
found in the remaining two equations.

Equation \refeq{rig-4} gives
\beql{rig-5}
  V^{(a}W^{b)}\del_bX^\mu-V^aV^b\del_bB^\mu = \epsilon^{ab}\del_b
	C^\mu,
\eeq
where $C^\mu$ is some spacetime vector.
Contracting this equation with $\del_aX_\mu$ gives
\begin{eqnarray}
  \nonumber
  2V^aW^b\gamma_{ab} - V^aV^b\beta_{ab} 
	&=& \epsilon^{ab}\del_bC^\mu\del_aX_\mu \\
  0 &=& \epsilon^{ab}\del_bC^\mu\del_aX_\mu,
\end{eqnarray}
which says that $\del_bC^\mu\del_aX_\mu,$ is symmetric,
i.e. $\del_aC^\mu = c\del_aX^\mu$, where $c$ is some constant.
Put into equation \refeq{rig-4} this gives
\begin{eqnarray}
  \nonumber
  V^{(a}W^{b)}\del_bX^\mu - V^aV^b\del_bB^\mu 
	&=& c\epsilon^{ab}\del_bX^\mu 
  \\ \nonumber
  \epsilon_{ca}\Rarr\hspace{0.8cm}
  \epsilon_{ca}\left(V^{(a}W^{b)}\del_bX^\mu - V^aV^b\del_bB^\mu\right)
	&=& c\del_cX^\mu \quad 
  \\ \nonumber
   V^c \Rarr\hspace{3.4cm}
  \epsilon_{ca}V^cV^bW^a\del_bX^\mu 
	&=& c \del_cX^\mu V^c 
  \\
   \Rarr\qquad
   c &=& \epsilon_{ab}V^aW^b.
\end{eqnarray}
Thus we have determined the constant $c$.
Using the relation 
$\epsilon^{ab}\epsilon_{cd}=\delta^a_d\epsilon^b_c-\epsilon^a_c\epsilon^b_d$
we can now write equuation \refeq{rig-5}
\[
  V^aW^b\del_bX^\mu + V^bW^a\del_bX^\mu-V^aV^b\del_bB^\mu
	= \del_cX^\mu V^cW^a - \del_dX^\mu V^aW^d,
\]
which gives
\beql{rig-6}
  V^b\del_bB^\mu = 2W^b\del_bX^\mu.
\eeq
This equation put into \refeq{rig-3} yields
\beql{rig-7}
  V^b\del_bX^\mu\del_aB^\mu = -W^b\gamma_{ab}.
\eeq
A contraction of equation \refeq{rig-6} with $\del_aX_\mu$, and using
\refeq{rig-7} gives us 
\beq
  V^b\del_bB^\mu\del_aX_\mu = -2V^b\del_bX^\mu\del_aB^\mu,
\eeq
which says that
$\det(\beta_{ab}+2\beta_{ba})=0$. This relation can (in 2D) be written
\beq
  \det(3\beta_{(ab)}) = -\hat{\beta}^2; \qquad \beta_{[ab]} 
	= \hat{\beta}\epsilon_{ab},
\eeq
where $\hat{\beta}=\beta_{01}-\beta_{10}$. 

%

\air
We have seen that method II does indeed lead to a sensible theory for
the tensionless limit of the rigid string, represented by the
field equations
(\ref{eq:rig-1}; \ref{eq:rig-2}; \ref{eq:rig-6}; \ref{eq:rig-7}).
The first two equations describe a tensionless Nambu-Goto string, which
we derived in section \ref{sec:NGstring}.
One special solution of the last two equations is the case where $W^a$
is parallel to $V^a$, and $\del_aB^\mu$ is parallel to $\del_aX^\mu$,
in which case we end up with exactly the same equations of motion as
for the tensionless Nambu-Goto string.

An interesting question that we do not go into in any more detail, is
what role the $B$ field actually plays in these equations.

\air
Note also that the action \refeq{rig-lim-action} can be made
conformally invariant in the same way as the tensionless Nambu-Goto
string, if we in addition demand that $B^\mu$ transforms like
$X^\mu$.

%% file: GRgravitation.tex
\label{chap:GR}

Consider a given manifold $S$ with metric $g_{\mu\nu}$. Suppose
furthermore that there exist a coordinate basis $\{x^\mu\}$ on the
manifold, and let comma denote partial derivative with respect
to the coordinates, i.e. $F_{,\mu} \equiv \frac{\del F}{\del
x^\mu}$.
Then we may define the
the \emph{Christoffel symbol} $\{^\alpha_{\mu\nu}\}$, 
the \emph{Riemann curvature tensor} $R^\alpha_{~\mu\beta\nu}$,
the \emph{Ricci tensor} $R_{\mu\nu}$ and 
the \emph{Ricci scalar} $R$ as follows:
\begin{eqnarray}
  \label{eq:defChristoffel}
  \{^\alpha_{\mu\nu}\} & \equiv & \12 g^{\alpha\beta} (g_{\beta\nu,\mu}
   + g_{\mu\beta,\nu} - g_{\mu\nu,\beta}),
  \\
  \label{eq:defRiemann}
  R^\alpha_{~\mu\beta\nu} & \equiv & \{^\alpha_{\mu\nu}\}_{,\beta}
    - \{^\alpha_{\mu\beta}\}_{,\nu} 
    + \{^\alpha_{\sigma\beta}\} \{^\sigma_{\mu\nu}\}
    - \{^\alpha_{\sigma\nu}\} \{^\sigma_{\mu\beta}\},
  \\
  \label{eq:defRicci}
  R_{\mu\nu} & \equiv & R^\alpha_{~\mu\alpha\nu},
  \\
  R & \equiv & g^{\mu\nu}R_{\mu\nu}.
\end{eqnarray}
The Christoffel symbol is very often written $\Gamma^\alpha_{\mu\nu}$,
and is an example of a \emph{connection}. However, it is \emph{not}
a tensor, so this notation can be very confusing. We write it in
the form above to emphasize this fact, but also because we will later
use the symbol $\Gamma^\alpha_{\mu\nu}$ with a different meaning.

Although $\{^\alpha_{\mu\nu}\}$ is not a tensor, it can be shown that
$R^\alpha_{~\mu\beta\nu}$ is, and thus has earned the right
to its name.

\section{The action}

Einstein's field equations (in vacuum) describing the dynamics of
spacetime are
\beql{EF}
  R_{\mu\nu} - \12 R g_{\mu\nu} + \Lambda g_{\mu\nu} = 0,
\eeq
and may
be found as the equations of motion derived from the 
\emph{Hilbert action} (see standard textbooks on gravitation,
e.g. \cite{wald:1984,misner:1973}) 
\beql{GRaction}
  S[g] = \frac{1}{\kappa} \int d^4x \sqrt{-g}(R(g)-2\Lambda).
\eeq
Einstein's constant $\kappa$ is defined as $\kappa \equiv
\frac{8\pi}{c^3}G_N$, where $G_N$ is Newton's gravitational constant.
$\Lambda$ is the cosmological constant, which may be thought of as
being related to the energy of vacuum.

This action contains metric fields $g_{\mu\nu}$, their
derivatives $\del g_{\mu\nu}$, and also their 
second derivatives $\del^2 g_{\mu\nu}$,
and was used by Hilbert to derive
Einstein's field equations only days after Einstein had
published his results.

It has later become clear that we may treat the connections
$\Gamma^\alpha_{\mu\nu}$ as independent field variables, and define
the Riemann tensor by means of $\Gamma^\alpha_{\mu\nu}$ instead of
$\{^\alpha_{\mu\nu}\}$. 
Equation \refeq{defRiemann} is then replaced by
\beql{defRiemannG}
   R^\alpha_{~\mu\beta\nu}  \equiv  
      \Gamma^\alpha_{\mu\nu,\beta}
    - \Gamma^\alpha_{\mu\beta,\nu} 
    + \Gamma^\alpha_{\sigma\beta} \Gamma^\sigma_{\mu\nu}
    - \Gamma^\alpha_{\sigma\nu} \Gamma^\sigma_{\mu\beta}.
\eeq
This gives the so-called \emph{Palatini action}
\cite{palatini:1919}
\beq
  S[g,\Gamma] = \frac{1}{\kappa} \int
	d^4x\sqrt{-g}(g^{\mu\nu}R_{\mu\nu}(\Gamma)-2\Lambda).
\eeq
Variation of $\Gamma$ and $g$ respectively give us
\begin{eqnarray}
  \delta \Gamma^\alpha_{\mu\nu} 
	& \Rightarrow & \Gamma^\alpha_{\mu\nu}= \{^\alpha_{\mu\nu}\},
  \\
  \label{eq:EFconstant}
  \delta g^{\mu\nu} & \Rightarrow &
	R_{\mu\nu}(\Gamma) - \12 g_{\mu\nu}
	g^{\rho\sigma}R_{\rho\sigma}(\Gamma) + \Lambda g_{\mu\nu} = 0.
\end{eqnarray}
We see that the relation between the connection and the metric,
$\Gamma^\alpha_{\mu\nu}=\{^\alpha_{\mu\nu}\}$,
comes out as a solution of the field equations. 
With the substitution $\Gamma\to\{\}$,
equation \refeq{EFconstant} is exactly the Einstein field equations
for vacuum, with a cosmological constant included. Details on these
calculations are found e.g. in \cite{misner:1973}.

\air
If we eliminate $\Gamma$ in $S[g,\Gamma]$ by solving its equation of
motion and substituting back, we just have to replace $\Gamma$ by
$\{\}$, thus recover the Hilbert action $S[g]$. 
In this sense we call $S[g,\Gamma]$ a \emph{parent action} for $S[g]$.

Alternatively,
we may eliminate the metric fields $g_{\mu\nu}$. Contraction with
$g^{\mu\nu}$ in \refeq{EFconstant} gives
\beq
  g^{\mu\nu}R_{\mu\nu}(g) = 4\Lambda.
\eeq
(If we worked with another spacetime dimension than 4 we would get
$\frac{2D}{D-2}\Lambda$ on the right hand side.)
Inserted back into equation \refeq{EFconstant} we find
\beql{LgRrelation}
  \Lambda g_{\mu\nu} = R_{\mu\nu}(\Gamma).
\eeq
As long as $\Lambda \not = 0$ we can eliminate $g_{\mu\nu}$ in
$S_\Lambda[g,\Gamma]$ and arrive at the 
\emph{Eddington-Schr\"odinger action}
\cite{eddington:1924,schrodinger:1947}
\beq
  S[\Gamma] = \frac{2}{\kappa\Lambda} \int d^4x
	\sqrt{-\det(R_{\mu\nu}(\Gamma)}).
\eeq
The actions $S[g]$ and $S[\Gamma]$ are \emph{dual},
i.e. they are both derivable from the same \emph{parent action}
$S[g,\Gamma]$.
All three actions will of course give us equivalent equations of motion,
so they are \emph{classically} equivalent.

\air
If we compare with the $p$-brane actions, we see that $S[\Gamma]$
resembles the Nambu-Goto form \refeq{stringNGaction},
while $S[g,\Gamma]$ 
is very similar to the Howe-Tucker action \refeq{howe-tucker}, and
partly the Polyakov action (c.f. section \ref{sec:weylstring}).
But there are important differences as well.

First, the strings were supposed to ``live'' in a background space,
i.e. a spacetime with some fixed metric. In the gravitational case, on
the other hand, this spacetime is precisely what is under study.
This aspect is in fact a major obstacle in the search for a quantum
field theory of gravitation.

Second, the $\Gamma^\alpha_{\mu\nu}$ 
fields in gravitation have a more complicated
nature than the position fields $X^\mu$ in the string models. 
For instance, $X^\mu$ are scalars under diffeomorphisms, while the
$\Gamma$'s are not even tensors.

\paragraph{The limit $\kappa\to\infty$}

{}From the definition of $\kappa$ we see that the limit
may be viewed as a   $c \to 0$ limit, or a $G_N \to \infty$ limit. 
This is the opposite of
the Newtonian limit, which can be thought of as $c \to \infty$. 
As the speed of light approaches
zero, lightcones will collapse into spacetime lines,
and points in space will be disconnected. So this limit leads to an
ultralocal field theory. 
These limits have been investigated e.g. in 
\cite{pilati:1982} and \cite{husain:1988} in the search for
a quantum description of gravity.

It is possible to make yet another interpretation of the 
$\kappa\to\infty$ limit by observing
(which is possible in the Hamiltonian formulation) that it is equal to
the so-called \emph{zero signature limit}. This limit represents some
intermediate stage between Euclidian space (signature $+1$) and Minkowski
space (signature $-1$). This viewpoint is taken by 
Teitelboim \cite{teitelboim:1980}. His paper also includes an
instructive discussion on what physical significance
this limit has. 

In the following we will see if we can use our well-developed methods
as a successful approach to this limit. General relativity as a theory
is quite different from the string models we have investigated so far,
and it is difficult to say on beforehand whether any of the two methods
will give anything of physical interest. But it is certainly worth a
try.

\section{Method I}

Of the actions $S[g,\Gamma]$, $S[g]$ and $S[\Gamma]$ only the latter
has the form that makes method I applicable. We will henceforth
consider this action.
Introducing the auxiliary field $\chi$ we can write a classically
equivalent action as
\begin{eqnarray}
  S[\chi,\Gamma] &=& \int d^4x \left[
	-\frac{4}{\Lambda^2}\det(R_{\mu\nu}(\Gamma)) \chi
	+\frac{1}{\kappa^2\chi} \right].
\end{eqnarray}
Elimination of $\chi$ will as usual give back the original action,
i.e. $\delta\chi \Rarr S[\Gamma]$.

However, we are interested in the $\kappa\to\infty$ limit, and the
action above then gives
\begin{eqnarray}
  S^0[\chi,\Gamma] &=& -\frac{4}{\Lambda^2} \int d^4x
	\det(R_{\mu\nu}(\Gamma)) \chi.
\end{eqnarray}
We could redefine the auxiliary field to get rid
of the factor $\frac{4}{\Lambda^2}$. But the form above makes it
manifestly clear that the exression is valid only for $\Lambda \neq
0$, so we keep it as it is.

The equations of motion are found by variations $\delta\chi$ and
$\delta\Gamma^\alpha_{\mu\nu}$:
\begin{eqnarray}
  \label{eq:GRI-lim3}
  \delta\chi &\Rarr&
	\frac{1}{\Lambda^2}\det(R_{\mu\nu}(\Gamma))=0,
  \\ \nonumber
  \delta\Gamma^\alpha_{\nu\mu} &\Rarr& \frac{1}{\Lambda^2} \left[
	-\nabla_\alpha(\chi{\cal RR^{\mu\nu}})
	+\nabla_\lambda(\chi{\cal RR^{\lambda\nu}})\delta^\mu_\alpha
  \right]=0,
%
\end{eqnarray}
where ${\cal R}\equiv \det(R(\Gamma))$,
${\cal R}^{\rho\sigma}=\frac{1}{3!}\frac{1}{{\cal R}}
\epsilon^{\rho\alpha\beta\gamma} \epsilon^{\sigma\mu\nu\lambda}
R_{\mu\alpha}R_{\nu\beta}R_{\lambda\gamma}$ is the inverse of
$R_{\rho\sigma}$,
and $\nabla_\alpha$ is the covariant derivative with $\Gamma$ as the
connection. 
The determinant ${\cal R}$ cancels in the combination
${\cal R}{\cal R}^{\mu\nu}$, so the second equation is non-trivial even
when $\det(R)=0$, which is imposed by the first equation.  

The relation \refeq{LgRrelation} connecting $\Gamma$ to the metric, 
$R_{\mu\nu}(\Gamma)=\Lambda g_{\mu\nu}$, cannot be used as 
the equivalence between $S[g,\Gamma]$ and $S[\chi,\Gamma]$ 
breaks when we take the limit $\kappa\to\infty$.
So even though we could solve the above equations for
$\Gamma^\alpha_{\mu\nu}$, the interpretations of what \emph{physics}
this limit represents would not be immediate. 

We do not elaborate more on this here, but turn instead to method II.
This method employes a much more general and
powerful formalism, and we have seen earlier that it may give rise to
several different limits. Our hope is that within this possible class
of limits we will find a physically significant one.

\section{Method II}

In the string case, the most straight forward way to arrive at
tensionless limits was found by starting from the Nambu-Goto
action. And because of its similarities to the Nambu-Goto action, we
might therefore expect the
action $S[\Gamma]$ to be the best starting point for deriving
$\kappa\to\infty$ limits in the gravity case. However, we will see
that the similarities are only formal.

The Lagrangian in $S[\Gamma]$ is
$L=\frac{2}{\kappa\Lambda}\sqrt{-\det(R_{\mu\nu}(\Gamma))}$. With the
definition of the Riemann tensor in terms of $\Gamma$ instead of
$\{\}$ we find the canonical conjugate momenta straight forwardly to
be
\begin{eqnarray}
  \nonumber
  \Pi_\alpha^{\mu\nu} & \equiv &
  \frac{\del L}{\del \Gamma^\alpha_{\mu\nu}}
	= \frac{2}{\kappa\Lambda}\sqrt{-\det(R)} 
	{\cal R}^{\sigma\rho} \frac{\del
  R_{\rho\sigma}}{\del \Gamma^\alpha_{\mu\nu,0}}
  \\
  & = & \frac{1}{\kappa\Lambda}\sqrt{-\det(R)}(\delta^0_\alpha 
	{\cal R}^{(\mu\nu)}
	-{\cal R}^{0(\mu}\delta^{\nu)}_\alpha),
\end{eqnarray}
where again ${\cal R}^{\mu\nu}$ is the \emph{inverse} of $R_{\mu\nu}$.

The naive Hamiltonian is found from
\begin{eqnarray}
  \nonumber
  h & = & \Pi_\alpha^{\mu\nu}\Gamma^\alpha_{\mu\nu,0} -
  \frac{2}{\kappa\Lambda}\sqrt{-\det(R)}
  \\ \label{eq:GR-ham-1}
  & = & \frac{2}{\kappa\Lambda}\sqrt{-\det(R)}(
	{\cal R}^{\mu\nu}\Gamma^0_{\mu\nu,0}
	-{\cal R}^{0\mu}\Gamma^\nu_{\mu\nu,0}
	-1).
\end{eqnarray}
This expression is not zero, and contains time derivatives 
both in $\det(R)$ and ${\cal R}^{\mu\nu}$. Still, the Hamiltonian is
known to be a function independent of time derivatives, so it is always
possible to eliminate time derivatives in favour of momenta.
However, the relation \refeq{GR-ham-1} for $h$ is clearly unwieldly
and makes the subsequent analysis very hard. It is thus gratifying
that there is another set of variables, the ADM variables, in which
the analysis becomes tractable.

\section{ADM approach (method II)}

We will now follow the fruitful approach originally made by
Arnowitt, Deser and Misner (ADM) \cite{arnowitt:1982}.

The starting point is the Hilbert action with zero cosmological
constant (for convenience),
\beql{hilbert}
  S=\frac{1}{\kappa}\int d^4x \sqrt{-g} R(g).
\eeq
The crucial point is the introduction of a new set of variables to
replace the ten metric components of $g_{\mu\nu}$.
In a Hamilton description we perform a space and time split. And
because of the rather intricate role of time in general relativity,
this is not as straight forwardly done as before. 
The way it is done, is by slicing the (4-dimensional) spacetime $S$ 
into spacelike (3-dimensional) hypersurfaces $\Sigma$. 
Spacetime may then be parameterized by means
of a continuous parameter, such that each value of this parameter
corresponds to
one of these hypersurfaces. This parameter is what we will call
\emph{time}, although it is not necessarily directly related to time
as measured by clocks.

Let $h_{ab}$ be the 3-dimensional  induced metric on $\Sigma$, and let
$h^{ab}$ be its inverse. Indices referring to $\Sigma$ (i.e. Latin
indices, $a,b$) are raised and lowered with this metric.
Now,
introduce the \emph{lapse function} $N$ and \emph{shift functions}
$N_a$ defined through
\beql{new-vars}
  g_{\mu\nu} = \left(
  \begin{array}{cc}
	N_aN^a-N^2	& N_b \\
	N_a^T		& h_{ab}
  \end{array} 
  \right);\qquad 
  \begin{array}{l}
	\mu,\nu=0,1,2,3;\\
	a,b=1,2,3.
  \end{array}
\eeq
The metric components $\{g_{\mu\nu}\}$ are now replaced by the set of
variables $\{h_{ab}, N, N_a\}$.
If we let $\vec{n}$ be a unit vector normal to the hypersurface
$\Sigma$, we find the extrinsic curvature $K_{ab}$ of $\Sigma$ as a
Lie derivative (c.f. appendix \ref{app:lie-derivative}) of the metric
in the direction of $\vec{n}$. 
This can again be shown to
depend on the 3-metric $h_{ab}$ and the shift functions in the following
manner \cite{wald:1984}:
\beql{K-h}
  K_{ab} = \12 \pounds_n h_{ab} 
         = \frac{1}{2N}(\dot{h}_{ab}-D_aN_b-D_bN_a),
\eeq
where $D_a$ denotes the covariant derivative associated with $h_{ab}$,
and the dot means differentiation with respect to the ``time''
parameter. 

The Gauss-Codazzi equation (c.f. appendix \ref{sec:curvature}) gives
the relation between intrinsic curvatures (the Ricci scalars) 
$R_S$ on $S$ and $R_\Sigma$ on $\Sigma$:
\beq
  R_S = R_\Sigma + K_{ab}K^{ab} - K^2,
\eeq
where $K\equiv h^{ab}K_{ab}$ and $K^{ab}=h^{ac}h^{bd}K_{cd}$.
From equation \refeq{new-vars} we find 
\beq
  \sqrt{-g}=N\sqrt{h}.
\eeq
Put together this means that we can rewrite the action
\refeq{hilbert} to
\beql{ADMaction}
  S=\frac{1}{\kappa}\int dx^0 d^3x 
	N\sqrt{h}(R + K_{ab}K^{ab} - K^2),
\eeq
where $R$ from now on refers to the intrinsic curvature on the
hypersurface, i.e. $R=R_\Sigma$.

We may now derive the canonical conjugate momenta:
\begin{eqnarray}
  P & \equiv & \frac{\del L}{\del \dot{N}} = 0,
  \\
  P^a & \equiv & \frac{\del L}{\del \dot{N_a}} = 0,
  \\ \label{eq:gravitation-momentum}
  \pi^{ab} & \equiv & \frac{\del L}{\del\dot{h}_{ab}} 
        = \frac{1}{\kappa}\sqrt{h}(K^{ab} - h^{ab}K).
\end{eqnarray}
$P$ and $P^a$ are identically zero, which means that $N$ and $N_a$ are 
non-dynamical variables. They play the role of Lagrange multipliers, as 
we will see explicitly later. 

Equation \refeq{gravitation-momentum} can be inverted to give an 
expression for $K_{ab}$ and $\dot{h}_{ab}$ (by use of \refeq{K-h}),
\begin{eqnarray}
  K_{ab} &=&  \frac{\kappa}{\sqrt{h}}
	(\pi_{ab} - \12 h_{ab}\pi),
  \\
  \dot{h}_{ab} &=& \frac{2N\kappa}{\sqrt{h}}
	(\pi_{ab} - \12 h_{ab}\pi)+D_aN_b+D_bD_a; 
	\quad \pi \equiv h^{ab}\pi_{ab},
\end{eqnarray}
which is easy to verify. The fact that we can obtain $\dot{h}_{ab}$ 
from $\pi_{ab}$
means that we do not have to search for constraints. This situation is very
similar to what was found for the Polyakov form of the string in section 
\ref{sec:weylstring}.

The Hamiltonian is now straight forward to calculate:
\begin{eqnarray}
  \nonumber
  H & \equiv & \pi^{ab}\dot{h}_{ab} - L
  \\
  & = & \tilde{N}\left[ \frac{1}{\sqrt{h}}(\pi^{ab}\pi_{ab} - \12\pi^2)
    -\frac{1}{\kappa^2}\sqrt{h}R \right] + 2\pi^{ab}D_aN_b,
%
%
\end{eqnarray}
where we have redefined $\tilde{N} \equiv \kappa N$. This is done to 
allow for the $\kappa\to\infty$ limit. 
A similar argument was made in section 
\ref{sec:weylstring} for the string case.

The last term in the Hamiltonian can be rewritten;
\beq
  2\pi^{ab}D_aN_b = 2\underbrace{D_a(\pi^{ab}N_b)}_{\to 0}
  - 2 N_bD_a\pi^{ab}.
\eeq
Again, the first term on the right hand side can be disregarded since
it is only a total derivative.
If we consider $\tilde{N}$ and $N_a$ as Lagrange 
multipliers we see that $H$ is just a sum of these constraints:
\begin{eqnarray}
  \Theta & = & \frac{1}{\sqrt{h}}(\pi^{ab}\pi_{ab} - \12\pi^2) 
           - \frac{1}{\kappa^2} \sqrt{h} R,
  \\
  \Theta^b & = & -2D_a\pi^{ab}.
\end{eqnarray}
Thus, we can write the Hamiltonian in this simple form
\beql{GRhamilton}
  H = \tilde{N} \Theta + N_b \Theta^b.
\eeq

\subsubsection{The limit $\kappa\to\infty$}

We see at once that we may take this limit simply by dropping the last
term in $\Theta$. 
And as mentioned in the beginning, this has the same effect as taking
the zero signature limit $\varepsilon \to 0$. The signature
$\varepsilon=\vec{n}\cdot\vec{n}$ of the spacetime metric only
influences on this term, and enters in such a way that taking
$\varepsilon \to 0$ removes the term proportional to $\sqrt{h}R$
\cite{teitelboim:1980}.

This limit has in turn been shown by Henneaux
\cite{henneaux:1979} to correspond to the four-dimensional action
\beql{e0GRaction}
  S= \int d^4x \Omega(x) ({\cal K}^{\alpha\beta}{\cal K}_{\alpha\beta}
     - {\cal K}^2); \quad \alpha,\beta=0,1,2,3.
\eeq
He does so by shoving that this action gives the same Hamilton 
formulation \refeq{GRhamilton} as the $\varepsilon\to 0$ limit of the
general relativity action.

The independent fields in the action \refeq{e0GRaction} are the
positive scalar density $\Omega(x)$ and the components of a symmetric
covariant tensor $g_{\alpha\beta}(x)$.
This ``metric'' $g_{\alpha\beta}$ is degenerate, i.e. 
$\det(g_{\alpha\beta})=0$, which means that it has only 9 independent
components. Together with $\Omega$ this gives 10 independent
fields, which is the same number as in the original action.

${\cal K}_{\alpha\beta}$ is the second fundamental tensor, defined as the 
Lie derivative in a unique direction $\vec{e}$,
\beq
{\cal K}_{\alpha\beta} = \12 \pounds_e g_{\alpha\beta}
  =\12 (e^\gamma g_{\alpha\beta,\gamma} 
         +e^\gamma_{,\alpha} g_{\gamma\beta}
         +e^\gamma_{,\beta} g_{\alpha\gamma}).
\eeq
And the vector field $\vec{e}$ is defined through
\beq
  {\cal G}^{\mu\nu} = \Omega^2e^\mu e^\nu; \qquad
  {\cal G}^{\alpha\beta} \equiv 
	\frac{1}{3!}\varepsilon^{\alpha\lambda\mu\nu}
	\varepsilon^{\beta\rho\sigma\tau}
	g_{\lambda\rho}g_{\mu\sigma}g_{\nu\tau},
\eeq
where ${\cal G}^{\mu\nu}$ is called the \emph{minor} of
$g_{\mu\nu}$. The vector $\vec{e}$ is completely determined from
$g_{\alpha\beta}$ and $\Omega$. 
It satisfies $g_{\alpha\beta}e^\beta=0$ and is thus orthogonal
to any other vector $v^\alpha$, i.e.  
$g_{\alpha\beta}e^\alpha v^\beta = 0$.

Since the metric is degenerate (i.e. the determinant vanishes), there
is no inverse $g^{\alpha\beta}$ satisfying
$g^{\alpha\beta}g_{\beta\gamma}=\delta^\alpha_\gamma$. However, the
class of symmetric tensors $G^{\alpha\beta}$ defined by
\beq
  G^{\alpha\beta}g_{\beta\gamma} = \delta^\alpha_\gamma -
	\theta_\gamma e^\alpha,
\eeq
where $\theta_\gamma$ is an arbitrary vector satisfying 
$\theta_\gamma e^\gamma=1$, can instead serve as an ``index raiser'' which
makes ${\cal K} = G^{\alpha\beta}{\cal K}_{\alpha\beta}$ and
${\cal K}^{\alpha\beta}{\cal K}_{\alpha\beta}$
well defined. 

For an elaboration of the ideas presented here, the reader should
consult \cite{henneaux:1979}.

\air
The conclusion of this section is that we \emph{are} able to find an
interesting limit by use of the phase space method. This is not as
straight forward as in the string cases, but by introduction of
the more suitable ADM coordinates, it can be done.

The most troublesome part has turned out to be the way back from phase space
to configuration space. This could not be done easily by elimination
of momenta and Lagrange multipliers as before. Instead, we adopted the
idea of first proposing a configuration space action, and then
showing that it gives the right Hamiltonian. 

Similar to the string models, we found the limit to correspond to a
degenerate geometry. In the present case, this means a non-Riemannian
space halfway between Euclidean and Minkowski space,
which corresponds to a theory of gravity based on local 
\emph{Caroll invariance}. 
(Normal gravitation is based on local Poincar\'e invariance.) 
The 10-parametric Caroll group was introduced by
Levy-Lebond \cite{levy-lebond:1965}.



%% file: ym.tex
We will now, very briefly, take a look on two models which are quite \
different from those we have studied thus far. The results we find
here will tell us something about the generality of the methods.  

\section{Yang-Mills theory}

In this section we will consider the simplest of all Yang-Mills
theories, namely Abelian $U(1)$ theory, which is the familiar Maxwell
theory of electrodynamics.

\subsection{The action}

The gauge field of electrodynamics is the electromagnetic potential,
denoted $A^\mu$. The electromagnetic field strength $F_{\mu\nu}$ can
then defined as 
$F_{\mu\nu} = \del_\mu A_\nu - \del_\nu A_\mu$.
And the action for free electrodynamics (i.e. without sources) is
\beql{ym-action}
  S = \frac{1}{g^2}\int d^4x 
	F^{\mu\nu}F_{\mu\nu}
    = \frac{1}{g^2}\int d^4x 
	\eta^{\mu\alpha}\eta^{\nu\beta}F_{\alpha\beta}F_{\mu\nu},
\eeq
where $g$ is the coupling constant, equal to the electric charge in
our case.
Note that this action is \emph{not} \emph{diff} invariant. (It is not
coupled to gravity.) 
The definition of $F_{\mu\nu}$ and a variation $\delta A^\mu$ 
together give the Maxwell equations.

From now on we will focus on the limit $g\to\infty$, which is a strong
coupling limit relevant at high energies. We want to find out whether
our methods can give any insight to this limit.

\paragraph{Method I} 
We can immediatly say that this method will not lead to anything
interesting, as the Lagrangian ${\cal L}=F^2$ in the action
\refeq{ym-action} does not have the proper form.  

%
%

\subsection{Method II}

To see the time dependence of the Lagrangian more explicitly, we rewrite
it (keep in mind that we use the Minkowski metric
$\eta^{\mu\nu}=(-,+,+,+)$),
\begin{eqnarray}
  \nonumber
  L&=& \frac{1}{g^2}\eta^{\mu\alpha}\eta^{\nu\beta}
	F_{\alpha\beta}F_{\mu\nu}
  \\
  &=& \frac{1}{g^2}\left(
	-2F_{0i}F_{0i} + F_{ij}F_{ij}\right); \quad
	i,j=1,2,3,
\end{eqnarray}
where repeated indices are to be summed over. 
The momenta are now found to be
\begin{eqnarray}
  P^0&\equiv& \frac{\del L}{\del(\del_0A_0)} = 0,
  \\ \label{eq:ym-momentum}
  P^i&\equiv& \frac{\del L}{\del(\del_0A_i)} =-\frac{4}{g^2}F_{0i}.
\end{eqnarray}
The first equation says that $A_0$ is a non-dynamical variable to be
treated as a Lagrange multiplier (which enforces Gauss' law). 
The dynamical variables are $A_i$,
and equation \refeq{ym-momentum} can be inverted to give an expression
for $\del_0A_i$:
\beq
  \del_0A_i = -\frac{g^2}{4}P^i + \del_iA_0.
\eeq 
The naive Hamiltonian is
\begin{eqnarray}
  \nonumber
  H&\equiv& P^i\del_0A_i - \frac{1}{g^2}F^{\mu\nu}F_{\mu\nu}
  \\
  &=& P^i\del_iA_0 - \frac{g^2}{8}P_iP^i - \frac{1}{g^2}F_{ij}F_{ij}.
\end{eqnarray}
This gives the phase space action
\beq
  S^{PS}=\int d^4x \left[
	P^iF_{0i} + \frac{g^2}{8}P^iP^i + \frac{1}{g^2}F_{ij}F_{ij}
	\right].
\eeq
Elimination of the momenta $P^i$ will of course give us the original
action, since we have no constraints.

The coupling constant $g$ appears both in the numerator and in the
denominator, which means that the $g\to\infty$ limit is not well
defined unless we can use a ``trick'' like in section
\ref{sec:weylstring} for eliminating $g$ in the numerator. However, in
the present case there is no Lagrange multiplier that can ``swallow''
the constant.
The conclusion is that we cannot use the same method to
define a $g\to\infty$ limit as we have applied earlier.

The assumption that the YM theory is Abelian is not essential.
For non-Abelian YM theories things will be more complicated,
but the problem of defining a $g\to\infty$ limit will certainly
remain. 

%% file: cs.tex
\section{Chern-Simons theory}

The calculations in this section are mainly based on 
\cite{kim:1998}

\subsection{The action}

The action for free Abelian Chern-Simons theory is
\beq
  S =  \frac{\kappa}{4}\int d^3x\epsilon_{\mu\nu\rho}A^\mu F^{\nu\rho}
	= \frac{\kappa}{2}\int d^3x \epsilon_{\mu\nu\rho}
	A^\mu\del^\nu A^\rho; \quad
	\mu,\nu,\rho = 0,1,2.
\eeq
Under diffeomorphisms $x^\mu\to\tilde{x}^\mu(x)$ we have the
transformations
\begin{eqnarray}
  d^3x &\to& d^3J^{-1}; \qquad\qquad J\equiv\det(\Lambda^a_b); 
	\quad \Lambda^a_b \equiv \frac{\del x^a}{\del\tilde{x}^b},
  \\
  \epsilon_{\mu\nu\rho}A^\mu\del^\nu A^\rho &\to& \epsilon_{\mu\nu\rho}
	\Lambda^\mu_\alpha\Lambda^\nu_\beta\Lambda^\rho_\gamma
	A^\alpha\del^\beta A^\gamma
	= J\epsilon_{\alpha\beta\gamma}A^\alpha\del^\beta A^\gamma,
\end{eqnarray}
which means that the action is \emph{diff} invariant. We may view the
Chern-Simons action as representing some kind of 3D gravitation.
The equations of motion found from a variation $\delta A^\alpha$ are
\beq
  \delta A^\alpha \Rarr \quad \epsilon_{\alpha\mu\nu}\del^\mu A^\nu=0.
\eeq

We now turn to the problem of finding a reasonable action describing the
$\kappa\to 0$ limit of this model. As for the Yang-Mills theory,
method I will fail to give anything interesting due to the form of the
action.
%
%

\subsection{Method II}
The canonical conjugate momenta to the fields $A^\mu$ are
\beq
  \Pi_\rho \equiv \frac{\del L}{\del(\del^0A^\rho)} 
	= \frac{\kappa}{2}\epsilon_{\mu 0\rho}A^\mu,
\eeq
which gives $\Pi_0=0$ and $\Pi_i=\frac{\kappa}{2}\epsilon_{ij}A^j$,
where $i,j$ are spatial indices.
This Legendre transformation from time derivatives to momenta is not
invertible, and we get the primary constraints:
\begin{eqnarray}
  \Theta_0 &\equiv& \Pi_0 \approx 0,
  \\
  \Theta_i &\equiv& \Pi_i - \frac{\kappa}{2}\epsilon_{ij}A^j \approx
  0; \quad i,j=1,2.
\end{eqnarray}

\air
The naive Hamiltonian is found to be
\begin{eqnarray}
  \nonumber
  H_{naive} &\equiv& \Pi_\rho\del^0A^\rho - \frac{\kappa}{2}
	\epsilon_{\mu\nu\rho}A^\mu\del^\nu A^\rho
	= -\frac{\kappa}{2}\epsilon_{\mu i\rho}A^\mu \del^i A^\rho
  \\
  &=&-\kappa A^0\epsilon_{ij}\del^i A^j,
\end{eqnarray}
where we have disregarded a total derivative. (Again, this expression
is in complete agreement with what we would get by using theorem
\ref{the:general} of section \ref{sec:zerohamilton}.)
The consistency conditions on the primary constraints are
\beq
  \{\Theta_\nu,H_{naive}\} + \lambda^\rho\{\Theta_\nu,\Theta_\rho\}
	\approx 0,
\eeq
and lead to
\begin{eqnarray}
  \kappa\epsilon_{ij}\del^i A^j &\approx& 0,
  \\
  \kappa\epsilon_{1j}\del^j A^0 - \lambda^2\kappa\epsilon_{12}
	& \approx & 0,
  \\
  \kappa\epsilon_{2j}\del^j A^0 + \lambda^1\kappa\epsilon_{12}
	&\approx & 0. 
\end{eqnarray}
We thus get one secondary constraint,
 \beq
  \Theta_3 = \kappa\epsilon_{ij}\del^i A^j,
\eeq
and the conditions on two of the Lagrange multipliers, 
$\lambda^i=\del^i A^0$. The consistency condition on the secondary
constraint $\Theta_3$ does not give any new constraints.

The total Hamiltonian can now be written
\beq
  H = -\kappa A^0\epsilon_{ij}\del^i A^j 
	+ \lambda^0\Pi_0
	+ \del^iA^0(\Pi_i - \frac{\kappa}{2}\epsilon_{ij}A^j)
	+ \lambda^3\kappa\epsilon_{ij}\del^i A^j,
\eeq
and the phase space action is found to be
\beq
  S^{PS}= \int d^3x \left[
	\Pi_i F^{0i} + \frac{\kappa}{2}\epsilon_{ij}A^0\del^i A^j
	+\Pi_0\del^0A^0-\Pi_0\lambda^0 
	- \lambda^3\kappa\epsilon_{ij}\del^i A^j \right].
\eeq

\air
Taking the $\kappa\to 0$ limit of this action gives
\beq
  S^{\kappa=0} = \int d^3x \left[
	\Pi_i F^{0i}
	+\Pi_0\del^0A^0-\Pi_0\lambda^0 \right].
\eeq
This action is linear in the momenta $\Pi_\mu$. Hence we cannot
eliminate the momenta and arrive at a sensible configuration space
action.  We must therefore conclude that this approach for studying
the $\kappa\to 0$ limit	does not work. 	

%% file: conclusion.tex

One aim of this thesis was to present two methods for deriving
tensionless limits of strings and the analogue in other models. This
has been done thoroughly by first presenting the most important
background theory in chapter \ref{chap:introduction}, and by going
through a series of examples in the subsequent chapters.
The reader will hopefully understand and be able to apply the ideas on
basis of this presentation.

One main result of the introductory chapter was the derivation of the
naive Hamiltonian for \emph{diff} invariant theories with only tensor 
fields. It was demonstrated that for such models the Hamiltonian is
constrained to be zero when the fields satisfy the field equations.

A second important goal of the thesis was to investigate how widely
applicable the methods
are. It was known from before that they work well for a number of
string models. The derivations in section \ref{sec:weylstring} and
\ref{sec:weylDbrane} revealed that the methods work perfectly also if
we start from the Weyl-invariant form of the bosonic string and
D-string actions. These are models that already at the very beginning
contain Lagrange multipliers (an auxiliary metric). And it is a noteworthy
result that the methods apply in such cases as well, which 
emphasizes their generality. 

The calculations for the rigid string also lead us to reasonable
actions for the tensionless limit. But in contrast the other string
models, we could not in this case use the two methods to derive the
\emph{same} form of the action. 

Applying the methods to general relativity turned out to be a much
more cumbersome task. Method I applied to the Eddington-Schr\"odinger
action, while we had to change to ADM coordinates for method II to
give expressions that we could handle in a reasonable way.

Finally the examples of chapter \ref{chap:othermodels} demonstrated
some of the limitations of the models. It was noted already in the
introduction that method I requires a specific form of the
Lagrangian. And when it comes to method II, it is of course not \emph{a
priori} given that it should work: We start with an action that is not
defined for the specific limit, and hope that through some
mathematical manipulations we can write an equivalent action that is
also defined in the limit. The most surprising is rather the fact that
it does indeed work so well for many theories.

\air
As general remarks we note that in the tensionless limit, the string
models naturally provide conformally invariant theories, and that the
geometries turn out to be degenerate.

\air
It would be interesting to know if there exist some crucial aspect
that determine whether method II works, and if so, \emph{what} this
is. It would also be of interest to know if it is possible to say
something general concerning the equality of method I and method II.
These are proposals for further investigation.

Another natural extension of this work, would be to look at
\emph{supersymmetric} models. This has been done in some extent
\cite{lindstrom:1991}, but still not very thoroughly.

%% file: densities.tex
\section{Tensor densities}
\label{sec:densities}
In this thesis we have not been very precise when talking about
tensors. We have often referred to quantities as being e.g. scalars
when they were really scalar \emph{densities}. The Lagrangian is one
example of this. The distinction has not been an important feature in
our calculations, but we will now give a brief description of the
difference between tensors and tensor densities.

For coordinate transformations $x^a \to \tilde{x}^a = \tilde{x}^a(x)$ we
define the Jacobi matrix
$J^a_b \equiv \frac{\del x^a}{\del \tilde{x}^b}$ 
and the Jacobi determinant 
$J\equiv \det(J^a_b)$.
The Jacobi matrix is the same as what we often call the
transformation matrix, denoted
$\Lambda^a_b=J^a_b$.

Scalars, vectors and second rank tensors are quantities that transform
as
\begin{eqnarray}
  \lefteqn{\mbox{scalar}} \hspace{4cm}
	&&\phi(x)~ \to \tilde{\phi}(\tilde{x})=\phi(x), \\
  \lefteqn{\mbox{vector}} \hspace{4cm}
	&&A_a(x) \to \tilde{A}_a(\tilde{x})=\Lambda^b_a A_b(x), \\
  \lefteqn{\mbox{2nd rank tensor}}\hspace{4cm}
	&&F_{ab}(x) \to \tilde{F}_{ab}(\tilde{x})
	=\Lambda^c_a\Lambda^d_b F_{cd}(x).\quad
\end{eqnarray}
On basis of this, we \emph{define} tensor densities to be quantities
that transform as tensors, but with an extra factor of the Jacobi
determinant. Thus, a tensor density \emph{of weight $n$} transforms as
\begin{eqnarray}
  \lefteqn{ \mbox{scalar density}} \hspace{4cm}
	&&\varphi(x)~ \to \tilde{\varphi}(\tilde{x})=J^n\varphi(x), \\
  \lefteqn{\mbox{vector density}} \hspace{4cm} 
	&&{\cal A}_a(x) \to \tilde{{\cal A}}_a(\tilde{x})
	=J^n\Lambda^b_a {\cal A}_b(x), \\
  \lefteqn{\mbox{2nd rank tensor density}} \hspace{4cm}
	&&{\cal F}_{ab}(x) \to \tilde{{\cal F}}_{ab}(\tilde{x})
	=J^n\Lambda^c_a\Lambda^d_b {\cal F}_{cd}(x).
\end{eqnarray}
Suppose we have the action $S=\int dx L(\phi^i)$. With a transformation
of the parameter $x$ the integral measure transforms as $dx\to
dxJ^{-1}$. For the action to be invariant, the Lagrangian $L$ must
then be a scalar density of weight 1 (i.e. $L\to JL$).

Other examples of scalar densities (of weight 1) are the square root
of the 
spacetime metric $\sqrt{-g}$ in general relativity, and the
square root of the induced metric $\sqrt{-\gamma}$, 
$\gamma_{ab}=\del_aX^\mu\del_bX_\mu$
(under world sheet reparameterizations, c.f. section \ref{sec:p-brane}).

%% file: determinants.tex
\section{Derivatives of determinants}
\label{sec:det-derivative}
Consider an $n\times n$ matrix $A_{ab}$. Write its inverse with upper
indices, i.e. $(A^{-1})_{ab} \equiv A^{ab}$ and
$A^{ab}A_{bc}=\delta^a_c$.
Introduce the set of matrices $(\tilde{A}_{ab})$, which are 
$(n-1)\times (n-1)$ 
matrices identical to $A$ except that row $a$ and column $b$ are
eliminated. (There are $n^2$ such matrices.)
The \emph{cofactor matrix} $C_{ab}$ is an 
$n\times n$
matrix which has as its elements the determinant of the 
$(\tilde{A}_{ab})$ matrices, i.e. $C_{ab}= \det(\tilde{A}_{ab})$.
Let us write, for short, $\det(A_{ab})=A$.
Then we can write
\begin{eqnarray}
  A^{ab} &=&(-1)^{a+b}\frac{C_{ba}}{A},
  \\
  A &=& \sum (\pm 1)\prod A_{ab} = (-1)^{a+c}\sum_{c=1}^n A_{ac} C_{ac}
	\qquad \mbox{independent of}~ a.
\end{eqnarray}
Then we easily see that 
$\frac{\del A}{\del A_{ab}} = (-1)^{a+b}C_{ab} = A A^{ba}$, which
gives the very useful results
\begin{eqnarray}
  \label{eq:det-deriv1}
  \frac{d}{dx} A &=& A A^{ab} \frac{dA_{ba}}{dx}
	= - A A_{ab} \frac{d A^{ba}}{dx},
  \\
  \label{eq:det-deriv2}
  \frac{d}{dx}\sqrt{-A} &=& \12 \sqrt{-A} A^{ab}\frac{dA_{ba}}{dx}
	= -\12 \sqrt{-A} A_{ab}\frac{dA^{ba}}{dx}.
\end{eqnarray}

%% file: vielbeins.tex
\section{On the vielbein formalism}
\label{sec:vielbeins}

Consider a D-dimensional manifold $S$, and suppose that there exist a
coordinate system $\{x^a\}$. Then we can introduce the coordinate
basis forms $\{\tilde{d}x^a\}$.  Any one-form on $S$ can then be described
by means of these basis forms, as $\tilde{V}=V_a\tilde{d}x^a$. We may
also define coordinate basis vectors, $\vec{u}_a\equiv
\frac{\del}{\del x^a}$, which then satisfy
$\tilde{d}x^a(\vec{u}_b)=\delta^a_b$.  

The metric tensor can then be written
\beql{metric-ab}
  \boldsymbol{g} = g_{ab}\tilde{d}x^a\otimes\tilde{d}x^b; 
	\qquad a,b=0,\dots,D-1.
\eeq
and we find
\beq
  \vec{u}_a\cdot\vec{u}_b = \boldsymbol{g}(\vec{u}_a,\vec{u}_b)
	=g_{cd}\tilde{d}x^c(\vec{u}_a)\tilde{d}x^d(\vec{u}_b)
	=g_{ab}.
\eeq

We know that, locally, a spacetime manifold behaves as flat space. So
it must be possible to introduce a new (position dependent) flat 
basis $\{\tilde{e}^A\}$, such that
\beql{metric-AB}
  \boldsymbol{g} = \eta_{AB}\tilde{e}^A\otimes\tilde{e}^B;
	\qquad A,B=0,\dots, D-1,	
\eeq
where $\eta_{AB}=diag(-,+,\dots,+)$ is the Minkowski metric.
Such a basis is often calleda \emph{tetrad} or an \emph{orthonormal
basis}.  
The new basis forms can be written as a linear combinations of the
coordinate basis forms, 
\beq
  \tilde{e}^A(x) = e_a^{~A}(x)\tilde{d}x^a,
\eeq
which defines the \emph{vielbein} $e_a^{~A}(x)$.
We see that the vielbein can be viewed as a transformation matrix
between the curved and flat bases. It can, however, not generally be
expressed as a Jacobi matrix.
Together with the equations \refeq{metric-ab} and \refeq{metric-AB},
this gives
\beq
  g_{ab} = \eta_{AB} e_a^{~A}e_b^{~B}.
\eeq
Writing $\det(g_{ab}) \equiv g$, $\det(e_a^{~A})\equiv e$ and
remembering that $\det(\eta_{AB})=-1$ this
gives immediately
\beq
  \sqrt{-g} = e.
\eeq

If we introduce the inverse vielbein $e_A^{~~a}$ satisfying
$e_A^{~~a}e_a^{~B}=\delta_A^B$ and $e_a^{~A}e_A^{~~b}=\delta_a^b$,
we can write $\tilde{d}x^a = e_A^{~~a}\tilde{e}^A$.
A one-form (or covariant tensors in general) can now be written either
in the coordinate or flat basis as $\tilde{V} = V_a\tilde{d}x^a =
V_A\tilde{e}^A$, and the components are related through the vielbeins:
\beq
  V_A = e_A^{~~b}V_b; \qquad
  V_a = e_a^{~B}V_B.
\eeq
Indices referring to the coordinate basis are often called \emph{Einstein
indices} (denoted here with small letters), while indices referring to
the flat basis are called \emph{Lorentz indices} (denoted with capital
letters).  

Vielbeins are often called \emph{vierbeins}, also in cases where they
are not 4-dimensional.
This formalism is particularly useful when one wants to consider
particles with spin in curved spaces.

%% file: lie.tex

\section{On Lie derivatives}
\label{app:lie-derivative}
The Lie derivative is in this thesis denoted $\pounds_{\vec n}$, and
represents a kind of generalized directional derivative. For a
more thorough presentation, the reader may consult \cite{wald:1984}.

The Lie derivative of a \emph{scalar} $f$ is defined as
\beq
  \pounds_{\vec n} f \equiv \vec{n}[f] = f_{,\mu} n^\mu,
\eeq
and gives the change in $f$ along $\vec{n}$. The Lie derivative of a
\emph{vector} $\vec{v}$ is
\beq
  \pounds_{\vec n} \vec{v} \equiv [\vec{n},\vec{v}] 
	= \nabla_{\vec{n}}\vec{v}
	- \nabla_{\vec{v}}\vec{n},
\eeq
and says something about how $\vec{v}$ is changed under a parallel
transport along $\vec{n}$. Here $\nabla$ may be any derivative operator.

In a coordinate basis $\{x^\alpha\}$ we may write the Lie derivative
of a \emph{1-form} field $\tilde{\sigma}$ and a \emph{second rank
tensor} $\boldsymbol{T}$ as
\begin{eqnarray}
  \pounds_{\vec{n}}\tilde{\sigma} &=& 
	(\sigma_{\alpha,\beta} n^\beta
	+\sigma_\beta n^\beta_{,\alpha})\tilde{d}x^\alpha,\\
  \pounds_{\vec{n}}\boldsymbol{T} &=&
	(T_{\alpha\beta,\mu}n^\mu
	+T_{\mu\beta}n^\mu_{,\alpha}
	+T_{\alpha\mu}n^\mu_{,\beta})
	\tilde{d}x^\alpha\otimes\tilde{d}x^\beta.
\end{eqnarray}  
It can be shown that this holds also if we exchange the partial
derivative ``$,$'' with some covariant derivative $\nabla$. Thus, we can write
in coordinate form
\beq
  \pounds_{\vec{n}}T_{\alpha\beta} = 
	(\nabla_\mu T_{\alpha\beta}) n^\mu
	+T_{\mu\beta}(\nabla_\alpha n^\mu)
	+T_{\alpha\mu}(\nabla_\beta n^\mu).
\eeq
If we let $\nabla$ be the covariant derivative associated with some
\emph{metric} $g_{\alpha\beta}$ (i.e. 
$\nabla_\mu n^\alpha = n^\alpha_{,\mu} + \{^\alpha_{\nu\mu}\}n^\nu$),
then the following result is not very difficult to prove:
\beq
  \pounds_{\vec{n}} g_{\alpha\beta}
	=\nabla_\alpha n_\beta + \nabla_\beta n_\alpha.
\eeq

%% file: curvature.tex
\section{The Gauss-Codazzi equation}
\label{sec:curvature}

The discussion here is mainly due to Wald \cite{wald:1984}.
Consider a $D$-dimensional manifold $S$ with a vector basis 
$\{\vec{e}_\mu; \mu=0,\dots,D-1\}$ and a corresponding form basis
$\{\tilde{\omega}^\mu\}$, such that 
$\tilde{\omega}^\mu(\vec{e}_\nu) = \delta^\mu_\nu$.
The metric tensor $\boldsymbol{g}$ on $S$ can then be written
\beq
  \boldsymbol{g}=g_{\mu\nu}\tilde{\omega}^\mu\otimes\tilde{\omega}^\nu;
	\qquad
	g_{\mu\nu} = \boldsymbol{g}(\vec{e}_\mu,\vec{e}_\nu)
	\equiv \vec{e}_\mu \cdot \vec{e}_\nu.
\eeq

\subsubsection{Induced metric}

Let $\Sigma$ be a $d$-dimensional hypersurface embedded in this
space. The basis vectors on this hypersurface are denoted 
$\{\vec{u}_a; a=0,\dots,d-1\}$, and  there are $d$ such vectors. 
The corresponding
basis forms are denoted $\{\tilde{w}^a\}$, and satisfy 
$\tilde{w}^a(\vec{u}_b)=\delta_b^a$.
The \emph{induced metric} ${\boldsymbol \gamma}$ on $\Sigma$ is then
\beq
  \boldsymbol{\gamma}=\gamma_{ab}\tilde{w}^a\otimes\tilde{w}^b;
	\qquad
	\gamma_{ab} = \boldsymbol{\gamma}(\vec{u}_a,\vec{u}_b)
	\equiv \vec{u}_a \cdot \vec{u}_b.
\eeq

Define $\{\vec{n}_i;i=d,\dots,D-1\}$ to be as set of $D-d$ 
orthonormal vectors perpendicular to 
$\Sigma$. This means $\vec{n}_i\cdot \vec{u}_a = 0$ and
$\vec{n}_i\cdot\vec{n}_j = \eta_{ij}$, where $\eta_{ij}$ is a diagonal
matrix with elements $\pm 1$, depending on
whether the the normal vectors are timelike ($-1$) or spacelike ($+1$). 

It is now evident that we can write any vector in $S$ as a linear
combination of $\vec{u}_a$ and $\vec{n}_i$. Especially, we have
$\vec{e}_\mu = \sigma_\mu^{~a}\vec{u}_a + \tau_\mu^{~i}\vec{n}_i$,
which gives
\beql{metric-decomp}
  g_{\mu\nu}=\vec{e}_\mu\cdot\vec{e}_\nu
	=\sigma_\mu^{~a}\sigma_\nu^{~b}\gamma_{ab}\
	+ \tau_\mu^{~i}\tau_\nu^{~j}\eta_{ij}
\eeq
and
\begin{eqnarray}
  \nonumber
  \vec{n}_i\cdot\vec{e}_\mu &=&\tau_\mu^{~j}\eta_{ji} \\
  \vec{u}_a\cdot\vec{e}_\mu &=&\sigma_\mu^{~b}\gamma_{ab}.
  \label{eq:decomp-1}
\end{eqnarray}
But we can also decompose by means of $\vec{e}_\mu$, and write 
$\vec{u}_a = u_a^{~\mu}\vec{e}_\mu$, 
$\vec{n}_i = n_i^{~\mu}\vec{e}_\mu$. With this decomposition we find
directly
\begin{eqnarray}
  \nonumber
  \vec{n}_i\cdot\vec{e}_\mu &=&n_i^{~\nu}g_{\nu\mu} = n_{i\mu} \\
  \vec{u}_a\cdot\vec{e}_\mu &=&u_a^{~\nu}g_{\nu\mu} = u_{a\mu}.
  \label{eq:decomp-2}
\end{eqnarray}
Equations \refeq{decomp-1} and \refeq{decomp-2} together give
$n_{i\mu}=\tau_\mu^{~j}\eta_{ji}$,
$u_{a\mu}=\sigma_\mu^{~b}\gamma_{ba}$. If we introduce $\eta^{ij}$ and
$\gamma^{ab}$ as the \emph{inverse} of $\eta_{ij}$ and $\gamma_{ab}$
respectively, we find
\beq
  \tau_\mu^{~i} = \eta^{ij}n_{j\mu} \equiv n^i_\mu; \qquad
  \sigma_\mu^{~a} = \gamma^{ab}u_{b\mu}.
\eeq
Put into the expression \refeq{metric-decomp} for $g_{\mu\nu}$, this
gives 
\beql{h-munu}
  g_{\mu\nu} = h_{\mu\nu} + \eta_{ij}n_\mu^i n_\nu^j; \qquad
	h_{\mu\nu}\equiv \gamma^{ab}u_{a\mu}u_{b\nu}.
\eeq
We will soon demonstrate that $h_{\mu\nu}$ is the induced metric.

We raise and lower Greek and Latin (middle alphabet) indices 
with $g_{\mu\nu}$ and $\eta_{ij}$, and their inverses $g^{\mu\nu}$ and
$\eta^{ij}$ respectively. Thus we have 
$h_{\mu\nu}=g_{\mu\nu}- n^i_{\mu} n_{i\nu}$,
$h^\alpha_\nu=\delta^\alpha_\nu - n^\alpha_i n^i_\nu$, and
$h^{\alpha\beta}=g^{\alpha\beta} - n^\alpha_i n^{i\beta}$.
Note that $h^{\mu\nu}$ is \emph{not} the inverse of $h_{\mu\nu}$.

We find immediately
\beql{h-in-Sigma}
  h_{\mu\nu}n_i^\nu = n_{i\mu} - n^j_\mu n_{j\nu} n_i^\nu = 0,
\eeq
which states that $\boldsymbol{h}$ is a tensor tangent to $\Sigma$.

Consider now an arbitrary tensor $\boldsymbol{T}$, which lies in
$\Sigma$. This means that it can be decomposed in two ways:
\beq
  \boldsymbol{T} = T_{\mu\nu}\tilde{\omega}^\mu \otimes\tilde{\omega}^\nu
	=T_{ab}\tilde{w}^a\otimes\tilde{w}^b.
\eeq
The relations between the components are
\beql{basis-change}
  T_{ab} = \boldsymbol{T}(\vec{u}_a,\vec{u}_b)
  = T_{\mu\nu}\tilde{\omega}^\mu(\vec{u}_a)\tilde{\omega}^\nu(\vec{u}_b)
  = T_{\mu\nu}u_a^{~\mu}u_b^{~\nu}.
\eeq
The tensor $\boldsymbol{h}$ is of this kind (i.e. tangent to $\Sigma$), 
so by use of \refeq{basis-change} and \refeq{h-munu} we can write
\beq
  h_{ab} = h_{\mu\nu}u_a^{~\mu}u_b^{~\nu}
	= \gamma^{cd}u_{c\mu}u_{d\nu}u_a^{~\mu}u_b^{~\nu}
	=\gamma^{cd}\gamma_{ac}\gamma_{bd} = \gamma_{ab}.
\eeq
This shows that $\boldsymbol{h}$ is indeed the induced metric, i.e. 
$\boldsymbol{h}=\boldsymbol{\gamma}$.

\air
Consider an arbitrary vector $\vec{v}=v^\mu \vec{e}_\mu$ in $S$.
Define 
$\vec{u} \equiv h^\mu_\nu v^\nu \vec{e}_\mu$, or in component form 
$u^\mu = h^\mu_\nu v^\nu=v^\mu - n_i^\mu n^i_\nu v^\nu$. 
Then we have
\beq
  \vec{u}\cdot\vec{n}_i = (v^\mu - n^j_\nu v^\nu n_j^\mu)\vec{e}_\mu
	\cdot \vec{e}_\rho n_i^\rho =0.
\eeq
In words, this means that the vector $\vec{u}$ is tangent to
$\Sigma$. Thus we may consider $h^\mu_\nu$ as a projection operator
from $S$ to $\Sigma$.

\air
If we have coordinate bases $\{\vec{e}_\mu = \frac{\del}{\del
x^\mu}\}$ on $S$ and $\{\vec{u}_a=\frac{\del}{\del\xi^a}\}$ on
$\Sigma$, we can write
\beq
  \vec{u}_a = \frac{\del}{\del\xi^a} 
	= \frac{\del x^\mu}{\del\xi^a}\frac{\del}{\del x^\mu}
	= \frac{\del x^\mu}{\del\xi^a}\vec{e}_\mu
	\equiv u_a^{~\mu}\vec{e}_\mu,
\eeq
which gives
\beq
  \gamma_{ab} = \vec{u}_a\cdot\vec{u}_b
	= u_a^{~\mu}u_b^{~\nu}g_{\mu\nu}
	=\frac{\del x^\mu}{\del\xi^a}\frac{\del
		x^\nu}{\del\xi^b}g_{\mu\nu}.
\eeq
This is the expression we use for the induced metric on the string
worldsheet.

\subsubsection{Extrinsic curvature}

We may define the \emph{extrinsic curvature} of the hypersurface $\Sigma$
as the Lie derivative of 
the metric in directions normal to $\Sigma$, i.e. along
$\vec{n}_i$. In mathematical terms,
\begin{eqnarray}
  \nonumber
  K_{\mu\nu}^i &\equiv& \12 \pounds_{\vec{n}_{i}}g_{\mu\nu}
	= \12\pounds_{\vec{n}_i}h_{\mu\nu}
  \\ \nonumber
  &=& \12(\nabla_\mu n^i_\nu + \nabla_\nu n^i_\mu)
  \\ 	
  &=& h_\mu^\rho\nabla_\rho n^i_\nu,
\end{eqnarray}
where $\nabla$ is the covariant derivative associated with
$g_{\mu\nu}$. From the above, we see that the extrinsic curvature is
symmetric, i.e. $K^i_{\mu\nu}=K^i_{\nu\mu}$, and, owing to the
projection factor $h_\mu^\rho$, tangent to $\Sigma$, i.e.
$K^i_{\mu\nu}n_j^\nu = 0$.

\subsubsection{Riemann curvature}

We now restrict to cases where $d=D-1$, i.e. there is only one normal
direction to the hypersurface $\Sigma$. The ``Minkowski metric''
$\eta_{ij}$ is then replaced by $\eta$, which is still $+1$ if the
normal vector is spacelike, and $-1$ if it is timelike.

The Riemann curvature tensor of $(S,g)$ is by definition given by
\beq
  R^\mu_{~\nu\alpha\beta}\omega_\mu 
	= \nabla_\alpha\nabla_\beta\omega_\nu
	-\nabla_\beta\nabla_\alpha\omega_\nu,
\eeq
where $\omega_\nu$ is any 1-form field on $S$. The corresponding
result for vectors $t^\mu$ is
\beq
  R^\mu_{~\nu\alpha\beta}t^\nu 
	= \nabla_\alpha\nabla_\beta t^\mu
	-\nabla_\beta\nabla_\alpha t^\mu.
\eeq
Similarly, if we let $\bar{\nabla}$ be the covariant derivative
associated with
$h_{\mu\nu}$, we can write the Riemann tensor of $(\Sigma,h)$ as
\beq
  \bar{R}^\mu_{~\nu\alpha\beta}\bar{\omega}_\mu
	=\bar{\nabla}_\alpha \bar{\nabla}_\beta\bar{\omega}_\nu
	-\bar{\nabla}_\beta \bar{\nabla}_\alpha\bar{\omega}_\nu,
\eeq
$\bar{\omega}_\nu$ now being a 1-form field on
$\Sigma$. (Quantities with a bar are all referring to $\Sigma$.)
The operator $\bar{\nabla}$ can be shown to be related to $\nabla$ in the
following simple way:
\beql{D-nabla}
  \bar{\nabla}_\alpha \bar{T}^{\mu\dots}_{~~~\nu\dots} = h^\mu_\rho\cdots
	 h_\nu^\sigma\cdots h_\alpha^\beta
	\nabla_\beta \bar{T}^{\rho\dots}_{~~~\sigma\dots},
\eeq
where $\bar{T}$ is some tensor on $\Sigma$. We define the Ricci 
tensor $R_{\mu\nu}$ and the Ricci scalar $R$ as
\begin{eqnarray}
  R_{\mu\nu} \equiv R^\alpha_{~\mu\alpha\nu};
	& & R \equiv g^{\mu\nu}R_{\mu\nu},
  \\
  \bar{R}_{\mu\nu} \equiv \bar{R}^\alpha_{~\mu\alpha\nu};
	& & \bar{R} \equiv g^{\mu\nu}\bar{R}_{\mu\nu}.
\end{eqnarray}
Let us now derive four useful relations.
\begin{eqnarray}
  \nonumber
  h_\alpha^\beta h_\mu^\nu \nabla_\beta  h_\nu^\rho 
	&=& h_\alpha^\beta h_\mu^\nu \nabla_\beta 
	(\delta_\nu^\rho -\eta n_\nu n^\rho)
  \\ \nonumber
  &=& -\eta h_\mu^\nu n^\rho \underbrace{h_\alpha^\beta\nabla_\beta n_\nu}
	_{=K^{\alpha\nu}}
	-\eta h_\alpha^\beta \underbrace{h_\mu^\nu n_\nu}_{=0}
	\nabla_\beta  n^\rho
  \\
  \label{eq:curv-1}
  &=&-\eta n^\rho h_\mu^\nu K_{\alpha\nu} 
	= -\eta n^\rho K_{\alpha\mu},
  \\[0.3cm]
  \nonumber
  h_\beta^\nu n^{\rho}\nabla_\nu\bar{\omega}_\rho
	&=& h_\beta^\nu\nabla_\nu(
	\underbrace{n^{\rho}\bar{\omega}_\rho}_{=0})
	- h_\beta^\nu\bar{\omega}_\rho \nabla_\nu n^{\rho}
	=-\bar{\omega}_\rho g^{\rho\mu}\underbrace{
	h_\beta^\nu\nabla_\nu n_\mu}_{K_{\beta\mu}}
  \\ 
  \label{eq:curv-2}
  &=&-g^{\rho\mu}K_{\beta\mu}\bar{\omega}_\rho 
	= -K_\beta^{\rho}\bar{\omega}_\rho,
  \\[0.3cm]
  \nonumber
  h^{\alpha\mu}h^{\beta\nu}R_{\alpha\beta\mu\nu} 
	&=& (g^{\alpha\mu}-\eta n^{\alpha} n^\mu)(g^{\beta\nu}-
	\eta n^{\beta} n^\nu) R_{\alpha\beta\mu\nu}
  \\
  \label{eq:curv-3}
  &=&R - 2\eta R_{\alpha\beta}n^\alpha n^{\beta},
  \\[0.3cm]
  \nonumber
   R_{\alpha\beta}n^{\alpha} n^\beta
	&=& R^\alpha_{~\beta\alpha\nu} n^{\beta} n^\nu
  \\ \nonumber
  &=& n^\nu (\nabla_\alpha \nabla_\nu 
	- \nabla_\nu \nabla_\alpha )n_\alpha
  \\ \nonumber
  &=& n^{\nu} \nabla_\alpha \nabla_\nu n^\alpha  
	- n^{\nu} \nabla_\nu\nabla_\alpha  n^\alpha 
  \\ \nonumber
  &=&\nabla_\alpha(n^{\nu}\nabla_\nu n^\alpha )
	- (\nabla_\alpha n^{\nu})(\nabla_\nu n^\alpha )
  \\[-0.1cm] \nonumber &&
     -\nabla_\nu(n^{\nu}\nabla_\alpha  n^\alpha )
	+ (\nabla_\nu n^{\nu})(\nabla_\alpha n^\alpha)
  \\[0.1cm]
  \label{eq:curv-4}
  &=& (K_\nu^{\nu})^2 - K_{\alpha\nu} K^{\alpha\nu}
	\underbrace{ 
	-\nabla_\nu(n^{\nu}\nabla_\alpha  n^\alpha )
	+\nabla_\alpha(n^{\nu}\nabla_\nu n^\alpha )
	}_{=t.d.},
\end{eqnarray}
where $t.d.$ is an abbreviation for total divergence. It will vanish
under integration.

Using the relation \refeq{D-nabla} between $\bar{\nabla}$ and $\nabla$ and
equation \refeq{curv-1}, we find
\beq
  \bar{\nabla}_\alpha\bar{\nabla}_\beta \bar{\omega}_\mu 
	=h_\alpha^\sigma h_\beta^\nu h_\mu^\rho
	\nabla_\sigma \nabla_\nu\bar{\omega}_\rho
	-\eta h_\mu^\rho K_{\alpha\beta} n^\nu \nabla_\nu\bar{\omega}_\rho
	-\eta h_\beta^\nu K_{\alpha\mu}n^\rho \nabla_\nu\bar{\omega}_\rho.
\eeq
Together with \refeq{curv-2} this gives
\beq
  \bar{R}^\alpha _{~\beta\mu\nu}= 
	h^\alpha_\sigma h^\tau_\beta h^\lambda_\mu h^\rho_\nu 
	R^\sigma_{~\tau\lambda\rho}
	-\eta K_{\mu\beta}K_{\nu}^{~~\alpha} 
	+ \eta K_{\nu\beta}K_{\mu}^{~~\alpha}.
\eeq
This equation is known as the \emph{Gauss-Codazzi equation}.
Using the equations \refeq{curv-3} and \refeq{curv-4} 
we end up with
\beql{intr-extr}
  \bar{R} = R - \eta(K_\alpha^{~\alpha})^2 
	+ \eta K_{\alpha\beta} K^{\alpha\beta} +t.d.,
\eeq
which gives the general relationship between the intrinsic
curvature $\bar{R}$ on $\Sigma$ and the extrinsic curvature
$K_{\mu\nu}$. 

\air
We said that the extrinsic curvatures are tensors tangent to the space
$\Sigma$, which means that they can be expressed by means of basis
forms $\tilde{w}^a$ on $\Sigma$ instead of basis forms
$\tilde{\omega}^\mu$ on $S$.  In mathematical terms,
\begin{eqnarray}
  \nonumber
  \boldsymbol{K} = K_{\mu\nu}\tilde{\omega}^\mu\otimes\tilde{\omega}^\nu
	= K_{ab}\tilde{w}^a\otimes\tilde{w}^b;
  &\quad& \mu,\nu=0,\dots,D-1;
  \\[-0.2cm]&& a,b=0,\dots,d-1=D-2.
\end{eqnarray}
On $\Sigma$ we use the \emph{induced} metric $\gamma_{ab}=h_{ab}$ and
its inverse $\gamma^{ab}$ to raise and lower indices. Thus we have
\begin{eqnarray}
  \nonumber
  Tr(\boldsymbol{K}) &=& K_\mu^{~\mu} 
	= g^{\mu\nu}K_{\mu\nu} 
	= \gamma^{ab}K_{ab}, \\
  \nonumber
  Tr(\boldsymbol{K}^2) &=& \eta K_{\mu\nu} K^{\mu\nu}
	= \eta g^{\mu\alpha}g^{\nu\beta}K_{\mu\nu}K_{\alpha\beta}
	= \eta\gamma^{ac}\gamma^{bd}K_{ab}K_{cd}.
\end{eqnarray}
This means altogether that equation \refeq{intr-extr} takes the form
\beql{curv-final}
  R = \bar{R} + \eta(K_a^{~a})^2 - \eta K_{ab} K^{ab} +t.d.,
\eeq
connecting quantities that refer to $S$ on the left hand side, and
quantities that refer to $\Sigma$ on the right hand side.

\air
In general relativity, a natural split is to let $\Sigma$ be a
spacelike 3-dimensional hypersurface. Then the normal
direction $\vec{n}$ is timelike, i.e. $\eta=-1$, which gives
\beq
  R = \bar{R}-K^2+K_{ab}K^{ab} + t.d.
\eeq
This result is used in chapter \ref{chap:GR} while doing the
space-time split necessary for a Hamilton description of general
relativity. 

\air
The results above can be generalized to situations where the dimension
difference between $S$ and $\Sigma$ is more than one, so that there
are several normal directions $\vec{n}_i$. In those cases equation
\refeq{curv-final} takes the more general form
\beq
  R = \bar{R} + \eta_{ij}K_a^{i~a} K_b^{j~b} 
	- \eta_{ij} K^i_{ab} K^{jab} +t.d.,
\eeq
which is a result that we apply in chapter \ref{chap:rigidstring} 
when investigating the rigid string.

%